\documentclass[11pt]{article}

\usepackage{latexsym,amsmath,epsfig}

\newcounter{subequation}[equation]
\makeatletter

\def\bcite{\@ifnextchar [{\@tempswatrue\@bcitex}{\@tempswafalse\@bcitex[]}}
\def\@bcitex[#1]#2{\if@filesw\immediate\write\@auxout{\string\citation{#2}}\fi
  \let\@bcitea\@empty
  \@bcite{\@for\@bciteb:=#2\do
    {\@bcitea\def\@bcitea{,\penalty\@m\ }%
     \def\@tempa##1##2\@nil{\edef\@bciteb{\if##1\space##2\else##1##2\fi}}%
     \expandafter\@tempa\@bciteb\@nil
     \@ifundefined{b@\@bciteb}{{\reset@font\bf ?}\@warning
       {Citation `\@bciteb' on page \thepage \space undefined}}%
     \hbox{\csname b@\@bciteb\endcsname}}}{#1}}
\def\@bcite#1#2{{#1\if@tempswa , #2\fi}}

\def\thesubequation{\theequation\@alph\c@subequation}
\def\@subeqnnum{{\rm (\thesubequation)}}
\def\slabel#1{\@bsphack\if@filesw {\let\thepage\relax
   \xdef\@gtempa{\write\@auxout{\string
      \newlabel{#1}{{\thesubequation}{\thepage}}}}}\@gtempa
   \if@nobreak \ifvmode\nobreak\fi\fi\fi\@esphack}
\def\subeqnarray{\stepcounter{equation}
\let\@currentlabel=\theequation\global\c@subequation\@ne
\global\@eqnswtrue
\global\@eqcnt\z@\tabskip\@centering\let\\=\@subeqncr
$$\halign to \displaywidth\bgroup\@eqnsel\hskip\@centering
  $\displaystyle\tabskip\z@{##}$&\global\@eqcnt\@ne
  \hskip 2\arraycolsep \hfil${##}$\hfil
  &\global\@eqcnt\tw@ \hskip 2\arraycolsep
  $\displaystyle\tabskip\z@{##}$\hfil
   \tabskip\@centering&\llap{##}\tabskip\z@\cr}
\def\endsubeqnarray{\@@subeqncr\egroup
                     $$\global\@ignoretrue}
\def\@subeqncr{{\ifnum0=`}\fi\@ifstar{\global\@eqpen\@M
    \@ysubeqncr}{\global\@eqpen\interdisplaylinepenalty \@ysubeqncr}}
\def\@ysubeqncr{\@ifnextchar [{\@xsubeqncr}{\@xsubeqncr[\z@]}}
\def\@xsubeqncr[#1]{\ifnum0=`{\fi}\@@subeqncr
   \noalign{\penalty\@eqpen\vskip\jot\vskip #1\relax}}
\def\@@subeqncr{\let\@tempa\relax
    \ifcase\@eqcnt \def\@tempa{& & &}\or \def\@tempa{& &}
      \else \def\@tempa{&}\fi
     \@tempa \if@eqnsw\@subeqnnum\refstepcounter{subequation}\fi
     \global\@eqnswtrue\global\@eqcnt\z@\cr}
\let\@ssubeqncr=\@subeqncr
\@namedef{subeqnarray*}{\def\@subeqncr{\nonumber\@ssubeqncr}\subeqnarray}
\@namedef{endsubeqnarray*}{\global\advance\c@equation\m@ne%
                           \nonumber\endsubeqnarray}

\makeatother

\DeclareFontFamily{OT1}{rsfs10}{}
\DeclareFontShape{OT1}{rsfs10}{m}{n}{ <-> rsfs10 }{}
\DeclareMathAlphabet{\mathscript}{OT1}{rsfs10}{m}{n}

\numberwithin{equation}{section}

\newcommand{\ns}{\normalsize}
\newcommand{\pt}{\partial}

\newcommand{\be}{\begin{equation}}
\newcommand{\ee}{\end{equation}}
\newcommand{\nn}{\nonumber}
\newcommand{\bea}{\begin{eqnarray}}
\newcommand{\eea}{\end{eqnarray}}
\newcommand{\bsea}{\begin{subeqnarray}} 
\newcommand{\esea}{\end{subeqnarray}}

\newcommand{\tr}{\textrm{tr}}
\newcommand{\mbf}[1]{\mathbf{#1}}
\newcommand{\w}{\wedge}
\newcommand{\Ds}{\not\!\!D}
\newcommand{\CC}{{\bf C}}
\newcommand{\ZZ}{{\bf Z}}

\def\a{\alpha}
\def\b{\beta}
\def\g{\gamma}
\def\c{\chi}
\def\d{\delta}
\def\e{\epsilon}

\def\z{\psi}

\def\k{\kappa}
\def\l{\lambda}
\def\m{\mu}
\def\n{\nu}
\def\o{\omega}
\def\p{\pi}

\def\r{\rho}
\def\s{\sigma}

\def\t{\tau}

\def\x{\xi}
\def\z{\zeta}
\def\w{\wedge}
\def\D{\Delta}

\def\G{\Gamma}
\def\J{\Psi}

\def\O{\Omega}

\def\cA{{\cal A}}
\def\cB{{\cal B}}
\def\cF{{\cal F}}
\def\cH{{\cal H}}
\def\cM{{\cal M}}
\def\cN{{\cal N}}
\def\cK{{\cal K}}
\def\cC{{\cal C}}
\def\cP{{\cal P}}
\def\cL{{\cal L}}

\def\Ib{\bar{I}}
\def\Jb{\bar{J}}
\def\Kb{\bar{K}}
\def\Lb{\bar{L}}
\def\vb{\bar{v}}
\def\ub{\bar{u}}
\def\bbar{\bar{b}}
\def\xib{\bar{\xi}}

\def\iz{i_0}
\def\jz{j_0}
\def\kz{k_0}
\def\ih{\hat{\imath}}
\def\jh{\hat{\jmath}}
\def\kh{\hat{k}}

\def\a{\alpha}
\def\b{\beta}
\def\g{\gamma}
\def\c{\chi}
\def\d{\delta}
\def\e{\epsilon}

\def\z{\psi}

\def\k{\kappa}
\def\l{\lambda}
\def\m{\mu}
\def\n{\nu}
\def\o{\omega}
\def\p{\pi}

\def\r{\rho}
\def\s{\sigma}

\def\t{\tau}

\def\x{\xi}
\def\z{\zeta}
\def\w{\wedge}
\def\D{\Delta}

\def\G{\Gamma}
\def\J{\Psi}

\def\O{\Omega}

\def\cA{{\cal A}}
\def\cF{{\cal F}}
\def\cM{{\cal M}}
\def\cK{{\cal K}}
\def\cC{{\cal C}}
\def\cP{{\cal P}}
\def\cL{{\cal L}}
\def\Ib{\bar{I}}
\def\Jb{\bar{J}}
\def\Kb{\bar{K}}
\def\Lb{\bar{L}}
\def\vb{\bar{v}}
\def\ub{\bar{u}}
\def\xib{\bar{\xi}}

\def\a{\alpha}
\def\b{\beta}
\def\g{\gamma}
\def\c{\chi}
\def\d{\delta}
\def\e{\epsilon}

\def\z{\psi}
\def\k{\kappa}
\def\l{\lambda}
\def\m{\mu}
\def\n{\nu}
\def\o{\omega}
\def\p{\pi}

\def\r{\rho}
\def\s{\sigma}
\def\t{\tau}

\def\x{\xi}
\def\z{\zeta}

\def\D{\Delta}

\def\G{\Gamma}
\def\J{\Psi}

\def\O{\Omega}

\def\cA{{\cal A}}
\def\cB{{\cal B}}
\def\cC{{\cal C}}
\def\cF{{\cal F}}
\def\cH{{\cal H}}
\def\cL{{\cal L}}
\def\cK{{\cal K}}
\def\cM{{\cal M}}
\def\cN{{\cal N}}
\def\cO{{\cal O}}
\def\cP{{\cal P}}

\begin{document}


\begin{titlepage}

\title{
   \vspace{-1cm}
   \hfill{\ns UPR-848T\\}
   \hfill{\ns \\[.5cm]}
   {\LARGE $\cN=1$ Supersymmetric Vacua in Heterotic
    M--Theory}}
\author{ 
   Burt A.~Ovrut
      \setcounter{footnote}{3}\thanks{Lectures presented at the Asia Pacific
Center for Theoretical Physics Third Winter School on ``Duality in Fields and
Strings" from January 21-February 5, 1999, Cheju Island, Korea.}
   \\[0.5cm]
   {\ns Department of Physics, University of Pennsylvania} \\
   {\ns Philadelphia, PA 19104--6396, USA}}

\date{}

\maketitle

\begin{abstract}

In the first lecture, we derive the five--dimensional 
effective action of strongly coupled
heterotic string theory for the complete $(1,1)$ sector of the theory
by performing a reduction, on a Calabi--Yau three--fold, of M--theory
on $S^1/Z_2$. The correct effective theory
is a gauged version of five--dimensional $\cN=1$ supergravity
coupled to Abelian vector multiplets, the universal hypermultiplet and
four--dimensional boundary theories with gauge and gauge matter
fields. The supersymmetric ground state of the
theory is a multi--charged BPS three--brane domain wall, which we
construct in general. In this first
lecture, we assume the ``standard'' embedding of the spin connection into the
$E_{8}$ gauge connection on one orbifiold fixed plane. In the second lecture,
we generalize these results to ``non--standard'' embeddings. That is,
we allow for
general $E_8\times E_8$ gauge bundles and for the presence of five-branes.
The five-branes span the four-dimensional uncompactified
space and are wrapped on holomorphic curves in the Calabi--Yau manifold.
Properties of these ``non--perturbative''vacua, as well as 
of the resulting low-energy theories,
are discussed. Characteristic features of the
low-energy theory, such as the threshold corrections to the gauge kinetic
functions, are significantly modified due to the presence of the
five-branes, as compared to the case of standard or non-standard
embeddings without five-branes. In the last lecture,
we review the spectral cover formalism for constructing both $U(n)$ and
$SU(n)$ holomorphic vector bundles on elliptically fibered Calabi--Yau
three--folds which admit a section. We discuss the allowed bases of these
three--folds and show that physical constraints eliminate Enriques surfaces
from consideration. Restricting the structure group to $SU(n)$,
we derive, in detail, a set of rules for the construction of three-family
particle physics theories with phenomenologically relevant gauge groups. 
We illustrate these ideas by
constructing several explicit three-family non-perturbative vacua.

\end{abstract}

\thispagestyle{empty}

\end{titlepage}


\section*{Introduction:}

Heterotic M--theory, first discussed by Ho\v rava and Witten~\cite{hw1,hw2,w},
holds great promise as the starting point for phenomenological investigations of low
energy particle physics and cosmology. In several papers~\cite{losw,add1}, the
five--dimensional effective action of Ho\v rava--Witten theory has been
constructed by dimensional reduction of $D=11$, $\cN=1$ supergravity on
Calabi--Yau three--folds. The resulting theory was shown to admit a pair of
BPS three--branes located at the orbifold fixed planes 
as its minimal static vacuum state. In subsequent work~\cite{us}, we
explored ``non-perturbative'' vacua, which consist of an arbitrary number of
BPS threebranes in addition to the orbifold fixed planes. Within the context
of both the minimal and non-perturbative vacua, we discussed the need for, and
analyzed, non-standard embedding; that is, we do not require that the
Calabi--Yau spin connection be embedded in one of the $E_{8}$ gauge groups.
Finally, in a series of papers~\cite{don1,usnew,usnewnew}, 
the generic mathematical structure of
non-standard embeddings were presented within the context of holomorphic
vector bundles. In these papers, we gave rules for the construction of three
family particle physics models with realistic unification groups. In the
present lectures we review this work, discussing the five--dimensional effective
theory and its three--brane BPS solutions in Lecture 1, non-perturbative vacua
and non-standard embedding in Lecture 2 and, finally, in Lecture 3 we discuss
the mathematics and physics of holomorphic vector bundles. 


\section*{Lecture 1: Heterotic M--Theory in Five Dimensions}


One of the phenomenologically most promising corners of the M--theory
moduli space, in addition to the weakly coupled heterotic string, is
the point described at low-energy by eleven-dimensional supergravity
on the orbifold $S^1/Z_2$ due to Ho\v rava and
Witten~\cite{hw1,hw2}. This theory gives the strongly coupled limit of
the heterotic string with, in addition to the bulk supergravity, two
sets of $E_8$ gauge fields residing one on each of the two
ten--dimensional fixed hyperplanes of the orbifold. It has been
shown~\cite{w} that this theory has phenomenologically interesting
compactifications on deformed Calabi--Yau three--folds times the
orbifold to four dimensions. Matching the 11--dimensional Newton
constant $\k$, the Calabi--Yau volume and the orbifold radius to the
known values of the Newton constant and the grand unification coupling
and scale leads to an orbifold radius which is about an order of
magnitude or so larger than the two other scales~\cite{w,bd}. This
suggests that, near this ``physical'' point in moduli space, the
theory appears effectively five--dimensional in some intermediate
energy regime. 

In previous papers~\cite{losw, add1} we have derived this five--dimensional
effective theory for the first time by
directly reducing Ho\v rava--Witten theory on a Calabi--Yau three--fold.
This calculation included all $(1,1)$ moduli as well as the universal
hypermultiplet.
We showed that a non--zero mode of the antisymmetric tensor field
strength has to be included for a consistent reduction from eleven to
five dimensions and that the correct five--dimensional effective
theory of strongly coupled heterotic string is given by a gauged
version of five--dimensional supergravity. A reduction of pure
eleven--dimensional supergravity on a Calabi--Yau
three--fold~\cite{CYred}, on the other hand, leads to a non--gauged
version of five--dimensional supergravity. Therefore, while this
provides a consistent low--energy description of M--theory on a smooth
manifold, it is not the correct effective theory for M--theory on
$S^1/Z_2$. The necessary additions are chiral four--dimensional
boundary theories with potential terms for the bulk moduli and, most
importantly, the aforementioned non--zero mode, living solely in the
Calabi-Yau three--fold, which leads to the gauging of the bulk supergravity. As
pointed out in ref.~\cite{losw, add1}, this theory is the correct starting
point for strongly coupled heterotic particle phenomenology as well as
early universe cosmology. Moreover, we have
shown that contact with four--dimensional physics should not be made
using flat space--time but rather via a domain--wall solution as the
background configuration. This domain wall arises as a BPS state of
the five--dimensional theory~\cite{losw, add1} and its existence is
intimately tied to the gauging of the theory. A reduction to four
dimensions on this domain wall has been performed in~\cite{add1, paschos} to
lowest non--trivial order. The result agrees with ref.~\cite{low1}
where the complete four--dimensional effective action to that order
has been derived directly from eleven dimensions. 

Various other aspects of the Ho\v rava--Witten description of strongly coupled
heterotic string theory have been addressed in the literature such as
the structure of the four--dimensional effective action, its relation
to 10--dimensional weakly coupled heterotic string, gaugino condensation,
and anomaly cancelation~[\bcite{low1}\,--\,\bcite{bkl}]. Aspects of
five--dimensional physics motivated by Ho\v rava--Witten theory and related
to particle phenomenology have been discussed in
ref.~\cite{bd,noy,sharpe,pes}. In refs.~\cite{lo,benakli2,benakli,low4}
five--dimensional early universe M--theory cosmology have been
investigated. Recently, aspects of five--dimensional physics have also
been discussed in ref.~\cite{elpp}. 

In this first lecture, we review the work presented in
ref.~\cite{losw, add1}.
Our central result is to obtain the five--dimensional effective theory of
strongly coupled heterotic string for all $(1,1)$ moduli fields and the
universal hypermultiplet, and construct
its fundamental BPS domain wall three--brane solutions. We show that,
in the bulk, this theory is indeed a form of gauged supergravity. 

Let us now summarize our conventions. We will
consider eleven-dimensional spacetime compactified on a Calabi-Yau space $X$,
with the subsequent reduction down to four dimensions effectively provided by
a double-domain-wall background, corresponding to an $S^1/Z_2$ orbifold. We
use coordinates $x^{I}$ with indices $I,J,K,\ldots = 0,\ldots ,9,11$ to
parameterize the full 11--dimensional space $M_{11}$. Throughout this paper,
when we refer to orbifolds, we will work in the ``upstairs'' picture
with the orbifold $S^1/Z_2$ in the $x^{11}$--direction. We choose the range
$x^{11}\in [-\pi\rho ,\pi\rho ]$ with the endpoints being identified. The
$Z_2$ orbifold symmetry acts as $x^{11}\rightarrow -x^{11}$. There then exist
two ten--dimensional hyperplanes fixed under the $Z_2$ symmetry which we
denote by $M_{10}^{(n)}$, $n=1,2$. Locally, they are specified by the
conditions $x^{11}=0,\pi\rho$. Upon reduction on a Calabi--Yau space to
five dimensions they lead to four--dimensional fixed hyperplanes $M_4^{(n)}$.
Barred indices $\bar{I},\bar{J},\bar{K},\ldots = 0,\ldots ,9$ are used for the
ten--dimensional space orthogonal to the orbifold. 
Upon reduction on the Calabi-Yau space we have a five-dimensional spacetime
$M_5$ labeled by indices $\a ,\b ,\g ,\ldots  = 0,\ldots ,3,11$. The
orbifold fixed planes become four-dimensional with indices
$\m,\n,\rho,\ldots = 0,\ldots ,3$. We use indices $A,B,C,\ldots =
4,\ldots 9$ for the Calabi--Yau space. Holomorphic and anti--holomorphic
indices on the Calabi--Yau space are denoted by $a,b,c,\ldots$ and
$\bar{a},\bar{b},\bar{c},\ldots$, respectively. The harmonic $(1,1)$--forms of
the Calabi--Yau space on which we will concentrate throughout this paper
are indexed by $i,j,k,\ldots = 1,\ldots ,h^{1,1}$.

The 11-dimensional Dirac--matrices $\G^I$ with $\{\G^I,\G^J\}=2g^{IJ}$
are decomposed as $\G^I = \{\g^\a\otimes\l ,{\bf 1}\otimes\l^A\}$
where $\g^\a$ and $\l^A$ are the five-- and six--dimensional Dirac
matrices, respectively. Here, $\l$ is the chiral projection matrix in
six dimensions with $\l^2=1$. Spinors in eleven dimensions are
Majorana with 32 real components throughout the paper. In five
dimensions we use symplectic--real spinors~\cite{c0}. 
Fields will be required to have a definite behavior under the $Z_2$
orbifold symmetry in $D=11$. We demand a bosonic field $\Phi$ to be
even or odd; that is, $\Phi (x^{11})=\pm\Phi (-x^{11})$. For a spinor
$\Psi$ the condition is $\G_{11}\Psi (-x^{11})=\pm\Psi (x^{11})$ and, depending
on the sign, we also call the spinor even or odd. The projection to one of
the orbifold planes leads then to a ten--dimensional Majorana--Weyl spinor
with definite chirality. Similarly, in five dimensions, bosonic fields will be
either even or odd, and there is a corresponding orbifold condition on
spinors. 


\section{Eleven--Dimensional Supergravity on an Orbifold}


In this section we briefly review the formulation of the low--energy
effective action of strongly coupled heterotic string theory as
eleven--dimensional supergravity on the orbifold $S^1/Z_2$ due to
Ho\v{r}ava and Witten~\cite{hw1,hw2}. 

The bosonic part of the action is given by
\begin{equation}
\label{action}
   S = S_{\rm SG}+S_{\rm YM}
\end{equation}
where $S_{\rm SG}$ is the familiar 11--dimensional supergravity action 
\begin{equation}
 S_{\rm SG} = -\frac{1}{2\k^2}\int_{M^{11}}\sqrt{-g}\left[ 
                    R+\frac{1}{24}G_{IJKL}G^{IJKL}
           +\frac{\sqrt{2}}{1728}\e^{I_1...I_{11}}
               C_{I_1I_2I_3}G_{I_4...I_7}G_{I_8...I_{11}} \right]
 \label{SSG}
\end{equation}
and $S_{\rm YM}$ describes the two $E_8$ Yang--Mills theories on the
orbifold planes, explicitly given by~\footnote{We note that there is a
debate in the literature about the precise value of the Yang--Mills
coupling constant in terms of $\k$. While we quote the original
value~\cite{hw2,deA} the value found in ref.~\cite{conrad} is
smaller. In the second case, the coefficients in the Yang-Mills
action~\eqref{SYM} and the Bianchi identity~\eqref{Bianchi} should
both be multiplied by $2^{-1/3}$. This potential factor will not be
essential in the following discussion as it will simply lead to a
redefinition of the five--dimensional coupling constants. In the
following, we will give the necessary modifications where
appropriate.} 
\begin{multline}
   \label{SYM}
   S_{\rm YM} = - \frac{1}{8\pi\k^2}\left(\frac{\k}{4\pi}\right)^{2/3}
        \int_{M_{10}^{(1)}}\sqrt{-g}\;\left[
           \tr(F^{(1)})^2 - \frac{1}{2}\tr R^2\right] \\
        - \frac{1}{8\pi\k^2}\left(\frac{\k}{4\pi}\right)^{2/3}
           \int_{M_{10}^{(2)}}\sqrt{-g}\;\left[
               \tr(F^{(2)})^2 - \frac{1}{2}\tr R^2\right]\; .
\end{multline}
Here $F_{\bar{I}\bar{J}}^{(n)}$ are the two $E_8$ gauge field strengths and
$C_{IJK}$ is the 3--form with field strength
$G_{IJKL}=24\,\partial_{[I}C_{JKL]}$. The above action has to be supplemented
by the Bianchi identity
\begin{equation}
 (dG)_{11\bar{I}\bar{J}\bar{K}\bar{L}} = -\frac{1}{2\sqrt{2}\pi}
    \left(\frac{\k}{4\pi}\right)^{2/3} \left\{ 
       J^{(1)}\d (x^{11}) + J^{(2)}\d (x^{11}-\pi\r )
       \right\}_{\bar{I}\bar{J}\bar{K}\bar{L}} \label{Bianchi}
\end{equation}
where the sources are defined by 
\begin{equation}
 J^{(n)}
    = {\rm tr}F^{(n)}\wedge F^{(n)} 
      - \frac{1}{2}{\rm tr}R\wedge R \; .\label{J}
\end{equation}
Note that, in analogy with the weakly coupled case, the boundary $\tr R^2$
terms in eq.~\eqref{SYM} are required by supersymmetry as pointed out in
ref.~\cite{low1}. Under the $Z_2$ orbifold symmetry, the field components
$g_{\bar{I}\bar{J}}$, $g_{11,11}$, $C_{\bar{I}\bar{J}11}$ are even, while
$g_{\bar{I}11}$, $C_{\bar{I}\bar{J}\bar{K}}$ are odd. The above action
is complete to order $\k^{2/3}$ relative to the bulk. Corrections, however, 
will appear as higher--dimension operators at order $\k^{4/3}$.

The fermionic fields of the theory are the 11--dimensional gravitino
$\J_I$ and the two 10--dimensional Majorana--Weyl spinors $\c^{(n)}$, located
on the boundaries, one for each $E_8$ gauge group. The components
$\J_{\bar{I}}$ of the gravitino are even while $\J_{11}$ is odd. The gravitino
supersymmetry variation is given by
\begin{equation} 
 \d\J_I = D_I\eta +\frac{\sqrt{2}}{288}\left(\G_{IJKLM}-8g_{IJ}\G_{KLM}
          \right)G^{JKLM}\eta +\; \cdots \label{susy}\; ,
\end{equation}
where the dots indicate terms that involve fermion fields. The spinor $\eta$
in this variation is $Z_2$ even.

The appearance of the boundary source terms in the Bianchi identity
has a simple interpretation by analogy with the theory of
$D$-branes. It is well known that the $U(N)$ gauge fields describing
the theory of $N$ overlapping $Dp$-branes encode the charges for
lower-dimensional $D$-branes embedded in the $Dp$-branes. For
instance, the magnetic flux $\tr F$ couples to the $p-1$-form
Ramond-Ramond potential, so describes $D(p-2)$-brane charge. Higher
cohomology classes $\tr F\wedge \cdots \wedge F$ describe the
embedding of lower-dimensional branes. Furthermore, if the $Dp$-brane
is curved, then the cohomology classes of the tangent bundle also
contribute. For instance $\tr R\w R$ induces $D(p-4)$-brane charge. 
We recall that in eleven dimensions it is M five-branes which are
magnetic sources for $G_{IJKL}$. Thus we can interpret the magnetic
sources in the Bianchi identity~\eqref{Bianchi} as five-branes
embedded in the orbifold fixed planes.


\section{The Five--Dimensional Effective Theory}


As mentioned above, matching of scales suggests that strongly
coupled heterotic string theory appears effectively five--dimensional in
some intermediate energy range. In this section we derive the
five--dimensional effective theory in this regime obtained by a
compactification on a Calabi--Yau three--fold. We expect that this
should lead to a theory with bulk $\cN=1$ five-dimensional supersymmetry
and four-dimensional $\cN=1$ supersymmetry on the orbifold fixed
planes. As we will see, doing this compactification 
consistently requires the inclusion of non--zero modes for the field strength
of the anti--symmetric tensor field. These non--zero modes appear in the
purely internal Calabi--Yau part of the anti--symmetric tensor field and
correspond to harmonic $(1,1)$ forms on the Calabi--Yau three--fold.
Consequently, to capture the complete structure of these non--zero modes, we
will have to consider the full $(1,1)$ sector of the theory. We will not, 
however, explicitly include the $(2,1)$ sector as it is largely unaffected by
the specific structure of Ho\v{r}ava--Witten theory. Instead, we comment on
the additions necessary to incorporate this sector along the way.

To make contact with the compactifications to four-dimensions
discussed by Witten~\cite{w}, one embeds the
spin-connection of the Calabi--Yau manifold in the gauge connection of
one of the $E_8$ groups breaking it to $E_6$. This is the so-called standard
embedding and, in this lecture, we will restrict our discussion to it. 
In general, this implies
that there is a non-zero instanton number on one of the orbifold planes.
From the discussion of the previous section, this can be interpreted as
including five-branes living in the orbifold plane in the
compactification. It is this additional element to the
compactification which introduces the non--zero mode and leads to much
of the interesting structure of the five-dimensional theory. We note
that the presence of five-brane charge is really unavoidable. Even
without exciting instanton number, the curvature of the Calabi-Yau three--fold
leads to an induced magnetic charge in the Bianchi
identity~\eqref{Bianchi}, forcing us to include non-zero modes.

We wish to emphasize that the standard embedding is not in any way required by
heterotic M--theory. In fact, unlike the case of weakly coupled heterotic
string theory where the choice of the standard embedding greatly simplifies
the vacuum, in M--theory this embedding is somewhat unnatural and is not
singled out in any way. We employ it in this lecture only because of its
familiarity and the fact that it was used in refs.~\cite{hw1, hw2}. The
following two lectures will be devoted to generalizing heterotic M--theory to
``non-standard'' embeddings.

\subsection{Zero Modes}

Let us now explain the structure of the zero mode fields used in the reduction
to five dimensions. We begin with the bulk. The background space--time
manifold is $M_{11}=X\times S_1/Z_2\times M_4$, where $X$ is
a Calabi--Yau three--fold and $M_4$ is four--dimensional Minkowski space.
Reduction on such a background leads to eight preserved supercharges and,
hence, to minimal $\cN=1$ supergravity in five dimensions. Due to the
projection condition, this leads to four preserved supercharges on the
orbifold planes implying four--dimensional $\cN=1$ supersymmetry on those
planes. Including the zero modes, the metric is given by
\begin{equation}
 ds^2 = V^{-2/3}g_{\a\b}dx^\a dx^\b +g_{AB}dx^Adx^B \label{metric}
\end{equation}
where $g_{AB}$ is the metric of the Calabi--Yau space $X$. Its K\"ahler form
is defined by~\footnote{Note here that we choose the opposite sign convention
as in ref.~\cite{w,low1} to conform with the literature on Calabi--Yau
reduction of 11--dimensional supergravity and type II
theories~\cite{CYred,bcf}.} $\o_{a\bar{b}}=ig_{a\bar{b}}$ and can be expanded
in terms of the harmonic $(1,1)$--forms $\o_{iAB}$, $i=1,\cdots ,h^{1,1}$ as
\begin{equation}
 \o_{AB} = a^i\o_{iAB}\; .\label{ai_def}
\end{equation}
The coefficients $a^i=a^i(x^\a )$ are the $(1,1)$ moduli of the Calabi--Yau
space. The Calabi--Yau volume modulus $V=V(x^\a )$ is defined by
\begin{equation}
 V=\frac{1}{v}\int_X\sqrt{^6g}\label{V_def}
\end{equation}
where $^6g$ is the determinant of the Calabi--Yau metric $g_{AB}$. In
order to make $V$ dimensionless we have introduced a coordinate volume
$v$ in this definition which can be chosen for convenience. The modulus
$V$ then measures the Calabi--Yau volume in units of $v$. The factor
$V^{-2/3}$ in eq.~\eqref{metric} has been chosen such that the metric
$g_{\a\b}$ is the five--dimensional Einstein frame metric. Clearly $V$
is not independent of the $(1,1)$ moduli $a^i$ but it can be expressed as
\begin{equation}
 V = \frac{1}{6}\cK (a)\; ,\quad \cK (a) = d_{ijk}a^ia^ja^k
\end{equation}
where $\cK (a)$ is the K\"ahler potential and $d_{ijk}$ are the Calabi--Yau
intersection numbers. Their definition, along with a more detailed account
of Calabi--Yau geometry, can be found in appendix A of ref.~\cite{add1}.

Let us now turn to the zero modes of the antisymmetric tensor field.
We have the potentials and field strengths,
\bea
 C_{\a\b\g}&,\quad& G_{\a\b\g\d}\nn \\
 C_{\a AB}=\frac{1}{6}\cA_\a^i\o_{iAB}
           &,\quad&G_{\a\b AB}=\cF_{\a\b}^i\o_{iAB} \\
 C_{abc} =\frac{1}{6}\x\O_{abc}&,\quad&G_{\a abc}=X_\a\O_{abc}\nn \; .
 \label{Czero}
\eea
The five--dimensional fields are therefore an antisymmetric
tensor field $C_{\a\b\g}$ with field strength $G_{\a\b\g\d}$, $h^{1,1}$
vector fields $\cA_\a^i$ with field strengths $\cF_{\a\b}^i$ and a
complex scalar $\x$ with field strength $X_\a$ that arises from the harmonic
$(3,0)$ form denoted by $\O_{abc}$. In the bulk the relations
between those fields and their field strengths are simply
\bea
 G_{\a\b\g\d} &=& 24\,\partial_{[\a}C_{\b\g\d ]} \nn \\
 \cF_{\a\b}^i &=& \partial_\a\cA_\b^i-\partial_\b\cA_\a^i \\
 X_\a &=& \partial_\a\x\nn \; .
\eea
These relations, however, will receive corrections from the boundary
controlled by the 11--dimensional Bianchi identity~\eqref{Bianchi}. We will
derive the associated five--dimensional Bianchi identities later.

\vspace{0.4cm}

Next, we should set up the structure of the boundary fields. The starting
point is the standard embedding of the spin connection in
the first $E_8$ gauge group such that
\begin{equation}
 \tr F^{(1)}\w F^{(1)} = \tr R\w R\; .
\label{condition}
\end{equation}
As a result, we have an $E_6$ gauge field $A_\a^{(1)}$ with field strength
$F_{\m\n}^{(1)}$ on the first hyperplane and an $E_8$ gauge field $A_\m^{(2)}$
with field strength $F_{\m\n}^{(2)}$ on the second hyperplane. In addition,
there are $h^{1,1}$ gauge matter fields from the $(1,1)$ sector on the first
plane. They are specified by
\begin{equation}
 A_b^{(1)} = \bar{A}_b+{\o_{ib}}^cT_{cp}C^{ip}
\end{equation}
where $\bar{A}_b$ is the (embedded) spin connection. Furthermore,
$p,q,r,\ldots =1,\ldots ,27$ are indices in the fundamental ${\bf 27}$
representation of $E_6$ and $T_{ap}$ are the $({\bf 3},{\bf 27})$
generators of $E_8$ that arise in the decomposition under the subgroup
$SU(3)\times E_6$. Their complex conjugate is denoted by
$T^{ap}$. The $C^{ip}$ are $h^{1,1}$ complex scalars in the
${\bf 27}$ representation of $E_6$. Useful traces for these generators
are $\tr (T_{ap}T^{bq})=\d_a^b\d_p^q$ and
$\tr (T_{ap}T_{bq}T_{cr})=\O_{abc}f_{pqr}$ where $f_{pqr}$ is the totally
symmetric tensor that projects out the singlet in ${\bf 27}^3$.

\subsection{The Nonzero Mode}

So far, what we have considered is similar to a reduction of pure
11--dimensional supergravity on a Calabi--Yau space, as for example
performed in ref.~\cite{CYred}, with the addition of gauge and gauge
matter fields on the boundaries. An important difference arises, however,
because the standard embedding~\eqref{condition}, unlike in the case
of the weakly coupled heterotic string, no longer leads to vanishing
sources in the Bianchi identity~\eqref{Bianchi}. Instead, there is a
net five-brane charge, with opposite sources on each fixed plane,
proportional to $\pm\tr R\w R$. The nontrivial components of
the Bianchi identity~\eqref{Bianchi} are given by 
\begin{equation}
 (dG)_{11ABCD} = -\frac{1}{4\sqrt{2}\pi}
    \left(\frac{\k}{4\pi}\right)^{2/3} \left\{ 
       \d (x^{11}) - \d (x^{11}-\pi\r )\right\} (\tr R\w R)_{ABCD}\; .
 \label{Bianchi1}
\end{equation}
As a result, the components $G_{ABCD}$ and $G_{ABC11}$ of the antisymmetric
tensor field are nonvanishing. More precisely, the above equation has to be
solved along with the equation of motion. 
\begin{equation}
 D_IG^{IJKL} = 0\; . \label{Geom}
\end{equation}
(Note that the Chern--Simons contribution to the antisymmetric tensor
field equation of motion vanishes if $G_{ABCD}$ and $G_{ABC11}$ are
the only nonzero components of $G_{IJKL}$.) 
The general solution of these equations is quite complicated and has
been given in ref.~\cite{low1} as an expansion in Calabi--Yau harmonic
functions. For the present purpose of deriving a 
five--dimensional effective action, we are only interested in the zero
mode terms in this expansion because the heavy Calabi--Yau modes decouple
as a result of the consistent Kaluza-Klein truncation to $D=5$. 
To work out the zero mode part of the solution, we note
that $\tr R\w R$ is a $(2,2)$ form on the Calabi--Yau space (since the
only nonvanishing components of a Calabi--Yau curvature tensor are
$R_{a\bar{b}c\bar{d}}$). Let us, therefore, introduce a basis $\n^i$,
$i=1,\cdots ,h^{2,2}=h^{1,1}$ of harmonic $(2,2)$ forms and
corresponding four--cycles $C_i$ such that
\begin{equation}
 \int_X \o_i\w \n^j = \d_i^j\; ,\quad \int_{C_i}\n^j = \d_i^j\; .
\end{equation}
The zero mode part $\left.\tr R\w R\right|_0$ of the source can then be
expanded as
\begin{equation}
 \left.\tr R\w R\right|_0 = -8\sqrt{2}\p \left(\frac{4\p}{\k}
                             \right)^{2/3}v\; \a_i\,\n^i \label{expans}
\end{equation}
where the numerical factor has been included for convenience. The
expansion coefficients $\a_i$ are
\begin{equation}
 \a_i =
 -\frac{\pi}{\sqrt{2}}\left(\frac{\k}{4\p}\right)^{2/3}\frac{1}{v^{2/3}}
 \b_i\; ,\qquad \b_i=-\frac{1}{8\pi^2}\int_{C_i}\tr R\w R \; .
 \label{alpha_def}
\end{equation}
Note that $\b_i$ are integers, characterizing the first Pontrjagin
class of the Calabi-Yau. It is then straightforward to see that the
zero mode part of the Bianchi identity~\eqref{Bianchi1} and the
equation of motion~\eqref{Geom} are 
solved by
\bea
 \left. G_{ABCD}\right|_0 &=& \a_i\,\n^i_{ABCD}\,\e (x^{11})=\frac{1}{4V}\a^i\,
                 {\e_{ABCD}}^{EF}\,\o_{iEF}\,\e (x^{11})\label{nonzero} \\
 \left.G_{ABC11}\right|_0 &=& 0\; .
\eea
Here $\e (x^{11})$ is the step function which is $+1$ for positive
$x^{11}$ and $-1$ otherwise. The index of the coefficient $\a^i$ in the
second part of the first equation has been raised using the metric
\begin{equation}
 G_{ij}(a) = \frac{1}{2V}\int_X\o_i\w (*\o_j) \label{CY_metric}
\end{equation}
on the $(1,1)$ moduli space. Note that, while the coefficients $\a_i$ with
lowered index are truly constants, as is apparent from
eq.~\eqref{alpha_def}, the coefficients $\a^i$ depend on the $(1,1)$
moduli $a^i$ since the metric~\eqref{CY_metric} does. From the
expansion~\eqref{expans} we can derive an expression for the boundary
$\tr F^2$ and $\tr R^2$ terms in the action~\eqref{SYM} which will be
essential for the reduction of the boundary theories. We have
\begin{equation} 
 \left.\tr R_{AB}R^{AB}\right|_0 = \left.\tr F^{(1)}_{AB}F^{(1)AB}\right|_0 =
  -4\sqrt{2}\p\left(\frac{4\p}{\k}\right)^{2/3}V^{-1}\a^i\o^{AB}\o_{iAB}
 \label{R2}
\end{equation}
while, of course
\begin{equation}
 \tr F^{(2)}_{AB}F^{(2)AB} = 0\; .
\end{equation}

The expression~\eqref{nonzero} for $G_{ABCD}$ with $\a_i$ as defined in
\eqref{alpha_def} is the new and somewhat unconventional ingredient in our
reduction. Using the terminology of ref.~\cite{gsw} we call this configuration
for the antisymmetric tensor field strength a nonzero mode. Generally,
a nonzero mode is defined as a nonzero internal antisymmetric tensor
field strength $G$ that solves the equation of motion. In contrast,
conventional zero modes of an antisymmetric tensor field, like those
in eq.~\eqref{Czero}, have vanishing field strength once the moduli fields
are set to constants. Since the kinetic term $G^2$ is positive for a nonzero
mode it corresponds to a nonzero energy configuration. Given that
nonzero modes, for a $p$--form field strength, satisfy
\begin{equation}
 dG=d^*G=0
\end{equation}
they correspond to harmonic forms of degree $p$. Hence, they can
be identified with the $p$th cohomology group $H^p(X)$ of the internal
manifold $X$. In the present case, we are dealing with a four--form
field strength on a Calabi--Yau three--fold $X$ so that the relevant
cohomology group is $H^4(X)$. The expression~\eqref{nonzero} is just an
expansion of the nonzero mode in terms of the basis $\{\n^i\}$ of
$H^4(X)$. The appearance of all harmonic $(2,2)$ forms shows that it is
necessary to include the complete $(1,1)$ sector into the low energy effective
action in order to fully describe the nonzero mode, as argued in the
beginning of this section. On the other hand, harmonic $(2,1)$ forms do not
appear here and are hence less important in our context. We stress that
the nonzero mode~\eqref{nonzero}, for a given Calabi--Yau space, specifies
a fixed element in $H^4(X)$ since the coefficients $\a_i$ are fixed in terms
of Calabi--Yau properties. In fact, they are related to the integers
$\b_i$ characterizing the first Pontrjagin class of the tangent
bundle. Thus we see that, correctly normalized, $G$ is in the integer
cohomology of the Calabi-Yau. This quantization condition has been
described in~\cite{wittq}

In a dimensional reduction of pure 11--dimensional supergravity, non--zero
modes can be considered as well but are usually dismissed as non--zero energy
configuration. Compactifications of 11--dimensional supergravity on various
manifolds including Calabi--Yau three-folds with non--zero modes have
been considered in the literature~\cite{llp}. The difference in our case is
that we are not free to turn off the non--zero mode. Its presence is
simply dictated by the nonvanishing boundary sources.

\subsection{The Five--Dimensional Effective Action}

Let us now summarize the field content which we have obtained above and
discuss how it fits into the multiplets of five--dimensional $\cN=1$
supergravity. The form of these multiplets and in particular the
conditions on the fermions is discussed in more detail
in appendix B of ref.~\cite{add1}. 
We know that the gravitational multiplet should contain
one vector field, the graviphoton. Thus since the reduction leads to
$h^{1,1}$ vectors, we must have $h^{1,1}-1$ vector multiplets. This
leaves us with the $h^{1,1}$ scalars $a^i$, the complex scalar $\xi$
and the three-form $C_{\a\b\g}$. Since there is one scalar in each
vector multiplet, we are left with three unaccounted for real scalars
(one from the set of $a^i$, and $\xi$) and the three-form. Together,
these fields form the ``universal hypermultiplet;'' universal because
it is present independently of the particular form of the
Calabi-Yau manifold. From this, it is clear that it must be the overall
volume breathing mode  $V=\frac{1}{6}d_{ijk}a^ia^ja^k$ that is the
additional scalar from the set of the $a^i$ which enters the universal
multiplet. The three-form may appear a little unusual, but one
should recall that in five dimensions a three-form is dual to a scalar
$\s$. Thus, the bosonic sector of the universal hypermultiplet consists
of the four scalars $(V,\s,\xi,\bar{\xi})$. 

The $h^{1,1}-1$ vector multiplet scalars are the remaining $a^i$. More
properly, since the breathing mode $V$ is already part of a
hypermultiplet it should be first scaled out when defining the shape
moduli 
\begin{equation}
 b^i=V^{-1/3}a^i\; .\label{bi_def}
\end{equation}
Note that the $h^{1,1}$ moduli $b^i$ represent only $h^{1,1}-1$
independent degrees of freedom as they satisfy the constraint
\begin{equation}
 \cK (b)\equiv d_{ijk}b^ib^jb^k =6\; .
\end{equation}
Alternatively, as described in appendix B of ref.~\cite{add1}, we can introduce
$h^{1,1}-1$ independent fields $\phi^x$ with $b^i=b^i(\phi^x)$. The
bosonic fields in the vector multiplets are then given by
$(\phi^x,b^x_i\cA^i_\a)$ ($b^x_i$ represents a projection onto the
$\phi^x$ subspace). Meanwhile the graviton and graviphoton of the
gravity multiplet are given by
$(g_{\a\b},\frac{2}{3}b_i\cA^i_\a)$. 

Therefore, in total, the five dimensional bulk theory contains a gravity
multiplet, the universal hypermultiplet and $h^{1,1}-1$ vector multiplets.
The inclusion of the $(2,1)$ sector of the Calabi--Yau space would lead
to an additional $h^{2,1}$ set of hypermultiplets in the theory. Since
they will not play a prominent r\^ole in our context they will not be
explicitly included in the following.

On the boundary $M^{(1)}_4$ we have an $E_6$ gauge multiplet
$(A_\m^{(1)},\c^{(1)})$ and $h^{1,1}$ chiral multiplets $(C^{ip},\eta^{ip})$
in the fundamental ${\bf 27}$ representation of $E_6$. Here $C^{ip}$ denote
the complex scalars and $\eta^{ip}$ the chiral fermions. The other boundary,
$M^{(2)}_4$, carries an $E_8$ gauge multiplet $(A_\m^{(2)},\c^{(2)})$ only.
Inclusion of the $(2,1)$ sector would add $h^{2,1}$ chiral multiplets in
the $\bf{\overline{\bf 27}}$ representation of $E_6$ to
the field content of the boundary $M_4^{(1)}$. Any even bulk field
will also survive on the boundary. Thus, in addition to the
four--dimensional part of the metric, the scalars $b^i$ together with
$\cA^i_{11}$, and $V$ and $\s$ survive on the boundaries. These pair
into $h^{1,1}$ chiral muliplets. 

After this survey we are ready to derive the bosonic part of the
five--dimensional effective action for the $(1,1)$ sector. Inserting the
expressions for the various fields from the previous subsection
into the action~\eqref{action}, using the formulae given in appendix A of
ref.~\cite{add1} and
dropping higher derivative terms we find
\begin{equation}
 S_5 = S_{\rm grav,vec}+S_{\rm hyper}+S_{\rm bound}+S_{\rm matter}
 \label{S5}
\end{equation}
with
\bsea
 S_{\rm grav,vec} &=& -\frac{1}{2\k_5^2}\int_{M_5}\sqrt{-g}\left[R+
                      G_{ij}\pt_\a b^i \pt^\a b^j +
                      \right.\nn \\
                   && \qquad\qquad\qquad\qquad \left.
                     G_{ij}\cF_{\a\b}^i\cF^{j\a\b}+\frac{\sqrt{2}}{12}
                      \e^{\a\b\g\d\e}d_{ijk}\cA_\a^i\cF_{\b\g}^j\cF_{\d\e}^k
                      \right]\\
 S_{\rm hyper} &=& -\frac{1}{2\k_5^2}\int_{M_5}\sqrt{-g}\left[
                   \frac{1}{2}V^{-2}\partial_\a V\partial^\a V
                   +2V^{-1}X_\a\bar{X}^\a
                   +\frac{1}{24}V^2G_{\a\b\g\d}G^{\a\b\g\d}
                   \right.\nn \\
                && \qquad\qquad\qquad\qquad
                    +\frac{\sqrt{2}}{24}\e^{\a\b\g\d\e}G_{\a\b\g\d}
                   \left(i(\x\bar{X}_\e-\bar{\x}X_\e )+
                   2\e (x^{11})\a_i\cA_\e^i\right)\nn\\
                &&\left.\qquad\qquad\qquad\qquad
                  +\frac{1}{2}V^{-2}G^{ij}\a_i\a_j\right]\qquad\\
 S_{\rm bound} &=& \frac{\sqrt{2}}{\k_5^2}\int_{M_4^{(1)}}\sqrt{-g}
                   \, V^{-1}\a_ib^i
                   -\frac{\sqrt{2}}{\k_5^2}\int_{M_4^{(2)}}\sqrt{-g}\,
                   V^{-1}\a_ib^i \\
 S_{\rm matter} &=& -\frac{1}{16\p\a_{\rm GUT}}
                   \sum_{n=1}^2\int_{M_4^{(n)}}\sqrt{-g}\, V\tr
                   {F_{\m\n}^{(n)}}^2\nn \\
                 && -\frac{1}{2\p\a_{\rm GUT}}
                    \int_{M_4^{(1)}}\sqrt{-g}\left[ G_{ij}(D_\m C)^i
                    (D^\m\bar{C})^j\right.\nn\\
                 &&\left.\qquad\qquad\qquad\qquad\qquad\qquad
                    +V^{-1}G^{ij}\frac{\partial W}
                    {\partial C^{ip}}\frac{\partial\bar{W}}
                    {\partial\bar{C}^j_p}+D^{(u)}D^{(u)}\right] \; .
                    \label{actparts}
\esea
All fields in this action that originate from the 11--dimensional
antisymmetric tensor field are subject to a nontrivial Bianchi identity.
Specifically, from eq.~\eqref{Bianchi} we have
\bsea
 (dG)_{11\m\n\r\s} &=& -\frac{\k_5^2}{4\sqrt{2}\pi\a_{\rm GUT}}\left\{ 
       J^{(1)} \d (x^{11})+ J^{(2)} \d (x^{11}-\pi\r )
       \right\}_{\m\n\r\s} \\
 (d\cF^i)_{11\m\n} &=& -\frac{\k_5^2}{4\sqrt{2}\pi\a_{\rm GUT}}J_{\m\n}^i \\
 (dX)_{11\m} &=& -\frac{\k_5^2}{4\sqrt{2}\pi\a_{\rm GUT}}J_\m \label{Bianchi5}
\esea
with the currents defined by
\bsea
 J_{\m\n\r\s}^{(n)}  &=& \left(\tr F^{(n)}\w F^{(n)}-\frac{1}{2}\tr R\w R
                         \right)_{\m\n\r\s} \\
 J_{\m\n}^i &=& -2iV^{-1}\G^i_{jk}\left((D_\m C)^{jp}(D_\n\bar{C})^k_p
                -(D_\m\bar{C})^k_p(D_\n C)^{jp}\right) \\
 J_\m &=& -\frac{i}{2}V^{-1}d_{ijk}f_{pqr}(D_\m C)^{ip}C^{jq}C^{kr}\; .
  \label{currents5}
\esea
The five--dimensional Newton constant $\k_5$ and the
Yang--Mills coupling $\a_{\rm GUT}$ are expressed in terms of
11--dimensional quantities as~\footnote{These relations are given
for the normalization of the 11--dimensional action as in eq.~\eqref{action}.
If instead the normalization of~\cite{conrad} is used the expression for 
$\a_{\rm GUT}$ gets rescaled to 
$a_{\rm GUT}=2^{1/3}\left(\k^2/2v\right)\left(4\p/\k\right)^{2/3}$ 
Otherwise the action and Bianchi identities are unchanged, except that in 
the expression~\eqref{alpha_def} for $\alpha_i$ the RHS is multiplied 
by $2^{1/3}$.}
\begin{equation}
 \k_5^2=\frac{\k^2}{v}\; ,\qquad \a_{\rm GUT} = \frac{\k^2}{2v}\left(
   \frac{4\p}{\k}\right)^{2/3}\; . \label{fdconst}
\end{equation}
We still need to define various quantities in the above action. The
metric $G_{ij}$ is given in terms of the K\"ahler potential $\cK$ as
\begin{equation}
 G_{ij}=-\frac{1}{2}\frac{\partial}{\partial b^i}\frac{\partial}
           {\partial b^j}\ln \cK\; .
\end{equation}
The corresponding connection $\G_{jk}^i$ is defined as
\begin{equation}
 \G_{jk}^i=\frac{1}{2}G^{il}\frac{\partial G_{jk}}{\partial b^l}\; .
\end{equation}
We recall that
\begin{equation}
 \cK = d_{ijk}b^ib^jb^k\; ,\label{kpot}
\end{equation}
where $d_{ijk}$ are the Calabi--Yau intersection numbers. All indices
$i,j,k,\cdots$ in the five--dimensional theory are raised and lowered with
the metric $G_{ij}$. A more explicit form of this metric can be found in
appendix A of ref.~\cite{add1}. We also recall that the fields $b^i$ are subject to the constraint
\begin{equation}
 \cK = 6 \label{b_cons}
\end{equation}
which should be taken into account when equations of motion are derived
from the above action. Most conveniently, it can be implemented by adding
a Lagrange multiplier term $\sqrt{-g}\l(\cK (b)-6)$ to the bulk action.
Furthermore, we need to define the superpotential
\begin{equation}
 W=\frac{1}{6}d_{ijk}f_{pqr}C^{ip}C^{jq}C^{kr}
\label{superpot}
\end{equation}
and the D--term
\begin{equation}
 D^{(u)} =G_{ij}\bar{C}^jT^{(u)}C^i
\label{Dterm}
\end{equation}
where $T^{(u)}$, $u=1,\ldots ,78$ are the $E_6$ generators in the fundamental
representation. The consistency of the above theory has been explicitly
checked by a reduction of the 11--dimensional equations of motion.

\vspace{0.4cm}

The most notable features of this action, at first sight, are the bulk
and boundary potentials for the $(1,1)$ moduli $V$ and $b^i$ that
appear in $S_{\rm hyper}$ and $S_{\rm bound}$. Those potentials involve
the five--brane charges $\a_i$, defined by eq.~\eqref{alpha_def}, that
characterize the nonzero mode. The bulk potential in the hypermultiplet
part of the action arises directly from the kinetic term $G^2$ of the
antisymmetric tensor field with the expression~\eqref{nonzero} for the
nonzero mode inserted. It can therefore be interpreted as the energy
contribution of the nonzero mode. The origin of the boundary potentials,
on the other hand, can be directly seen from eq.~\eqref{R2} and
the boundary actions~\eqref{SYM}. Essentially, they arise because the
standard embedding leads to nonvanishing internal boundary actions due
to the crucial factor $1/2$ in front of the $\tr R^2$ terms. This is in
complete analogy with the appearance of nonvanishing sources in the
internal part of the Bianchi identity which led us to introduce the
nonzero mode.


\section{Relation to Five-Dimensional Supergravity Theories}


As we have argued in the previous section, the five-dimensional effective
action~\eqref{S5} should have $\cN=1$, $D=5$ supersymmetry in the bulk
and $\cN=1$, $D=4$ supersymmetry on the boundary. In this section, we will
rewrite the action in a supersymmetric form. This will allow us to
complete the action~\eqref{actparts} to include fermionic terms and
give the supersymmetry transformations. One thing we will not do is
complete the supersymmetry transformations to include the bulk and
boundary couplings, but we assume a consistent completion is possible,
as in eleven dimensions. 

Of particular interest is the presence of potential terms
in the bulk theory. Such terms are forbidden unless the theory is
gauged; that is, unless some of the fields are charged under Abelian
gauge fields $\cA_\a$. In order to identify the supersymmetry
structure of the theory in hand, we derived, in appendix B of ref.~\cite{add1},
the general
form of gauged $D=5$, $\cN=1$ supergravity with charged hypermultiplets, borrowing
heavily from the work of G\"unaydin \textit{et al.}~\cite{GST1,GST2}
and Sierra~\cite{Sierra}, and from the general theory of gauged $D=4$,
$\cN=2$ supergravity as given, for instance, in~\cite{andetal}. 

Let us start by giving the $\cN=1$ structure of the four-dimensional
boundary theory. As discussed above, we have a set of chiral multiplets
with scalar components $C^{ip}$, together with vector multiplets with
gauge fields $A_\m^{(i)}$. (The vectors live on both boundaries, but
the chiral matter lives only on the $E_6$ boundary.) In addition, the
scalars from the bulk $(A,\s)$ and $(b^i,\cA^i_{11})$ also form chiral
multiplets. From the form of the theory on the boundaries we can give
explicitly the functions determining the $\cN=1$ theory. We have already
given the form of the superpotential and the $D$-term on the $E_6$
boundary in equations~\eqref{superpot} and~\eqref{Dterm}. It is also easy
to read off the K\"ahler potential on the $E_6$ boundary and the
gauge kinetic functions on either fixed plane. We find, without care
to correct normalizations,
\begin{equation}
   K = G_{ij}C^{ip}{\bar C}^i_p \qquad 
   f^{(n)} = V + i\s
\label{Kfs}
\end{equation}
The appearance of $\s$ in the gauge kinetic function is not
immediately apparent from the action~\eqref{S5}. However, it is easy
to show that on making the dualization of $C_{\a\b\g}$ to $\s$, which
is described in more detail below, the magnetic source in the Bianchi
indentity~\eqref{Bianchi5} for $C_{\a\b\g}$, becomes an electric
source for $\s$. The result is that the gauge kinetic terms in the
boundary action are modified to 
\begin{equation}
  -\frac{1}{16\p\a_{\rm GUT}} \sum_{n=1}^2 
        \int_{M_4^{(n)}}\sqrt{-g}\, \left[ 
          V \tr F_{\m\n}^{(n)}F^{(n)\m\n} 
          - \frac{\s}{2}\e^{\m\n\r\s}\tr F_{\m\n}^{(n)}\tr
             F_{\r\s}^{(n)} \right]
\end{equation}
One notes that the expressions~\eqref{Kfs} include dependence on the
bulk fields $b^i$ and $V$, evaluated on the appropriate
boundary. Further, we are considering the bulk multiplets as
parameters, as their dynamics comes from bulk kinetic terms. 

Now let us turn to the bulk theory. Our goal will be to identify the
action~\eqref{actparts} with the bosonic part of the general gauged
theory discussed in appendix B of ref.~\cite{add1}. The gauged theory is
characterized by a special Riemannian manifold $\cM_V$ describing the
vector multiplet sigma-model, a quaternionic manifold $\cM_H$
describing the hypermultiplet sigma-model, and a set of Killing vectors
and prepotentials on $\cM_H$. These are the structures we must
identify in the action~\eqref{actparts}.

We start by concentrating on the hypermultiplet structure. We have
argued that, after dualizing the three-form potential $C_{\a\b\g}$ to a
scalar $\s$, the fields $(V,\s,\xi,\xib)$ represent the scalar components of a
hypermultiplet. Concentrating on the kinetic terms let us make the
dualization explicit. We find
\begin{equation}
   G_{\a\b\g\d} = \frac{1}{\sqrt{2}}V^{-2}{\e_{\a\b\g\d}}^\e\left\{
      \pt_\e\s - i\left( \x\pt_\e\xib -\xib\pt_\e\x \right)
      - 2\e (x^{11})\a_i\cA^i_\e \right\} \; .
\end{equation}
The kinetic terms can then be written in the form
\begin{equation}
   h_{uv} D_\a q^u D^\a q^v 
\label{Hke}
\end{equation}
where $q^u=(V,\s,\x,\xib)^u$ and 
\begin{equation}
   D_\a q^u = \left(\pt_\a V,\, \pt_\a\s - 2\e (x^{11})\a_i\cA^i_\a,\, 
                  \pt_\a\x,\, \pt_\a\xib \right)^u
\label{covDer}
\end{equation}
and the metric is given by
\begin{equation}
   h_{uv} dq^u dq^v = \frac{1}{4V^2} dV^2 
       + \frac{1}{4V^2} \left[d\s+i(\x d\xib-\xib d\xi)\right]^2 
       + \frac{1}{V} d\xi d\xib \; .
\end{equation}
This reproduces the well-known result that the universal multiplet
classically parameterizes the quaternionic space
$\cM_H=SU(2,1)/U(2)$~\cite{quat}. 

In what follows, we would like to have an explicit realization of the
quaternionic structure of $\cM_H$. A review of quaternionic geometry
is given in appendix B of ref.~\cite{add1}. We will now give expressions for the
quantities defined there, following a discussion given
in~\cite{strom}. Since we have a single hypermultiplet, the holonomy
of $\cM_H$ should be $SU(2)\times Sp(2)=SU(2)\times SU(2)$. To
distinguish these, we will refer to the first factor as
$SU(2)$ and the second as $Sp(2)$. Defining the symplectic matrix
$\O_{ab}$ such that $\O_{12}=-1$, we have the vielbein 
\begin{equation}
   V^{Aa} = \frac{1}{\sqrt{2}}
       \left( \begin{array}{cc} 
          u & \vb \\ v & -\ub 
       \end{array} \right)^{Aa}
\end{equation}
where we have introduced the one-forms
\begin{equation}
   u = \frac{d\xi}{\sqrt{V}} \qquad 
   v = \frac{1}{2V} \left( dV + id\s + \x d\xib - \xib d\x \right)
\end{equation}
and their complex conjugates $\ub$ and $\vb$. We find that the $SU(2)$
connection is given by
\begin{equation}
   {\o^A}_B = {\left( \begin{array}{cc} 
          \frac{1}{4}(v-\vb) & -u \\ \ub & -\frac{1}{4}(v-\vb)
        \end{array} \right)^A}_B
\end{equation}
while the $Sp(2)$ connection is 
\begin{equation}
   {\D^a}_b = {\left( \begin{array}{cc}
          -\frac{3}{4}(v-\vb) & 0 \\ 0 & \frac{3}{4}(v-\vb) 
        \end{array} \right)^a}_b \; .
\end{equation}
The triplet of K\"ahler forms is given by 
\begin{equation}
   {K^A}_B = {\left( \begin{array}{cc}
           \frac{1}{2}(u \wedge \ub - v \wedge \vb)  &  u \wedge \vb \\
           v \wedge \ub  &  - \frac{1}{2}(u \wedge \ub - v \wedge \vb)
         \end{array} \right)^A}_B \; .
\end{equation}
With these definitions, one finds that the coset space
$SU(2,1)/U(2)$, satisfies the conditions for a
quaternionic manifold. 

So far our discussion has ignored the most important aspect of the
hypermultiplet sigma-model. We note that the kinetic terms
in~\eqref{Hke} were in terms of a modified derivative~\eqref{covDer},
which included the gauge fields $\cA^i_\a$. It appears that the
hypermultiplet is charged under a $U(1)$ symmetry. Comparing with our
discussion of gauged supergravity given in appendix B of ref.~\cite{add1}, 
we see that this
is indeed the case. The coset space $\cM_H$ admits an Abelian
isometry generated by the Killing vector
\begin{equation}
   k = \pt_\s = iV^{-1} \left( \pt_v - \pt_{\vb} \right) \; .
\end{equation}
In general, we can write the modified derivative~\eqref{covDer} in the
covariant form 
\begin{equation}
   D_\a q^u = \pt_\a q^u + g \cA^i_\a k_i^u
\end{equation}
with
\begin{equation}
   g k_i = -2\e (x^{11})\a_i k = -2i\e (x^{11})\a_i V^{-1} \left( \pt_v -
           \pt_{\vb} \right)
\; .
\end{equation}
(Note that the gauge coupling is absorbed in $\a_i$.) For consistency,
the $k^i_u$ should be writable in terms of a triplet of prepotentials. 
This is indeed the case and we find the
prepotentials 
\begin{equation}
   g{{\cP_i}^A}_B = {\left( \begin{array}{cc} 
          - \frac{1}{4}i\e (x^{11})\a_i V^{-1} & 0 \\ 0 &
          \frac{1}{4}i\e (x^{11})\a_i V^{-1}
        \end{array} \right)^A}_B \; .
\end{equation}
Thus it appears that the $\s$-component of the hypermultiplet is
charged under each Abelian gauge field $\cA_\a^i$, with a charge
proportional to $\a_i$. In particular, we can write the covariant
derivative as
\begin{equation}
   D_\a \s = 
     \pt_\a \s +\frac{1}{4\sqrt{2}\p}\left(\frac{4\p}{\k_5}\right)^{2/3}
     \a_{\rm GUT}\,\e (x^{11})\b_i\cA^i_\a
\end{equation}
where $\b_i$ are integers characterizing the first Pontrjagin class of
the Calabi-Yau. 

If this interpretation is correct, the rest of the action should
coincide with the general form for gauged supergravity given in
appendix B of ref.~\cite{add1}. 
It is clear that the vector multiplets are already in the
correct form. Comparing the bosonic action~\eqref{actparts} with the
general form given in (B.25) of ref.~\cite{add1}, 
we see that the gravitational and vector
kinetic terms exactly match. (In the appendix of ref.~\cite{add1}, we have set the
five-dimensional gravitational coupling $v/\k^2$ to unity.) The
structure of the metric $G_{ij}$ is identical, as is the appearance of
Chern-Simons couplings. The compactification gives an interpretation
of the numbers $d_{ijk}$ in the K\"ahler potential (B.7) of ref.~\cite{add1}
and~\eqref{kpot}. They are the Calabi-Yau intersection numbers. 

The final check of this identification is to calculate the form of
the potential. We have in general, from (B.29) of ref.~\cite{add1},
\begin{equation}
\begin{split}
   g^2V &= - 2 g^2 G_{ij} \tr \cP^i\cP^j + 4 g^2 b_i b_j \tr \cP^i\cP^j
          + \frac{g^2}{2} b^i b^j h_{uv} k^u_i k^v_j \\
     &= \frac{1}{4}V^{-2}G^{ij}\a_i\a_j \; ,
\end{split}
\end{equation}
exactly matching the derived potential. 

Thus, we can conclude that the bulk effective action is described by a
set of Abelian vector multiplets coupled to a single charged
hypermultiplet. The vector sigma-model manifold $\cM_V$ has the
general form described in appendix B of ref.~\cite{add1}, 
but now the $d_{ijk}$ in the
K\"ahler potential have the interpretation as Calabi-Yau
intersection numbers. The hypermultiplet manifold $\cM_H$ is the coset
space $SU(2,1)/U(2)$. A $U(1)$ isometry, corresponding to
the shift symmetry of the dualized three-form, is gauged. The charge
of the hypermultiplet scalar field under each Abelian vector field
$A_\a^i$ is given by $\a_i$. 

The appearance of gauged supergravity when non-zero modes are included
has been seen before in the context of type II compactifications on
Calabi-Yau manifolds to four-dimensions~\cite{sp,michelson}. It is
natural to ask why this gauging arises. The appearance of a
potential term is easy to interpret. We have included a non-zero
four-form field strength $G_{IJKL}$ on four-cycles of the
Calabi-Yau. These contribute an energy proportional to the square of
the field strength. For fixed total charge $\a_i$ (the integral of $G$
over a cycle), the energy is reduced the larger the four-cycle. Thus
it is no longer true that all points in Calabi-Yau moduli space have
the same energy. As an example we see that the potential naturally
drives the Calabi-Yau to large volume, minimizing the $G^2$ energy. 

From the five-dimensional point of view, once we have a potential
term, the theory must be gauged if it is to remain supersymmetric. We see
that it is the dual of the five-dimensional three-form which is
gauged. This arises because of the Chern-Simons term in eleven
dimensions. Turning on non-zero modes, this term acts as an electric
source for the five-dimensional three-form, though dependent on the
gauge fields $\cA^i$. Dualizing, the invariance $\s\to
\s+\mathrm{const}$ is a reflection of an absence of local electric
charge. Thus it not surprising that the effect of the electric
Chern-Simons terms is to modify this to a local gauge symmetry. We note
that from this argument it can only ever be the five-dimensional
three-form which becomes gauged by non-zero modes, whatever
particular compactification to a $\cN=1$ five-dimensional theory is
considered.

We end this section by giving the the specific form of the fermionic
supersymmetry variations. These are calculated using the general forms
given in (B.30),(B.31) and (B.32) of appendix B in
ref.~\cite{add1}, together
with the explicit expressions for the vielbein, connections, Killing
vectors and prepotentials given above. We find
\bsea
\d \psi_\a^A &=& \nabla_\a\e^A 
     + \frac{\sqrt{2}i}{8}
          \left({\g_\a}^{\b\g}-4\d_\a^\b\g^\g\right)b_i\cF_{\b\g}^i\e^A
     - {{P_\a}^A}_B \e^B \nn \\ && \qquad
     - \frac{\sqrt{2}}{12}V^{-1}b^i\a_i\g_\a\,\e (x^{11}){{\t_3}^A}_B\e^B 
     \\
\d \l^{xA} &=& b_i^x\left(-\frac{1}{2}i\g^\a\partial_\a b^i\e^A
             -\frac{1}{2\sqrt{2}}\g^{\a\b}\cF_{\a\b}^i\e^A
             -\frac{i}{2\sqrt{2}}V^{-1}\a^i\e (x^{11}){{\t_3}^A}_B\e^B
             \right) \\
\d \z^a &=& 
     - i {{Q_\a}^A}_B \g^\a \e^B
     - \frac{i}{\sqrt{2}}b^i\a_i V^{-1}\e (x^{11}){{\t_3}^a}_B\e^B
\label{susy5}
\esea
where $\t_i$ with  $i=1,2,3$ are the Pauli spin matrices and we have
the matrices
\begin{equation}
\begin{aligned}
   {{P_\a}^A}_B &= {\left( \begin{array}{cc} 
           \frac{\sqrt{2}i}{96}V \e_{\a\b\g\d\e}G^{\b\g\d\e} & 
           V^{-1/2}\pt_\a\x \\
           - V^{-1/2}\pt_\a{\bar\x} & 
           - \frac{\sqrt{2}i}{96}V \e_{\a\b\g\d\e}G^{\b\g\d\e}
        \end{array} \right)^A}_B \\
   {{Q_\a}^A}_B &= {\left( \begin{array}{cc}
           \frac{\sqrt{2}i}{48}V \e_{\a\b\g\d\e}G^{\b\g\d\e} 
               - \frac{1}{2}V^{-1}\pt_\a V &
           V^{-1/2}\pt_\a\x \\
           V^{-1/2}\pt_\a{\bar\x} & 
           \frac{\sqrt{2}i}{48}V \e_{\a\b\g\d\e}G^{\b\g\d\e} 
               + \frac{1}{2}V^{-1}\pt_\a V 
        \end{array} \right)^A}_B
\end{aligned}
\end{equation}

\section{The Domain Wall Solution}

In this section, we would like to find the simplest BPS solutions
of the five--dimensional theory, including the coupling to
the potential terms induced by the nonzero mode. As we will
see, these solutions provide the appropriate background for a reduction to
four dimensions and can therefore be viewed as the ``vacua'' of the theory.
After a general derivation of the solutions, we will discuss
several limiting cases of interest. 

\subsection{The General Solution}

Let us first simplify the discussion somewhat by concentrating on the
fields which are essential. Since we would like to find solutions that
couple to the bulk potential terms we should certainly keep the
hypermultiplet scalar $V$ (the Calabi--Yau breathing mode) and the
vector multiplet scalars $b^i$ (the shape moduli). It turns out that those
fields plus the five--dimensional metric are already sufficient. The
action~\eqref{S5} can be consistently truncated to this reduced field
content leading to
\bea
 2\k_5^2 S_5 &=& - \int_{M_5}\sqrt{-g}\left[R+G_{ij}
         \partial_\a b^i\partial^\a b^j+\frac{1}{2}V^{-2}\partial_\a V
         \partial^\a V+\frac{1}{2}V^{-2}G^{ij}\a_i\a_j+\l (\cK - 6)
         \right] \nn\\
      && \qquad\qquad
         + 2\sqrt{2}\int_{M_4^{(1)}}\sqrt{-g}\, V^{-1}\a_ib^i - 2\sqrt{2}
         \int_{M_4^{(2)}}\sqrt{-g}\,V^{-1}\a_ib^i\; .
 \label{S5_red}
\eea
Note that we have explicitly added the Lagrange multiplier term which
ensures the constraint~\eqref{b_cons} on $b^i$.
For a finite Calabi--Yau volume $V$, that is, for an uncompactified
internal space, the potential terms in this action do not vanish and, hence,
flat space is not a solution of the theory. Therefore, the question arises
of what the ``vacuum'' state of the theory is. A clue is provided by the
fact that cosmological--type potentials in $D$ dimensions generally couple
to $D-2$ branes. This is well known from the eight--brane~\cite{8brane}
which appears as a solution of the massive extension of type IIA
supergravity~\cite{romans} in ten dimensions. There, the eight--brane
couples to a cosmological--type potential which consists of a single
``cosmological'' constant multiplied by a certain power of the dilaton.
A way to understand to appearance of an eight--brane in this context
is to dualize the cosmological constant to a nine--form antisymmetric
tensor field which, according to the usual counting, should couple
to an $8+1$--dimensional extended object. A systematic study of $D-2$
brane solutions in various dimensions using a generalized Scherk--Schwarz
reduction can be found in ref.~\cite{dom}. The present case is
somewhat more complicated in that it involves $h^{1,1}$ scalar fields (as
opposed to just the dilaton) and, correspondingly, $h^{1,1}$ constants
$\a_i$ (as opposed to just one cosmological constant). Still, we can take
a lead from the massive IIA example and dualize each of the constants
$\a_i$ to a four--form antisymmetric tensor field. This would leave us
with a theory that contains $h^{1,1}$ such antisymmetric tensor fields
and, hence, a corresponding number of different types of three--branes
that couple to those. The constants
$\a_i$ can then be identified as the charges of these different types
of three--branes. Since those constants are fixed in terms of the
underlying theory (and are generically nonzero) one
cannot really look for a ``pure'' solution which carries only one type
of charge. Instead, what we are looking for is a multi--charged three--brane
which is a mixture of the various different types as specified by
the charges $\a_i$. Clearly, the transverse space for a three--brane
in five--dimensions is just one--dimensional. Given that the boundary
source terms necessarily introduce dependence on the $x^{11}$ coordinate,
this one--dimensional space can only be in the direction of the orbifold.

\vspace{0.4cm}

     From the above remarks it is now clear that the proper Ansatz for
the type of solutions we are looking for is given by
\bea
 ds_5^2 &=& a(y)^2dx^\m dx^\n\eta_{\m\n}+b(y)^2dy^2 \nn \\
 V &=& V(y) \label{3_ans}\\
 b^i &=& b^i(y)\nn\; ,
\eea
where we use $y=x^{11}$ from now on. The equations of motion
derived from the action~\eqref{S5_red} still contains the Lagrange
multiplier $\l$. It can be eliminated using
eqs. (A.19)--(A.21) from appendix A of ref.~\cite{add1}.
A solution to the resulting equations of the form~\eqref{3_ans} is still
somewhat hard to find, essentially due to the complication caused by the
inclusion of all $(1,1)$ moduli and the associated K\"ahler structure. The
trick is to express the solution in terms of certain functions $f^i=f^i(y)$
which are only implicitly defined rather than trying to find fully explicit
formulae. It turns out that those functions are fixed by the equations
\begin{equation}
 d_{ijk}f^jf^k=H_i\; ,\quad H_i=2\sqrt{2}k\a_i|y|+k_i\label{beta_def}
\end{equation}
where $k$ and $k_i$ are arbitrary constants. Then the solution can be written
as
\bea
 V &=&\left(\frac{1}{6}d_{ijk}f^if^jf^k\right)^2 \nn \\
 a &=&\tilde{k}V^{1/6}\nn\\
 b &=& kV^{2/3} \\
 b^i &=&V^{-1/6}f^i\nn
 \label{solution}
\eea
where $\tilde{k}$ is another arbitrary constant. We should check that
this solution is indeed a BPS state of the theory; that is, that it
preserves four of the eight supercharges. For the reduced field content,
the supersymmetry transformations~\eqref{susy5} lead to the following
Killing spinor equations
\bsea
 \d\psi_\m^A=0 & \quad :\quad &
    \g_\m\left(\frac{a'}{a}\g_{11}\e^A-\frac{\sqrt{2}b}{6V}
                 b^i\a_i\, \e (y)\, {{\t_3}^A}_B \e^B\right) = 0 \\
 \d\psi_{11}^A=0 & \quad :\quad &
    {\e^A}'-\frac{\sqrt{2}b}{12V}b^i\a_i\g_{11}\, \e (y) \,
                   {{\t_3}^A}_B \e^B = 0 \\
 \d\l^{xA}=0 & \quad :\quad &
    {b^i}'\g_{11}\e^A+\frac{b}{\sqrt{2}V}\left(
               \a^i-\frac{2}{3}b^j\a_jb^i\right)
               \e (y)\,{{\t_3}^A}_B \e^B = 0 \\
 \d\z^a = 0 & \quad :\quad &
    \frac{V'}{V}\g_{11}\e^A-\frac{\sqrt{2}b}{V}b^i\a_i\,\e (y)\,
              {{\t_3}^A}_B \e^B = 0\; ,
 \label{killing}
\esea
where the prime denotes the derivative with respect to $y$. These
equations are satisfied for the solution~\eqref{solution} if the
spinor $\e^A$ takes the form
\begin{equation}
 \e^A = a^{1/2}\e^A_0\; ,\quad \g_{11}\e^A_0 = {(\t_3)^A}_B \e^B_0\; ,
\end{equation}
where $\e^A_0$ is a constant spinor. As a result, the solution preserves
indeed four supercharges.

As can be seen from eq.~\eqref{beta_def} the solution is described in terms
of $h^{1,1}$ linear functions $H_i$. This follows the general pattern of
$p$--brane solutions coupled to $n$ different charges which can be
expressed in terms of $n$ harmonic functions on the transverse space. In our
case the number of charges $\a_i$ is precisely
$h^{1,1}$ and the transverse space is just one--dimensional leading to
linear functions. Generally, elementary brane solutions have singularities
at the location of the branes which have to be supported by brane worldvolume
theories. The pure bulk theory does not impose any restrictions on the
number and locations of these singularities. Correspondingly, if we would
just consider the bulk part of the action~\eqref{S5_red} we could place an
arbitrary number of  parallel three--branes anywhere on the orbifold.
However, the theory~\eqref{S5_red} involves two four--dimensional boundary
actions which provide source terms that should be matched. This is possible,
in the present case, because the height of the boundary potentials
in~\eqref{S5_red} is set by the three--brane charges $\a_i$. If we decide
that the solution should have no further singularities other than those
matched by the two boundaries we arrive at the specific form of the
harmonic functions $H_i$ in eq.~\eqref{beta_def}. In fact, we have
\begin{equation}
 {H_i}'' = 4\sqrt{2}k\a_i(\d (y)-\d (y-\p\r ))\; ,
\end{equation}
indicating sources at the orbifold planes $y=0,\p\r$. 
Recall that we have restricted the range of $y$ to $y\in [-\p\r ,\p\r ]$
with the endpoints identified. This explains the second delta--function
at $y=\p\r$ in the above equation.

In conclusion, the solution~\eqref{solution} represents a multi--charged
double domain wall (three--brane) solution with the two walls located
at the orbifold planes. It preserves four--dimensional Poincar\'e invariance
as well as four of the eight supercharges and has therefore the correct
properties to make contact with four--dimensional $\cN=1$ supergravity.
More precisely, those theories should arise as a dimensional reduction of
the five--dimensional theory on the domain wall background. In this sense,
the solution~\eqref{solution} can be viewed as the vacuum state of the
five--dimensional theory. From the perspective of the four--dimensional 
theory the domain
wall solution plays an interesting r\^ole. It is oriented precisely
in the four uncompactified dimensions and carries the physical gauge
and gauge matter fields. Therefore, at low energy four--dimensional space--time
gets identified with the three--brane worldvolume. In this sense, our
Universe lives on the worldvolume of a three--brane. Finally, 
we would like to discuss some physically relevant limiting examples
of the general solution.

\subsection{Universal Solution}

In ref.~\cite{losw} we have presented a related three--brane solution which
was less general in that it involved the universal Calabi--Yau
modulus $V$ only. Clearly, we should be able to recover this solution
from eq.~\eqref{solution} if we consider the specific case $h^{1,1}=1$.
Then we have $d_{111}=6$ and it follows from eq.~\eqref{beta_def} that
\begin{equation}
 f^1 = \left(\frac{\sqrt{2}}{3}k\a_1|y|+k_1\right)^{1/2}\; .
\end{equation}
Inserting this into eq.~\eqref{solution} provides us with the explicit
solution in this case which is given by
\bea
 a &=& a_0H^{1/2} \nn \\
 b &=& b_0H^2\qquad\qquad H=\frac{\sqrt{2}}{3}\a |y|+c_0\; ,\quad \a =\a^1 
 \label{u_sol}\\
 V &=& b_0H^3\nn \; .
\eea
The constant $a_0$, $b_0$ and $c_0$ are related to the integration constants
in eq.~\eqref{solution} by
\begin{equation}
 a_0=\tilde{k}k^{1/2}\; ,\qquad b_0=k^3\; ,\qquad c_0=\frac{k_1}{k}\; .
\end{equation}
Eq.~\eqref{u_sol} is indeed exactly the solution that was found in
ref.~\cite{losw}. It still represents a double domain wall. However, in
contrast to the general solution it couples to one charge $\a =\a^1$ only.
Geometrically, it describes a variation of the five--dimensional metric and
the Calabi--Yau volume across the orbifold. The form of the
solution~\eqref{u_sol} is typical for brane solutions that couple to one charge
and, in fact, fits into the general scheme of domain walls in various
dimensions~\cite{dom}.

\vspace{0.5cm}

One may ask if a structure as simple as the above universal solution is,
in some way, also part of the general solution~\eqref{solution} even
if $h^{1,1}>1$. To see that this is indeed the case, we define constants
$\bar{\a}^i$ and $\a$ by
\begin{equation}
 d_{ijk}\bar{\a}^j\bar{\a}^k=\frac{2}{3}\a_i\; ,\qquad 
 \a = 9\left(\frac{1}{6}d_{ijk}\bar{\a}^i\bar{\a}^j\bar{\a}^k\right)^{2/3}\; .
 \label{ab_def}
\end{equation}
In addition, we choose the following special values for the integration
constants $k_i$ in eq.~\eqref{beta_def}
\begin{equation}
 k_i=6kc_0\frac{\a_i}{\a}
\end{equation}
where $c_0$ is an arbitrary constant. Thanks to this specific choice, we
can easily solve~\eqref{beta_def} for $f^i$. Inserting the result into
eq.~\eqref{solution} gives the explicit solution
\bea
 a &=& a_0H^{1/2} \nn \\
 b &=& b_0H^2\; ,\qquad H=\frac{\sqrt{2}}{3}\a |y|+c_0 \label{u_sol1}\\
 V &=& b_0H^3\nn \\
 b^i &=& 3\a^{-1/2}\bar{\a}^i\nn\; .
\eea
As before, $a_0$ and $b_0$ are constants expressed in terms of the integration
constants in ~\eqref{solution} as
\begin{equation}
 a_0=\tilde{k}k^{1/2}\; ,\qquad b_0=k^3\; .
\end{equation}
Hence, for arbitrary values of $h^{1,1}$, we have identified a special case
of the general solution~\eqref{solution} where the fields $a$, $b$ and $V$
behave in exactly the same way as in the universal solution~\eqref{u_sol}.
The charge $\a$ which appears in this special solution is now a complicated
function of the various charges $\a_i$ in the way defined by
eq.~\eqref{ab_def}. In addition, the shape moduli $b^i$ are constant.
Consequently, for this special solution the metric and the Calabi--Yau
volume vary as in the universal solution while the shape of the Calabi--Yau
space is fixed.

\subsection{Another Simple Example}

A nontrivial example where the domain wall solution can be obtained
explicitly is provided by
\begin{equation}
 h^{1,1}=3\; ,\qquad d_{123}=1\; ,
\end{equation}
and $d_{ijk}=0$ otherwise. The K\"ahler potential is then given by
\begin{equation}
 \cK = 6\,b^1b^2b^3\label{STU} \; .
\end{equation}
In a four--dimensional effective theory the real fields $b^i$ are
promoted to complex scalars. Then the K\"ahler potential~\eqref{STU}
is associated with the coset space $\left[ SU(1,1)/U(1)\right]^3$
\cite{cfg} and describes the STU--model.
Due to the simple structure of intersection numbers eq.~\eqref{beta_def}
can be easily solved for the functions $f_i$ resulting in
\begin{equation}
 f^i  = (H_1H_2H_3)^{1/2}H_i^{-1}\; .
\end{equation}
Inserting into eq.~\eqref{solution} then gives the explicit solution
\bea
 V &=& (H_1H_2H_3)^{-1} \nn \\
 a &=& \tilde{k}(H_1H_2H_3)^{-1/6}\qquad\qquad H_i=2\sqrt{2}k\a_i |y|+k_i\\
 b &=& k(H_1H_2H_3)^{-2/3}\nn \\
 b^i  &=& (H_1H_2H_3)^{2/3}H_i^{-1}\nn
\eea
for $i=1,2,3$. As before $k$, $\tilde{k}$ and $k_i$ denote constants.


\section*{Lecture 2: Non-Standard Embedding and Five--Branes}


To make contact with low-energy physics, one of the central issues in
string theory has been to find vacua leading to chiral
four-dimensional theories with $\cN=1$ supersymmetry. In recent years,
the new understanding of the non-perturbative behavior of string
theory has broadened the scope for approaching these issues. Specifically, 
the inclusion of brane states, that is, vacua with
non-trivial form-fields, increases the class of possible backgrounds
giving a chiral $\cN=1$ theory in four dimensions, and has raised the
possibility of gauge interactions arising from the brane world-volume
theory itself.  

In this second lecture, we will consider a class of eleven-dimensional
M--theory vacua based on the strongly coupled limit of the $E_8 \times
E_8$ heterotic string, as described by Ho\v rava and
Witten~\cite{hw1,hw2}. At low energy, these are compactifications of
eleven-dimensional supergravity on an $S^1/Z_2$ orbifold, with $E_8$
gauge fields at each of the two orbifold fixed planes. Following
Witten~\cite{w}, we can further compactify on a Calabi--Yau three-fold
to give a chiral $\cN=1$ theory in four-dimensions. Essentially, all
the early discussions of the low-energy properties of
compactifications~\cite{hor}--\cite{BDDR} were limited to the
standard embedding, where the Calabi--Yau spin connection is embedded
in one of the $E_8$ gauge groups. In~\cite{us}, we considered the general
configuration leading to $\cN=1$ supersymmetry, where, first, we allowed
for general gauge bundles, and, second, included five-branes, states 
which are essentially non-perturbative in heterotic string
theory. The possibility of such generalizations was first put forward
by Witten~\cite{w}. Recently, non--standard embedding gauge threshold
corrections of orbifold models have been computed in ref.~\cite{stieb} and
have, in the large radius limit, been compared to the expressions calculated
from Ho\v rava--Witten theory. Gauge thresholds of non-standard
embeddings in the strongly coupled limit have also been discussed
in~\cite{benakli}. A toy model of gauge fields coming from five-branes
close to the orbifold planes has been presented
in~\cite{BDDR}. In this lecture, we review the results of~\cite{us}.

The $\cN=1$ vacua we will discuss have the following structure. One
starts with the spacetime $M_{11}= S^1/Z_2\times X\times M_4$, where
$X$ is a Calabi--Yau three-fold and $M_4$ is flat Minkowski space.
As in the weakly coupled limit, to preserve the
four supercharges, arbitrary holomorphic $E_8$ gauge bundles over $X$
(satisfying the Donaldson--Uhlenbeck--Yau condition) are allowed on
each plane. In particular, there is no requirement that the
spin-connection of the Calabi--Yau space be embedded in the gauge connection
of one of the $E_8$ bundles. This generalization is what is meant by
non-standard embedding, and has a long history in the phenomenology of
weakly coupled strings (for early discussions see
refs.~\cite{w0,ww,gsw}). In addition, one can add five-branes, located
at points throughout the orbifold interval. The five-branes will
preserve some supersymmetry, provided the branes are wrapped
on holomorphic two-cycles within $X$ and otherwise span the flat
Minkowski space $M_4$~\cite{w}. 

Both the gauge fields and the five-branes are magnetic sources for the
four-form field strength $G$ of the bulk supergravity, and so excite a
non-zero $G$ within the compact $S^1/Z_2\times X$ space. This has two
effects. First, since the space is compact, there can be no net
magnetic charge, for there is nowhere for the flux to ``escape''. Thus,
there is a cohomological condition that the sum of the sources must be
zero. Secondly, the non-zero form field enters the Killing spinor
equation and so, to preserve supersymmetry, the background geometry
must have a compensating distortion~\cite{w}. This leads to a
perturbative expansion of the supersymmetric background. Such an
expansion is familiar in non-standard embeddings in the weakly
coupled heterotic string~\cite{w0,ww,gsw}. In the strongly coupled
limit, it appears even for the standard embedding. From this point of
view, the generalization to include non-standard embedding and
five-branes is very natural.

Having found the vacuum as a perturbative solution, one is then
interested in the form of the low-energy theory of the massless
excitations around this compactification. It is well known that, in
the standard embedding, to match the low-energy Newton constant and
grand unified parameters, one needs to take a Calabi--Yau manifold of size
comparable to the eleven-dimensional Planck length, with the orbifold an
order of magnitude or so larger. Thus, it is natural to consider
effective actions both in five dimensions, where only $X$ is
compactified, and four, which is appropriate to momenta below the
orbifold scale. For the standard embedding, the four-dimensional action
has been calculated to leading non-trivial order~\cite{hp,low1}. Although
the expansion is completely non-perturbative, it turns out that, to
this order, the form of the effective action is identical to the large
radius Calabi--Yau limit of the one-loop effective action calculated in the
weak limit. There are threshold corrections in the gauge couplings as
well as in the matter field K\"ahler potential. In five dimensions,
because of the non-zero mode of $G$, the theory is a form of gauged
supergravity in the bulk, coupled to gauge theories on the fixed
planes~\cite{losw,add1}. There is no homogeneous background solution
but, rather, the correct vacuum is a BPS domain wall solution,
supported by sources on the fixed planes and a potential in the bulk.

Calculating the modifications to the low-energy effective actions due
to non-standard embedding and five-branes will be the main point of
this lecture. Our results can be summarized as follows. In section two, we
discuss the expansion of the background solution, the cohomology condition
on the five-brane and orbifold magnetic sources and the constraints on
the zeroth-order background to preserve supersymmetry. We then give the
solution to first order. Expanding in terms of eigenfunctions on the
Calabi--Yau three-fold, we show that the main contribution comes from the
massless modes. Sections three and four discuss the low-energy actions
in the case of non-standard embedding and inclusion of five-branes
respectively. This requires an analysis of the theory on the
five-brane world-volume, which is given in section 4.2. In summary, we
find 
\begin{itemize}
\item For non-standard embeddings, in the absence of five-branes, the
      five-dimensional action has the same form as in the standard
      embedding case both in the bulk 
      and on the orbifold planes. However, the values of the gauge coupling
      parameters, related to the gauging of the bulk supergravity, depend on
      the form of the non-standard embedding.
\item The non-standard embedding allows many different breaking
      patterns for the $E_8$ groups. In particular, it is no longer
      necessary that the visible sector be broken to $E_6$. Rather,
      more general gauge groups $G^{(1)}, G^{(2)}\subset E_8$ and
      corresponding gauge matter can occur on the respective orbifold
      planes.
\item In the presence of five branes, the form of the bulk
      five-dimensional action between any pair of neighboring branes
      is the same as in the case of standard embedding. The
      four-dimensional fixed-plane theories also have the same
      form and couplings to the bulk fields. However, there are additional
      four-dimensional theories, arrayed throughout the orbifold and again
      coupling to the bulk fields, which arise from the five-brane
      world-volume degrees of freedom. 
\item In the conventional picture, the five-brane worldvolume theories
      provide new hidden sectors. Generically, the theory for a single
      five-brane is $\cN=1$
      supersymmetric with $g$ $U(1)$ vector multiplets, together with a
      universal chiral multiplet and a set of chiral fields parameterizing
      the moduli space of holomorphic genus $g$ two-cycles in $X$. This gauge
      group can be enhanced when five-branes overlap or when the embedding
      of a single fivebrane degenerates. In general, the total rank of the
      gauge group remains unchanged.  
\item The presence of five-branes also allows for new types of $E_8\times
      E_8$ breaking patterns, beyond those associated with non-standard
      embeddings alone. This is because the presence of five-brane sources
      leads to a wider range of solutions satisfying the zero cohomology
      condition.  
\item Reducing to four dimensions, the effective action is modified
      with respect to the standard embedding case. For pure
      non-standard embeddings, both the gauge and K\"ahler threshold
      corrections are identical in form to those of the standard
      embedding. However, the presence of the five-branes
      significantly modifies these corrections so that, for instance,
      both $E_8$ sectors can get threshold corrections of the same sign.  
\end{itemize}
The new threshold corrections due to the five-branes have no analog in
the weakly coupled limit since, first, the branes are
non-perturbative and, second, the corrections depend on the positions
of the five-branes across the orbifold, moduli which simply do not
exist in the weakly coupled limit. Similarly, the appearance of new
gauge groups due to five-branes is a non-perturbative 
effect. Finally, we note, it appears that there is 
a constraint on the total rank of the full gauge group from orbifold
fixed planes and five-branes, which arises from positivity constraints
in the magnetic charge cohomology condition. 

\section{Vacua with Non-Standard Embedding and Five-Branes}

In this section, we are going to construct generalized heterotic M--theory
vacua appropriate for a reduction of the theory to $\cN =1$
supergravity theories in both five and four dimensions.
To lowest order (in the sense explained below), these vacua have the usual
space-time structure $M_{11}=S^1/Z_2\times X\times M_4$ where $X$ is a
Calabi--Yau three-fold and $M_4$ is four-dimensional Minkowski space.
As compared to the vacua constructed to date, we will allow for two
generalizations. First, we will not restrict ourselves to embedding the
Calabi--Yau spin connection into a subgroup $SU(3)\subset E_8$ but,
rather, allow for general (supersymmetry preserving) gauge field
sources on the orbifold hyperplanes. Secondly, we will allow for the
presence of five-branes that stretch across $M_4$ and wrap around a
holomorphic curve in $X$. As we will see, the inclusion of five-branes
makes it much easier to satisfy the necessary constraints. Therefore,
their inclusion is essential for a complete discussion 
non-standard embeddings, and leads to a considerable increase in the
number of such vacua.

\subsection{Expansion Parameters}

Before we proceed to the actual computation, let us explain
the types of corrections to the lowest order background that one expects.
For the weakly coupled heterotic string, it is well known
that non-standard embeddings lead to corrections to the Calabi--Yau
background. They can be computed perturbatively~\cite{w0,ww,gsw} as a
series in
\begin{equation}
 \e_W = \frac{\a '}{v_{10}^{1/3}}\label{ew0}
\end{equation}
where $v_{10}$ is the Calabi--Yau volume measured in terms of the
ten-dimensional Einstein frame metric. At larger string coupling,
one also gets contributions from string loops. Thus the full solution
is a double expansion involving both $\e_W$ and the string coupling constant. 

On the other hand, in the strongly coupled heterotic string, it has
been shown that, even in the case of the standard embedding, there are
corrections originating from the localization of the
gauge fields to the ten-dimensional orbifold planes~\cite{w,low1}.
Again, these corrections can be organized in a double
expansion. However, one now uses a parameterization appropriate to the
strongly coupled theory. The 11-dimensional Ho\v rava--Witten
effective action has an expansion in $\k$, the 11-dimensional
Newton constant. For the compactification on $S^1/Z_2\times X$, there
are two other scales, the size of the orbifold interval $\p\r$ and the
volume $v$ of the Calabi--Yau threefold, each measured in the
11-dimensional metric. Solving the equations of motion and
supersymmetry conditions for the action to order $\k^{2/3}$, one finds
the correction to the background is a double expansion, linear, at
this order, in the parameter 
\begin{equation}
 \e_S = \left(\frac{\k}{4\p}\right)^{2/3}
        \frac{\p\r}{v^{2/3}}\label{es}
\end{equation}
but to all orders in
\begin{equation}
 \e_R = \frac{v^{1/6}}{\p\r}. \label{er}
\end{equation}
It is natural to use the same expansion for the background with
non-standard embedding and the inclusion of five-branes. As we will show
explicitly, the solution to the order $\k^{2/3}$ can be obtained as an
expansion in eigenfunctions of the Calabi--Yau Laplacian. It turns out
that the zero-eigenvalue, or ``massless'', terms in this expansion are
precisely of order $\e_S$, while the massive terms are of order
$\e_R\e_S$. Therefore, although one could expect corrections to
arbitrary order in $\e_R$, to leading order in $\e_S$ only the
zeroth-order and linear terms in $\e_R$ contribute. 

Clearly, for the above expansion to be valid both
$\e_S$ and $\e_R$ should be small. Let us briefly discuss the
situation at the physical point, that is, at the values of $\k$, $v$
and $\r$ that lead to the appropriate values for the four-dimensional Newton
constant and the grand unification coupling parameter and scale. There,
both the 11--dimensional Planck length $\k^{2/9}$, as well as the
Calabi--Yau radius $v^{1/6}$,
are of the order $10^{-16}\mbox{ GeV}^{-1}$ while the orbifold radius
is an order of magnitude or so larger. Inserting this into
eq.~\eqref{es} and \eqref{er} shows that $\e_S$ is of order one~\cite{bd}
while $\e_R$ is an order of magnitude or so smaller. At the physical
point, therefore, we have 
\begin{equation}
   \e_R\ll\e_S=O(1)\; .
\end{equation}
Consequently, neglecting higher-order terms in $\e_S$ might not provide a
good approximation at the physical point. It is, however, the best one
can do at the moment given that M--theory on $S^1/Z_2$ is only known
as an effective theory to order $\k^{2/3}$. On the other hand, in
fact, higher-order terms in $\e_R$ should be strongly
suppressed and can be safely neglected. 

It is interesting to note how this strong coupling expansion is
related to the weak coupling expansion with non-standard
embedding. Writing $\e_W$ in terms of 11-dimensional quantities,
one finds 
\begin{equation}
 \e_W = \left(\frac{\k}{4\p}\right)^{2/3}\frac{1}{\p^2\r v^{1/3}}\label{ew}
\end{equation}
and hence
\begin{equation}
 \e_W=\frac{1}{\p}\e_R^2\e_S \; .\label{ew_es}
\end{equation}
Let us try to make this relation plausible. In the weak coupling limit,
the orbifold becomes small. Hence, one expects to extract the weak coupling
part of the full background by performing an orbifold average. We recall
that the massive terms in the full background are of order $\e_R\e_S$.
In addition, we will find that those massive modes decay exponentially
as one moves away from the orbifold planes, at a rate set by the Calabi--Yau
radius $v^{1/6}$. Therefore, when performing the average, one picks up
another factor of $\e_R$ leading to $\e_R^2\e_S$ as the order of the averaged
massive terms. This is in perfect agreement with the expectation,
\eqref{ew_es}, from the weakly coupled heterotic string\footnote{There is no
such comparison for the massless modes as they correspond to trivial
integration constants on the weakly coupled side which can be absorbed
into a redefinition of the moduli. This will be explained in detail later on.}.

\subsection{Basic Equations and Zeroth-Order Background}

\begin{figure}[t]
   \centerline{\psfig{figure=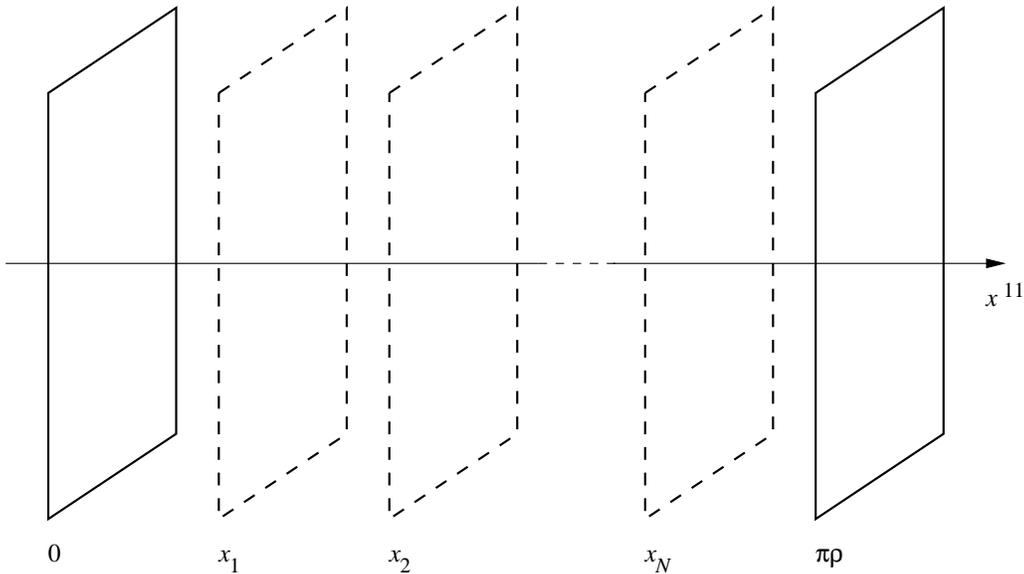,height=3in}}
   \caption{Orbifold interval with boundaries at $0$, $\p\r$ and $N$
            five-branes at $x_1,\dots ,x_N$. The mirror interval from
            $0$ to $-\p\r$ is suppressed in this diagram.}
   \label{fig1}
\end{figure}

The M--theory vacuum is given in the 11-dimensional limit by
specifying the metric $g_{IJ}$ and the three-form $C_{IJK}$ with field
strength $G_{IJKL}=24\,\partial_{[I}C_{JKL]}$. To the order
$\k^{2/3}$, the set of equations to be solved consists of the Killing spinor
equation
\begin{equation}
 \d\Psi_I = D_I\eta +\frac{\sqrt{2}}{288}
            \left(\G_{IJKLM}-8g_{IJ}\G_{KLM}\right)G^{JKLM}\eta = 0\; ,
 \label{killing2}
\end{equation}
for a Majorana spinor $\eta$, the equation of motion for $G$
\begin{equation}
 D_IG^{IJKL} = 0 \label{Geom2}
\end{equation}
and the Bianchi identity\footnote{Here we are using the normalization
given in ref.~\cite{hw2}. Conrad~\cite{conrad} has argued that the correct
normalization is smaller. In that case, the coefficient of the
right-hand side of the Bianchi identity~\eqref{G} and eqns.~\eqref{S}
and~\eqref{cB} below are all multiplied by $2^{-1/3}$. Furthermore, the
definition of $\e_S$ in eqn.~\eqref{es} should also be multiplied by
$2^{-1/3}$.}
\bea
 (dG)_{11\Ib\Jb\Kb\Lb} &=& 2\sqrt{2}\p\left(\frac{\k}{4\p}
                           \right)^{2/3}\left[J^{(0)}\d (x^{11})+J^{(N+1)}
                           \d (x^{11}-\p\r )+\right.\nn \\
                       &&\left.\qquad\qquad\qquad\qquad\frac{1}{2}
                         \sum_{n=1}^NJ^{(n)}(\d (x^{11}-x_n)+\d (x^{11}+x_n))
                           \right]_{\Ib\Jb\Kb\Lb}\; .\label{G}
\eea
Here the sources $J^{(0)}$ and $J^{(N+1)}$ on the orbifold planes are as
usual given by
\begin{equation}
\begin{aligned}
 J^{(0)} &= -\frac{1}{8\p^2}\left.\left(\tr F^{(1)}\wedge F^{(1)} 
      - \frac{1}{2}\tr R\wedge R\right)\right|_{x^{11}=0} \; , \\
 J^{(N+1)} &= -\frac{1}{8\p^2}\left.\left(\tr F^{(2)}\wedge F^{(2)} 
      - \frac{1}{2}\tr R\wedge R\right)\right|_{x^{11}=\p\r} \; .
\end{aligned}
\label{J2} 
\end{equation}
We have also introduced $N$ additional sources $J^{(n)}$,
$n=1,\dots ,N$. They come from $N$ five-branes located at
$x^{11}=x_1,\dots ,x_N$ where $0\leq x_1\leq\dots\leq x_N\leq\p\r$
(see fig.~\ref{fig1}). Note that each five-brane at $x^{11}=x_n$ has
to be paired with a mirror five-brane at $x^{11}=-x_n$ with the same
source since the Bianchi identity must be even under the $Z_2$
orbifold symmetry. Our normalization is such that the total source of
each pair is $J^{(n)}$. The structure of these five-brane sources
will be discussed below. 
We are interested in finding solutions of these equations that preserve
$3+1$-dimensional Poincar\'e invariance and admit a Killing spinor $\eta$
corresponding to four preserved supercharges and, hence, $\cN =1$
supersymmetry in four dimensions.

The usual procedure to find such solutions is to solve the equations
perturbatively. One starts by choosing a space
$S^1/Z_2\times X\times M_4$, where $X$ is a Calabi--Yau three-fold with a
Ricci-flat metric $g_{AB}$, admitting a Killing spinor $\eta^{({\rm CY})}$.
To lowest order, the solution, denoted in the following by $(0)$, is then
given by
\begin{equation}
\begin{aligned}
 {ds^{(0)}}^2 \equiv g_{IJ}^{(0)}dx^Idx^J &= \eta_{\m\n}dx^\m dx^\n
     +g_{AB}dx^Adx^B+(dx^{11})^2 \\
 G_{IJKL}^{(0)} &= 0 \\
 \eta^{(0)} &= \eta^{({\rm CY})}\; .
\label{sol0} 
\end{aligned}
\end{equation}
Note that it is consistent, to this order, to set the antisymmetric
tensor field to zero since the sources in the Bianchi identity are
proportional to $\k^{2/3}$ and, hence, first order in $\e_S$.

One must also ensure that the theories on the orbifold planes preserve
supersymmetry. This leads to the familiar constraint, following from
the vanishing of the supersymmetry variation of the gauginos, that
\begin{equation}
\G^{\Ib\Jb}F_{\Ib\Jb}^{(1)}\eta |_{x^{11}=0}=
\G^{\Ib\Jb}F_{\Ib\Jb}^{(2)}\eta |_{x^{11}=\p\r}=0\; .\label{gau_killing}
\end{equation}
As discussed in~\cite{gsw}, this implies that each $E_8$ gauge field
is a holomorphic gauge bundle over the Calabi--Yau three-fold, satisfying the
Donaldson--Uhlenbeck--Yau condition. The holomorphicity implies that
$F^{(1)}_{AB}$ and $F^{(2)}_{AB}$ are (1,1)-forms. It follows that, since
$R_{AB}$ for a Calabi--Yau three-fold is also a (1,1)-form, the orbifold
sources $J^{(0)}$ and $J^{(N+1)}$, defined by eq.~\eqref{J2}, are closed
$(2,2)$-forms.  

For the five-brane world-volume theory to be supersymmetric, the branes
must be embedded in the Calabi--Yau space in a particular way~\cite{w}. To
preserve Lorentz invariance in $M_4$, they must span the  $3+1$-dimensional
uncompactified space. The remaining spatial dimensions must then be
wrapped on a two-cycle in the Calabi--Yau space. The condition of
supersymmetry implies that the cycle is a holomorphic
curve~\cite{w,bbs,vb}. As we will show in section 4.2, in such a situation,
we preserve four supercharges on the five-brane worldvolume corresponding
to $\cN =1$ supersymmetry in four dimensions. Since the five-branes are
magnetic sources for $G$, they enter the right-hand side of the
Bianchi identity~\eqref{G} as source terms, which should be localized
on the five-brane world-volumes. The delta function in $x^{11}$ gives
the localization in the orbifold direction, while the four-forms
$J^{(n)}$ must give the localization of the $n$-th five-brane on the
two-cycle $\cC_2^{(n)}$. Explicitly, for any two-cycle $\cC_2$, one
can introduce a delta-function four-form $\d(\cC_2)$, defined in the
usual way, such that for any two-form $\chi$,
\begin{equation}
   \int_X \chi \wedge \d(\cC_2) = \int_{\cC_2} \chi \; ,
\end{equation}
so that $\d(\cC_2)$ is localized on $\cC_2$. In general, we would
expect that $J^{(n)}$ is proportional to $\d(\cC_2^{(n)})$. In fact,
the correct normalization of the five-brane magnetic
charge~\cite{DMW,wittq} implies that the two are equal, that is
\begin{equation}
   J^{(n)} = \d(\cC_2^{(n)}) \; .
\label{Jdef}
\end{equation}
Since the cycles are holomorphic, $J^{(n)}$, like the orbifold
sources, are closed (2,2)-forms. 

There is one further condition which the five-branes and the
fields  on the orbifold planes must satisfy. This is a cohomology condition
on the Bianchi identity~\cite{w}. Consider integrating the
identity~\eqref{G} over a five-cycle which spans the orbifold interval
together with an arbitrary four-cycle $\cC_4$ in the Calabi--Yau three-fold.
Since $dG$ is exact, this integral must vanish. Physically this is the
statement that there can be no net charge in a compact space, since there
is nowhere for the flux to ``escape''. Performing the integral over the
orbifold, we derive, using \eqref{G}, the condition
\begin{equation}
   - \frac{1}{8\p^2}\int_{\cC_4}\tr F^{(1)}\wedge F^{(1)}
      - \frac{1}{8\p^2}\int_{\cC_4}\tr F^{(2)}\wedge F^{(2)}
      + \frac{1}{8\p^2}\int_{\cC_4}\tr R \wedge R
      + \sum_{n=1}^{N} \int_{\cC_4}J^{(n)}
      = 0\; . \label{chargecon}
\end{equation}
Hence, the net magnetic charge over $\cC_4$ is zero. Equivalently, this
implies that the sum of the sources must be cohomologically trivial, that is 
\begin{equation}
   \left[\sum_{n=0}^{N+1}J^{(n)}\right] = 0\; .\label{coh}
\end{equation}

\begin{figure}[t]
   \centerline{\psfig{figure=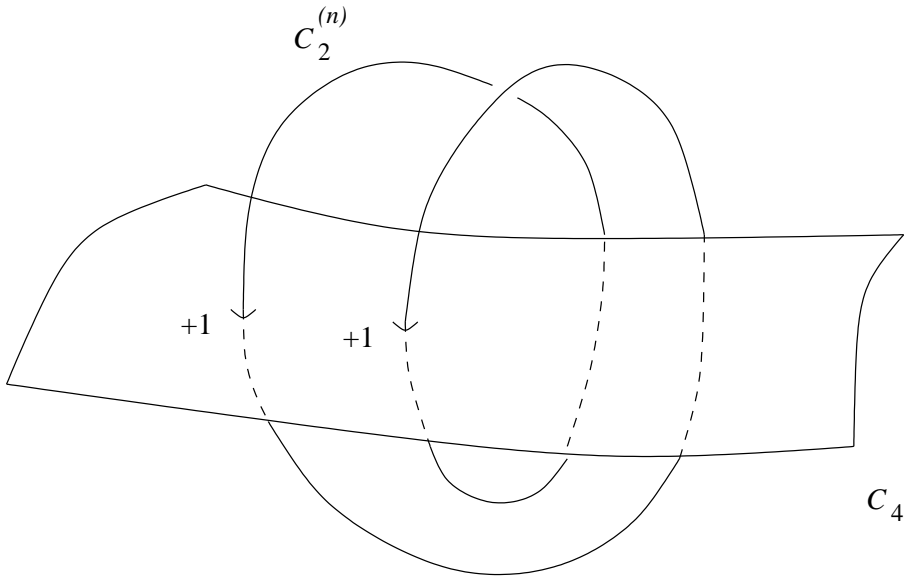,height=3in}}
   \caption{Intersection of a five-brane wrapped on the holomorphic
   cycle $\cC_2^{(n)}$ and a four-cycle $\cC_4$. In this example 
   the five-brane contributes two units of magnetic charge on $\cC_4$.}
   \label{intersect}
\end{figure}

Let us now return to the normalization of the five-brane
charges. We note that in equation~\eqref{chargecon} the first three
terms are all integers. They are topological invariants, giving the
instanton numbers (second Chern numbers) of the two $E_8$ bundles and
the instanton number (first Pontrjagin number) of the tangent bundle of
the Calabi--Yau three-fold. Hence, the above constraint shows that
$n_5(\cC_4)=\sum_{n=1}^{N}\int_{\cC_4}J^{(n)}$ must also be an
integer. In fact, with the normalization given in eqn.~\eqref{Jdef}, each
$\int_{\cC_4}J^{(n)}$ is an integer. It is also a topological
invariant, giving the intersection number~\cite{GH} of the $n$-th
brane, on the two-cycle $\cC_2^{(n)}$, with the four-cycle
$\cC_4$. This can be understood as follows (see
fig.~\ref{intersect}). 
The two cycles naturally intersect at points in
the Calabi--Yau manifold. Thus in $\cC_4$, the five-brane appears as a
set of point-like magnetic charges located at each intersection. The
net contribution of the five-brane to the magnetic charge on $\cC_4$
is then the sum of the point charges, which is precisely the
intersection number. Given the normalization of~\eqref{Jdef}, each
intersection contributes one unit of magnetic charge. We also note
that, for a holomorphic curve $\cC_4$, since $\cC_2^{(n)}$ is
holomorphic, it is a theorem~\cite{GH} that the intersection number is
always positive. This is related to the fact that only five-branes and
not anti-five-branes are allowed if we are to preserve
supersymmetry. In summary, the main point is that the normalization of
the five-brane charge is such that each five-brane intersection with
$\cC_4$ and each gauge instanton on the orbifold plane carry the same
amount of magnetic charge~\cite{DMW,wittq}. 

We can then rewrite the cohomology condition~\eqref{chargecon} on a
particular holomorphic four-cycle $\cC_4$ as 
\begin{equation}
   n_1(\cC_4) + n_2(\cC_4) + n_5(\cC_4) = n_R(\cC_4)
\label{ccon}
\end{equation}
which states that the sum of the number of instantons on the two $E_8$
bundles and the sum of the intersection numbers of each five-brane
with the four-cycle $\cC_4$, must equal the instanton number for the
Calabi--Yau tangent bundle, a number which is fixed once the
Calabi--Yau geometry is chosen.

In summary, we see that to define the zeroth-order background we must
specify the following data
\begin{itemize}
\item
a Calabi--Yau three-fold $X$,
\item
two holomorphic vector bundles over $X$, one for each fixed plane,
satisfying the Donaldson--Uhlenbeck--Yau condition. In general, there
is no constraint that these bundles correspond to the embedding of the
spin-connection in the gauge connection,
\item
a set of five-branes, each spanning the uncompactified $3+1$ dimensional
space and wrapping a holomorphic two-cycle in the Calabi--Yau space,
\item
the sum of the five-branes magnetic charges and the instanton numbers
from the gauge bundles, must equal the tangent space instanton number
of $X$, as in equation~\eqref{ccon},
\end{itemize}
We can then proceed to calculate the first-order corrections to the
background. 

\subsection{First-Order Background}

As an expansion in $\e_S$, we write the bulk fields and the Killing spinor as 
\begin{equation}
 \begin{aligned}
    g_{IJ} &= g_{IJ}^{(0)}+g_{IJ}^{(1)} \\
    C_{IJK} &= C_{IJK}^{(0)}+C_{IJK}^{(1)}\\
    \eta &= \eta^{(0)}+\eta^{(1)}\; .
  \end{aligned}
 \label{decomp1}
\end{equation}
where the index $(0)$ refers to the uncorrected background, given
in \eqref{sol0}, and the index $(1)$ to the corrections to first order 
in $\e_S$. 

Expanding to this order in $\e_S$, we get for the Killing spinor equation
\eqref{killing2}
\begin{multline}
 \d\Psi_I = D_I^{(0)}\eta^{(1)}-\frac{1}{8}\left(D_J^{(0)}
              g_{KI}^{(1)}-D_K^{(0)}g_{JI}^{(1)}\right)
                \G^{JK}\eta^{(0)} \\
              +\frac{\sqrt{2}}{288}\left(
                \G_{IJKLM}-8g_{IJ}^{(0)}\G_{KLM}\right)
                G^{(1)JKLM}\eta^{(0)} = 0
 \label{S0}
\end{multline}
and for the equation of motion for $G$ \eqref{Geom2} and the
Bianchi identity \eqref{G}
\begin{equation}
\begin{aligned}
 D_I^{(0)}G^{(1)IJKL} &= 0 \\
 (dG^{(1)})_{11\Ib\Jb\Kb\Lb} &= 2\sqrt{2}\p\left(\frac{\k}{4\p}
                                 \right)^{2/3}\left[ J^{(0)}\d (x^{11})+
                                 J^{(N+1)}\d (x^{11}-\p\r) \right.\\
                             &\qquad\qquad\qquad\qquad \left.+\frac{1}{2}
                                 \sum_{n=1}^{N}J^{(n)}(\d (x^{11}-x_n)+
                                 \d (x^{11}+x_n)) \right]_{\Ib\Jb\Kb\Lb}\;
                               . \label{S}
\end{aligned}
\end{equation}
First, we note that the only nonvanishing components of the
antisymmetric tensor $G^{(1)}$ are $G^{(1)}_{a\bar{b}c\bar{d}}$ and
$G_{a\bar{b}c11}^{(1)}$. This follows from the Bianchi identity for
$G^{(1)}$ in eq.~\eqref{S} and the fact that all sources $J^{(n)}$ are
$(2,2)$ forms. For $G^{(1)}$ of this form, the Killing spinor equation
has been analyzed in ref.~\cite{w}. It has been shown in that paper
that the corrections first order in $\e_S$ to the metric and Killing spinor
should have the structure 
\begin{equation}
 g^{(1)}_{\m\n} = b\eta_{\m\n}\; ,\qquad g^{(1)}_{AB} = h_{AB}\; \qquad
 g^{(1)}_{11,11} = \g\; ,\qquad \eta^{(1)} = \psi\eta^{(0)}
\end{equation}
with orbifold and Calabi--Yau dependent functions $b$, $h_{AB}$, $\g$
and $\psi$. Furthermore, in \cite{w} a consistent set of differential
equations has been derived from eq.~\eqref{S0} which determines
$b$, $h_{AB}$, $\g$ and $\psi$ in terms of $G^{(1)}$. An
explicit solution for these differential equations in terms of the dual
antisymmetric tensor $\cB$ defined by
\begin{equation}
 \cH = d\cB = *G^{(1)}
\end{equation}
was presented in ref.~\cite{low1}. In the following, we adopt the harmonic
gauge, $d^*\cB = 0$. Then, since the sources in the Bianchi
identity~\eqref{S} are $(2,2)$ forms, the only nonvanishing components
of $\cB$ are 
\begin{equation}
 \cB_{\m\n\r\s a\bar{b}} = \e_{\m\n\r\s}\cB_{a\bar{b}}
\end{equation}
with $\cB_{a\bar{b}}$ a $(1,1)$ form on the Calabi--Yau space. Using the
results of ref.~\cite{low1}, the Killing spinor equation \eqref{S0} is
solved by
\begin{equation}
\begin{aligned}
  h_{a\bbar} &= \sqrt{2}i \left( \cB_{a\bbar} 
     - \frac{1}{3}\o_{a\bbar}\cB \right) \\
  b &= \frac{\sqrt{2}}{6} \cB  \\
  \g &= -\frac{\sqrt{2}}{3} \cB \\
  \psi &= -\frac{\sqrt{2}}{24} \cB  \\
  G_{ABCD}^{(1)} &= \frac{1}{2}\e_{ABCDEF}\partial_{11}\cB^{EF}  \\
  G_{ABC11}^{(1)} &= \frac{1}{2}\e_{ABCDEF}\partial^D\cB^{EF} 
\end{aligned}
\label{sol} 
\end{equation}
where $\cB = \o^{AB}\cB_{AB}$ and $\o_{a\bar{b}}=-ig_{a\bar{b}}$ is the 
K\"ahler form. We have, therefore, explicitly expressed the complete
background in terms of the $(1,1)$ form $\cB_{a\bar{b}}$. All that
remains then is to determine this $(1,1)$ form, which can be done following
the methods given in ref.~\cite{low1}. In the harmonic gauge, which
implies
\begin{equation}
 D_A^{(0)}\cB^{AB} = 0\; ,
\end{equation}
$\cB_{AB}$ is determined from eq.~\eqref{S} by solving
\begin{multline}
  \left( \D_X + D_{11}^2 \right) \cB_{AB} = 
       2\sqrt{2}\pi\left(\frac{\k}{4\pi}\right)^{2/3}
          \left[ *_X J^{(0)} \d(x^{11}) + *_X J^{(N+1)} \d(x^{11}-\p\r)
          \right. \\ \left.
       + \frac{1}{2}\sum_{n=1}^{N} *_X J^{(n)} 
          \left( \d(x^{11}-x_n) + \d(x^{11}+x_n) \right)
          \right]_{AB}\; .\label{cB}
\end{multline}
where $\D_X$ is the Laplacian and $*_X$ the Hodge star operator on the
Calabi--Yau space. Essentially, this is the equation for a potential
between a set of charged plates positioned through the orbifold
interval at the fixed planes and the five-brane locations. The charge
is not uniform over the Calabi--Yau space. To find a solution,
following ref.~\cite{low1} we introduce eigenmodes $\p_{i\,a\bar{b}}$
of this Laplacian with eigenvalues $-\l_i^2$ so that 
\begin{equation}
  \D_X \p_{i\,a\bbar} = - \l_{i}^2 \p_{i\,a\bbar} \; .
\end{equation}
Generically, $\l_i$ is of order $v^{-1/6}$. The metric on the space of
eigenmodes
\begin{equation}
 G_{ij} = \frac{1}{2v}\int_X\p_i\wedge (*\p_j)
\end{equation}
is used to raise and lower $i$-type indices.
Particularly relevant are the massless modes with $\l_i=0$, which are
precisely the $h^{1,1}$ harmonic $(1,1)$ forms of the Calabi--Yau space.
We will also denote these harmonic $(1,1)$ forms by $\o_{iAB}$. In the
following, in order to distinguish between massless and massive modes,
we will use indices $\iz,\jz,\kz,\dots = 1,\dots,h^{1,1}$ for the former
and indices $\ih,\jh,\kh,\dots$ for the latter, while we continue
to use $i,j,k,\dots$ for all modes. Let us now expand the sources in
terms of the eigenfunctions as
\begin{equation}
 *_X J^{(n)} = \frac{1}{2v^{2/3}} \sum_i\b_{i}^{(n)}\p^i
              \label{Jexp}
\end{equation}
where
\begin{equation}
 \b_i^{(n)} = \frac{1}{v^{1/3}} \int_X\p_i\wedge J^{(n)}\; .
\end{equation}
If we introduce four-cycles $\cC_{4\iz}$ dual to the harmonic $(1,1)$
forms $\o_{\iz}$, we can write for the massless modes
\begin{equation}
 \b_{\iz}^{(n)} = \int_{\cC_{4\iz}}J^{(n)}\; .
\label{bzdef}
\end{equation}
Specifically, it follows from \eqref{J2} that $\b_{\iz}^{(0)}$ and
$\b_{\iz}^{(N+1)}$ represent the instanton numbers of the gauge fields on
the orbifold planes minus half the instanton number of the tangent bundle
and, hence, it would appear, are in general half-integer. However,
since $M_{11}$ must be a spin manifold (since it must admit spinors),
the tangent bundle instanton number must be divisible by
two~\cite{wittq} and so $\b_{\iz}^{(0)}$ and $\b_{\iz}^{(N+1)}$ are, in
fact, integer. Furthermore, $\b_{\iz}^{(n)}$, $n=1,\dots ,N$ are the
five-brane charges, given by the intersection number of each
five-brane with the cycle $\cC_{4\iz}$, and are also integers. Let us
also expand $\cB_{AB}$ in terms of eigenfunctions as 
\begin{equation}
 \cB_{AB} = \sum_ib_i\p_{AB}^i\label{bser}
\end{equation}
Then inserting this expansion, together with the expression~\eqref{Jexp}
for the sources, into eq.~\eqref{cB}, it is straightforward to obtain
\bea
 \left(\partial_{11}^2-\l_i^2\right)b_i &=& \frac{\sqrt{2}\e_S}{\r}
     \left[ \b_i^{(0)}\d (x^{11})
     +\b_i^{(N+1)}\d (x^{11}-\p\r )\right.\nn \\
     &&\qquad\left. + \frac{1}{2}\sum_{n=1}^{N}\b_i^{(n)}(\d (x^{11}-x_n)
       +\d (x^{11}+x_n))\right]
     \label{b}
\eea

It is then easy to solve these equation to give an explicit
solution for the massive and massless modes. We note that the size of
the sources is set by $\e_S/\r$ which, from eq.~\eqref{es},
is independent of the size of the orbifold. We first solve
eq.~\eqref{b} for the massive modes, that is, for $\l_i\neq 0$. In
terms of the normalized orbifold coordinates
\begin{equation}
 z=\frac{x^{11}}{\p\r}\; ,\qquad z_n=\frac{x_n}{\p\r}\; ,\quad n=1,\dots ,N
 \; ,
\end{equation}
$z_0=0$ and $z_{N+1}=1$, we find
\begin{multline}
 b_{\ih} = \frac{\p\e_S}{\sqrt{2}}\d_{\ih}\left[
              \left(\sum_{m=0}^n c_{\ih ,m}\b_{\ih}^m\right)
              \sinh(\d_{\ih}^{-1}|z|)
           \right. \\ \left.  
              + \left( \sum_{m=n+1}^{N+1}s_{\ih ,m}\b_{\ih}^{(m)}
                   - \frac{c_{\ih ,N+1}}{s_{\ih ,N+1}}
                     \sum_{m=0}^{N+1}c_{\ih ,m}\b_{\ih}^{(m)}
              \right)\cosh (\d_{\ih}^{-1}|z|)
           \right]
\label{massive}
\end{multline}
in the interval
\begin{equation}
 z_n\leq |z|\leq z_{n+1}\; ,\nn
\end{equation}
for fixed $n$, where $n=0,\dots ,N$. Here we have defined
\begin{equation}
 \d_{\ih} = \frac{1}{\p\r\l_{\ih}}\; ,\qquad c_{\ih ,n} = \cosh
            (\d_{\ih}^{-1}z_n)\; ,
 \qquad s_{\ih ,n} = \sinh (\d_{\ih}^{-1}z_n)\; .
\end{equation}
Note that, since the eigenvalues $\l_{\ih}$ are of order $v^{-1/6}$, the
quantities $\d_{\ih}$ defined above are of order $\e_R$. Therefore,
as already stated, the size of the massive modes is set by $\e_R\e_S$.

We now turn to the massless modes. First note that, in order to have a
solution of~\eqref{b}, we must have 
\begin{equation}
 \sum_{n=0}^{N+1}\b_{\iz}^{(n)} = 0\; .\label{coh1}
\end{equation}
However, from the definition~\eqref{bzdef}, we see that this is, of
course, none other than the cohomology condition~\eqref{coh} described
above, and so is indeed satisfied. Integrating eq.~\eqref{b} for
$\l_i=0$ we then find~\cite{low1}
\begin{equation}
 b_{\iz} = \frac{\p\e_S}{\sqrt{2}}\left[\sum_{m=0}^n
       \b_{\iz}^{(m)}(|z|-z_m)-\frac{1}{2}\sum_{m=0}^{N+1}(z_m^2-2z_m)
       \b_{\iz}^{(m)}\right]
\label{massless}
\end{equation}
in the interval
\begin{equation}
 z_n\leq |z|\leq z_{n+1}\; ,\nn
\end{equation}
for fixed $n$, where $n=0,\dots ,N$. As already discussed, the massless
modes are of order $\e_S$ and, unlike for the massive modes, no additional
factor of $\e_R$ appears.

It is important to note that there could have been an arbitrary
constant in the zero-mode solutions. However, such a constant can always be
absorbed into a redefinition of the Calabi--Yau zero modes or,
correspondingly, the low energy fields. Consequently, in the
solution~\eqref{massless} we have fixed the constant by taking the
orbifold average of the solution to be zero. This will be important later in
deriving low-energy effective actions. 

Before we discuss the implications of these equations in detail, let us
summarize our results. We have constructed heterotic M--theory backgrounds
with non-standard embeddings including the presence of
bulk five-branes. We started with a standard Calabi--Yau background
with gauge fields and five-branes to lowest order and showed that
corrections to it can be computed in a double expansion in $\e_S$ and
$\e_R$. Explicitly, we have solved the problem to linear
order in $\e_S$ and to all orders in $\e_R$. We found the massive modes
to be of order $\e_R\e_S$ while the massless modes are of order $\e_S$.
Therefore, although one could have expected corrections of arbitrary power
in $\e_R$, we only find zeroth- and first-order contributions at the 
linear level in $\e_S$. Concentrating on the leading order massless modes,
in each interval between two five-branes, $z_n\leq |z|\leq z_{n+1}$, the
massless modes vary linearly with a slope proportional to the total charge
$\sum_{m=0}^n\b_{\iz}^{(m)}$ to the left of the interval. (Note that the total
charge to the right of the interval has the same magnitude but opposite
sign due to eq.~\eqref{coh1}.) At the five-brane
locations, the linear pieces match continuously but with kinks which
lead to the delta-function sources when the second derivative is
computed. (A specific example is given in section 4.1,
see fig.~\ref{fig2}.) Similar kinks appear for the massive modes which,
however, vary in a more complicated way between each pair of
five-branes.

\section{Backgrounds Without Five-Branes}

In this section, we will restrict the previous general solutions to the
case of pure non-standard embedding without additional five-branes and
discuss some properties of such backgrounds and the resulting low-energy
effective actions in both four and five dimensions.

\subsection{Properties of the Background}

To specialize to the case without five-branes, we set $N=0$ and recall
that $z_0=0$ and $z_1=1$. Also, the vanishing cohomology condition~\eqref{coh1}
implies that we have only one independent charge
\begin{equation}
 \b_{\iz} \equiv \b_{\iz}^{(0)}=-\b_{\iz}^{(1)}\; 
\label{onebeta}
\end{equation}
per mode.
Using this information to simplify eq.~\eqref{massless}, we find for the
massless modes
\begin{equation}
 b_{\iz} = \frac{\p\e_S}{\sqrt{2}}\b_{\iz} 
    \left(|z|-\frac{1}{2}\right)\label{massless0}\; .
\end{equation}
In the same way, we obtain from eq.~\eqref{massive} for the massive modes
\begin{equation}
 b_{\ih} = \frac{\p\e_S}{\sqrt{2}}\d_{\ih}\left[ 
           (\b_{\ih}^{(0)}-\b_{\ih}^{(1)})
           \frac{\sinh (\d_{\ih}^{-1}(|z|-1/2))}{2\cosh (\d_{\ih}^{-1}/2)}-
           (\b_{\ih}^{(0)}+\b_{\ih}^{(1)})
           \frac{\cosh (\d_{\ih}^{-1} (|z|-1/2))}{2\sinh (\d_{\ih}^{-1} /2)}
           \right]\; . \label{massive0}
\end{equation}
Note that, unlike for the massless modes, here we have no relation between
the coefficients $\b_{\ih}^{(0)}$ and $\b_{\ih}^{(1)}$. Let us compare these
results to the case of the standard embedding~\cite{low1}. We see that the
massless modes solution is, in fact, completely unchanged in form from the
the standard embedding case, though the parameter $\b_{\iz}$ can be
different. This is a direct consequence of the cohomology
condition~\eqref{coh1} which, for the simple case without
five-branes, tells us that the instanton numbers on the two orbifold
planes always have to be equal and opposite. There is no similar condition
for the massive modes and we therefore expect a difference from the
standard embedding case. Indeed, the standard embedding case is obtained
from eq.~\eqref{massive0} by setting $\b_{\ih}^{(0)}+\b_{\ih}^{(1)}=0$ so that
the second term vanishes. As was noticed in ref.~\cite{low1}, the
first term in eq.~\eqref{massive0} vanishes at the middle of the interval
$z=1/2$ for all modes. Hence, for the standard embedding, at this point
the space-time background receives no correction and, in particular,
the Calabi--Yau space is undeformed. We see that the second term in
eq.~\eqref{massive0} does not share this property. Therefore, for
non-standard embeddings, there is generically no point on the orbifold
where the space-time remains uncorrected.

Furthermore, we see that the massive modes depend on the combination
$\d_{\ih}^{-1}z$ only. Therefore, in terms of the normalized orbifold
coordinate $z$ (the orbifold coordinate $x^{11}$), the massive modes
indeed fall off exponentially with a scale set by $\d_{\ih}$ (by
$v^{1/6}$). In fact, as might be expected, we see that this part of
the solution is essentially independent of the size of the
orbifold. Averaging the above expression for the massive modes over
the orbifold, one should pick up the corresponding weak coupling
correction. Clearly, as a consequence of the exponential fall-off, 
the averaging procedure leads to an additional suppression by $\e_R$.
Given that the order of a heavy mode is $\e_R\e_S$, we conclude that
its average is of the order $\e_R^2\e_S$. According to eq.~\eqref{ew_es},
this is just $\e_W$ and, hence, the expected weak coupling expansion parameter.

\subsection{Low-Energy Effective Actions}

What are the implications of the above results for the low-energy effective
action? Since the orbifold is expected to be larger than the Calabi--Yau
radius, it is natural to first reduce to a five-dimensional effective
theory consisting of the usual $3+1$ space-time dimensions and
the orbifold and, subsequently, reduce this theory further down to four
dimensions. First, we should explain how a background
appropriate for a reduction to $\cN =1$ supersymmetry in four dimensions
can be used to derive a sensible $\cN =1$ theory in five
dimensions~\cite{low1}\footnote{By $\cN=1$ in five dimensions we mean
a theory with eight supercharges. In four dimensions, $\cN=1$ means a
theory with four supercharges.}. The point is that, as we have seen, the
background can be split into massless and massive eigenmodes. Reducing
from eleven to five dimensions on an undeformed Calabi--Yau
background, these correspond to massless moduli fields and heavy
Kaluza--Klein modes. Working to linear order in $\e_S$, the heavy
modes completely decouple from the massless modes and so can
essentially be dropped. The background then appears as a particular
solution to the five-dimensional effective action, where the moduli
depend non-trivially on the orbifold direction. Thus, in summary, to
derive the correct five-dimensional action, we need only keep the
massless modes in a reduction on an undeformed Calabi--Yau space. However,
a similar procedure is not possible for the topologically non-trivial
components $G^{(1)}_{ABCD}$ of the antisymmetric tensor field strength. Such a
configuration of the internal field strength is not a modulus, but
rather a non-zero mode. As a consequence, the proper five-dimensional
theory is obtained as a reduction on an undeformed Calabi--Yau background
but including non-zero modes for $G$. It is these non-zero modes which
introduce all the interesting structure into the theory, notably, that
in the bulk we have a gauged supergravity and that the theory admits no
homogeneous vacuum. In the case at hand, the precise structure of the
non-zero mode can be directly read off from the background as
presented.

Let us now briefly review the results of such a reduction for the
standard embedding as presented
in ref.~\cite{losw,add1} and discussed in Lecture 1. 
It was found that the five-dimensional
effective action consists of a gauged $\cN =1$ bulk supergravity
theory with $h^{1,1}-1$ vector multiplets and $h^{2,1}+1$
hypermultiplets coupled to four-dimensional $\cN =1$ boundary
theories. The field content of the orbifold plane at $x^{11}=0$
consists of an $E_6$ gauge multiplet and $h^{1,1}$ and $h^{2,1}$ 
chiral multiplets, while the plane at $x^{11}=\p\r$ carries $E_8$
gauge multiplets only. The gauging of the bulk supergravity is with
respect to a $U(1)$ isometry in the universal hypermultiplet coset 
space with the gauge field being a certain linear combination of the
graviphoton and the vector fields in the vector multiplets. The gauging
also leads to a bulk potential for the $(1,1)$ moduli.
In addition, there are potentials for the $(1,1)$ moduli confined to the
orbifold planes which have opposite strength. As we have mentioned,
the characteristic features of this theory, such as the gauging and
the existence of the potentials, can be traced back to the existence
of the non-zero mode. Furthermore, the vacuum solution of this
five-dimensional theory, appropriate for a reduction to four
dimensions, was found to be a double BPS domain wall with the two
worldvolumes stretched across the orbifold planes.  

Which of the above features generalize to non-standard embeddings?
The spectrum of zero mode fields in the bulk will, of course, be unchanged.
Due to the nonstandard embedding, we can have more general gauge multiplets
with groups $G^{(1)}, G^{(2)}\subset E_8$ on the orbifold planes and
also corresponding observable and hidden sector matter transforming
under these groups. We are interested in the effective action up to
linear order in $\e_S$. It is clear that, as above, to this order, the
massive part of the background completely decouples from the low-energy
effective action since the massless and massive eigenfunction on the
Calabi--Yau space are orthogonal~\cite{low1}. Hence, the 
form of the effective action to linear order in $\e_S$ is completely
determined by the massless part of the background. On the other hand,
due to the cohomology condition~\eqref{coh1}, the form of the massless
part of the background corrections is same as in the standard
embedding case, as we have just shown. Hence, in deriving the
five-dimensional effective action for non-standard
embedding, we use the same non-zero mode in the reduction as for the
standard embedding. This will lead to gauging and bulk and boundary
potentials exactly as in the standard embedding case. 

Let us explain these last facts in some more detail. First, we
identify the non-zero mode of $G$ in the case of non-standard
embedding. Inserting the mode~\eqref{massless0} into the expansion for
$\cB_{AB}$, eq.~\eqref{bser}, we can use eq.~\eqref{sol} to compute
the four-form field strength $G^{(1)}$. While the massless part of
$G^{(1)}_{ABC11}$ vanishes, we find for the massless part of
$G^{(1)}_{ABCD}$
\begin{equation}
 G^{(1)} = \frac{1}{2V}*\o_{\iz}\a^{\iz}\e (x^{11})\label{nonzero2}
\end{equation}
where $V$ is the Calabi--Yau volume modulus defined by
\begin{equation}
 V=\frac{1}{2\p\r v}\int_{X\times S^1/Z_2}\sqrt{^6g}\label{V}
\end{equation}
and we have introduced the parameter
\begin{equation}
 \a_{i_0} = \frac{\sqrt{2}\e_S}{\r}\b_{i_0}\; .\label{alpha}
\end{equation}
to conform with the notation of \cite{losw,add1}. Furthermore, $\e (x^{11})$
is the stepfunction which is $+1$ for positive $x^{11}$ and $-1$ otherwise. 
Eq.~\eqref{nonzero2} is precisely the non-zero mode we have mentioned
above. Note that $V$ measures the orbifold average of the Calabi--Yau
volume in units of $v$. In general, the parameters $\a_{i_0}$ depend
on the choice of both the tangent and the gauge bundles. Explicitly,
from eqs.~\eqref{J2}, \eqref{bzdef} and the cohomology
condition~\eqref{coh1}, we have, for general embeddings, 
\begin{equation}
\begin{aligned}
    \a_{i_0} &= - \frac{\e_S}{4\sqrt{2}\p^2\r}\int_{\cC_{4i_0}}\left(
          \tr F^{(1)}\wedge F^{(1)} - \frac{1}{2}\tr R\wedge R\right) \\
             &= \frac{\e_S}{4\sqrt{2}\p^2\r}\int_{\cC_{4i_0}}\left(
          \tr F^{(2)}\wedge F^{(2)} - \frac{1}{2}\tr R\wedge R\right)\; .
\end{aligned}
\end{equation}
In the case of the standard embedding, the tangent bundle and one of the
$E_8$ gauge bundles are identified, while the other gauge bundle is
taken to be trivial, so that this reduces to 
\begin{equation}
 \a_{i_0} = - \frac{\e_S}{8\sqrt{2}\p^2\r}\int_{\cC_{4i_0}}\tr R\wedge R\; .
\end{equation}
This is the relation given in ref.~\cite{add1}. The point is that the
expression for the non-zero mode~\eqref{nonzero2} has the same form for
both standard and non-standard embeddings. All that changes are the
values of the parameters $\a_{i_0}$.  

Now let us demonstrate how the gauging of the bulk supergravity arises
in the case of non-standard embedding. Consider the five-dimensional
three-form zero--mode $C_5$, with field strength $G_5$, and the part of the
11--dimensional three-form that leads to the $h^{1,1}$ vector fields
$\cA^{\iz}$, namely $C=\cA^{\iz}\wedge\o_{\iz}$. Inserting these two
fields, together with the non-zero mode~\eqref{nonzero2}, into the
Chern--Simons term in the eleven-dimensional supergravity
action~\cite{add1} leads to 
\begin{equation}
 \int_{M_{11}}C\wedge G\wedge G \sim 
     \int_{M_5}\e (x^{11})\a_{\iz}\cA^{\iz}\wedge G_5\; .
 \label{gauging}
\end{equation}
The three-form $C_5$ can be dualized to a scalar in five dimensions,
which becomes one of the four scalars $q^u$ in the universal
hypermultiplet. Then, the above term directly causes the gauging of
the isometry in the hypermultiplet coset space that corresponds to the
axionic shift in the dual scalar. The gauging is with respect to the
linear combination $\a_{\iz}\cA^{\iz}$. Explicitly, we
find~\cite{add1} that the universal hypermultiplet kinetic term is of
the form
\begin{equation}
   \int_{M_5} \sqrt{-g} h_{uv} D_\a q^u D^\a q^v
\end{equation}
with the covariant derivative 
\begin{equation}
   D_\a q^u = \pt_\a q^u + \e (x^{11}) \a_{i_0}\cA^{i_0}_\a k^u
\end{equation}
where $k^u$ is a Killing vector in the hypermultiplet sigma-model
manifold, pointing in the direction of the axionic shift. We see that,
since the non-zero mode~\eqref{nonzero2} had the same form for both
standard and non-standard embeddings, the gauging of the supergravity
also has the same form. The only difference is in the values of the
charges $\a_{i_0}$. 

Similarly, the bulk potential should have the same form in the
standard and non-standard embedding cases. Inserting the
non-zero mode~\eqref{nonzero2} into the kinetic term $G\wedge *G$ of
the four-form field strength in the eleven-dimensional supergravity
action leads to a bulk potential for the volume modulus $V$ and the
other $(1,1)$ moduli. More precisely, 
one finds
\begin{equation}
 \int_{M_{11}}G^{(1)}\wedge *G^{(1)} \sim \int_{M_5}\sqrt{-g}
   V^{-2}\a_{i_0}\a_{j_0}\tilde{G}^{i_0j_0}
\label{potential}
\end{equation}
where
\begin{equation}
 \tilde{G}_{i_0j_0}=V^{2/3}G_{i_0j_0}
\end{equation}
is a renormalized metric that depends on the Calabi--Yau shape moduli
(see ref.~\cite{add1} for details). Note that it follows from
supersymmetry that such a potential must arise when an isometry of the
universal hypermultiplet sigma-model manifold is gauged. 

The potentials on the orbifold planes arise from the ten-dimensional
actions on the planes, with the internal gauge fields and
curvature inserted. Using identities of the form 
\begin{equation}
   \int_X \o\wedge \tr R \wedge R \sim \int_X \sqrt{-g}\tr R^2
\end{equation}
we find
\begin{equation}
 \sum_{n=1}^2\int_{M_{10}^{(n)}}\sqrt{-g}\left(\tr (F^{(n)})^2-\frac{1}{2}
     \tr R^2\right)
  \sim \int_{M^{(1)}_4}\sqrt{-g}V^{-1}\a_{i_0} b^{i_0} 
       -  \int_{M^{(2)}_4}\sqrt{-g}V^{-1}\a_{i_0} b^{i_0}
\label{YM}
\end{equation}
where $b^{i_0}$ are the K\"ahler shape moduli defined by the expansion
of the K\"ahler form $\o=V^{1/3}b^{i_0}\o_{i_0}$. 
As for the standard embedding case, the potentials come out with
opposite strength, again a consequence of the cohomology
condition~\eqref{onebeta}, $\b_{i_0}^{(0)}=-\b_{i_0}^{(1)}$.

In summary, we conclude that the five-dimensional effective
action derived in ref.~\cite{losw,add1} for the standard embedding
is, in fact, much more general and applies, with appropriate adjustment
of the boundary field content and the charges $\a_{i_0}$, to any
Calabi--Yau-based non-standard embedding without additional
five-branes. Furthermore, the double domain wall vacuum solution of
the five-dimensional theory is unchanged, since it does not depend on
the field content on the orbifold planes. 

The four-dimensional theory is
obtained as a reduction on this domain wall. Hence, the
four-dimensional effective action will be unchanged in the case of
non-standard embeddings without five-branes, except for the possibility of more
general gauge groups and matter multiplets. One further new feature, 
in the case of non-standard embedding, is the possibility 
of gauge matter on the hidden orbifold plane. In this case,
the threshold-like correction to the matter part of the K\"ahler
potential will be different for observable and hidden sectors in the
same way the gauge kinetic functions of the two sectors differ. 

To be more concrete, let us consider the universal case with moduli
$S$ and $T$, gauge fields of $G^{(1)}\times G^{(2)}\subset E_8\times
E_8$ and corresponding gauge matter $C^{(1)}$ and $C^{(2)}$,
transforming under $G^{(1)}$ and $G^{(2)}$, respectively. Then, we have
for the K\"ahler potential and the gauge kinetic functions
\bea
 K &=& -\log (S+\bar{S})-3\log(T+\bar{T})+Z_1|C^{(1)}|^2+Z_2|C^{(2)}|^2\nn \\
 Z_1 &=& \frac{3}{T+\bar{T}}+\frac{\p\e_S\b}{S+\bar{S}}\nn \\
 Z_2 &=& \frac{3}{T+\bar{T}}-\frac{\p\e_S\b}{S+\bar{S}} 
    \label{gkf}\\
 f^{(1)} &=&S+\p\e_S\b T\nn \\
 f^{(2)} &=&S-\p\e_S\b T\nn\; .
\eea
where $\b$ is the single instanton charge, of the type defined in
eqn.~\eqref{onebeta}, corresponding to the universal K\"ahler deformation.
For vacua based on the standard embedding, it was pointed out in
ref.~\cite{w} that, if $\b>0$ so that the smaller of the two
couplings corresponds to the observable sector, then, fitting this to
the grand unification coupling, the larger coupling is of order one at the
``physical'' point. Hence, gaugino condensation in the hidden sector
appears to be a likely scenario. We have just shown that, in fact, this
statement continues to apply to all Calabi--Yau based non-standard
embedding vacua without additional bulk five-branes, provided $\b>0$,
since the gauge kinetic functions are completely unchanged. Gaugino
condensation, therefore, appears to be a generic possibility for such
vacua. 

\section{Backgrounds with Five-Branes}

Let us now turn to the much more interesting case of non-standard
embeddings with five-branes in the bulk. We will concentrate on the
massless modes, since, as above, it is these modes which will determine the
low-energy action.  

\subsection{Properties of the Background}

\begin{figure}[t]
   \centerline{\psfig{figure=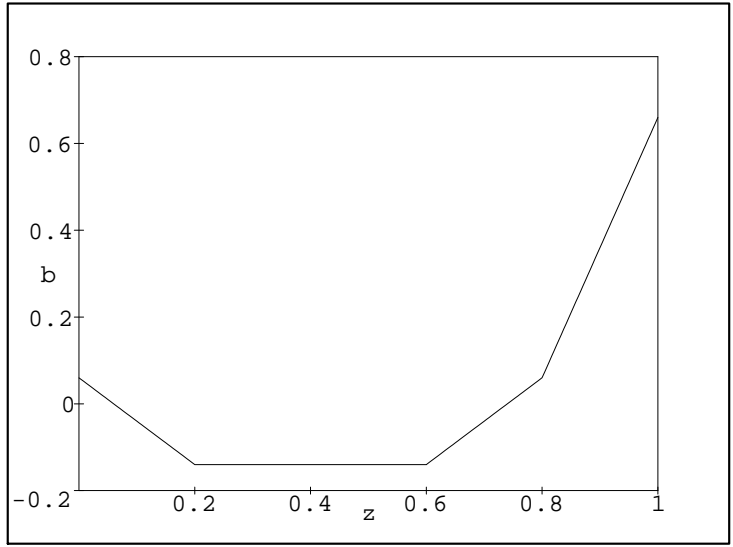,height=3in}}
   \caption{Orbifold dependence of a massless mode
   $(\sqrt{2}/\p\e_S)b$ for four five-branes at
   $(z_1,z_2,z_3,z_4)=(0.2,0.6,0.8,0.8)$ with charges
   $(\b^{(1)},\b^{(2)},\b^{(3)},\b^{(4)})=(1,1,1,1)$ and 
   instanton numbers $(\b^{(0)},\b^{(4)})=(-1,-3)$.} 
   \label{fig2}
\end{figure}

The general solution~\eqref{massless} for the massless modes shows a linear
behaviour for each interval between two five-branes. The slope, however,
varies from interval to interval in a way controlled by the five-brane
charges. The same statement applies to the variation of geometrical
quantities, like the Calabi--Yau volume, across the orbifold. Let us consider
an example for a certain massless mode $b$. Four five-branes with charges
$(\b^{(1)},\b^{(2)},\b^{(3)},\b^{(4)})=(1,1,1,1)$ are positioned at
$(z_1,z_2,z_3,z_4)=(0.2,0.6,0.8,0.8)$. Note that the third and fourth
five-brane are coincident. The instanton numbers on the orbifold
planes are chosen to be $(\b^{(0)},\b^{(4)})=(-1,-3)$. Note that the
total charge sums up to zero as required by the cohomology
constraint~\eqref{coh1}. The orbifold dependence of
$(\sqrt{2}/\p\e_S)b$ is depicted in fig.~\ref{fig2}. 
It is clear that the additional five-brane charges introduce much more
freedom as compared to the case without five-branes. For example, while
in the latter case one always has $b(0)=-b(1)$ leading to equal, but
opposite, gauge threshold corrections, the example in fig.~\ref{fig2}
shows that $b(0),b(1)>0$ is possible. One, therefore, expects the thresholds
in the low-energy gauge kinetic functions to change. This will be
analyzed in a moment. Another interesting phenomenon in the above example
is that the mode is constant between the first and second five-brane.
This is a direct consequence of our choice of the charges which sum up
to zero both to the left and the right of this interval. If such a
property is arranged for all massless modes, the Calabi--Yau volume
remains exactly constant throughout this interval.

\subsection{Five-Branes on Calabi--Yau Two-Cycles}

The inclusion of five-branes not only generalizes the types of
background one can consider, but also introduces new degrees of
freedom into the theory, namely, the dynamical fields on the
five-branes themselves. In this section, we will consider what
low-energy fields survive on one of the five-branes when it is wrapped
around a two-cycle in the Calabi--Yau three-fold. 

In general, the fields on a single five-brane are as
follows~\cite{GT,KM}. The 
simplest are the bosonic coordinates $X^I$ describing the embedding of
the brane into 11-dimensional spacetime. The additional bosonic
field is a world-volume two-form potential $B$ with field strength $H=dB$
satisfying a generalized self-duality condition. For small
fluctuations, the duality condition simplifies to the conventional
constraint $H=*H$. These degrees of freedom are paired with spacetime
fermions $\theta$, leading to a Green--Schwarz type action, with
manifest spacetime supersymmetry and local
kappa-symmetry~\cite{BLNPST,APPS}. (As usual, including the self-dual
field in the action is difficult, but is possible by either including
an auxiliary field or abandoning a covariant formulation.) For a
five-brane in flat space, one can choose a gauge such that the
dynamical fields fall into a six-dimensional massless tensor multiplet
with $(0,2)$ supersymmetry on the brane
world-volume~\cite{Kallosh,CKvP}. The multiplet has five scalars
describing the motion in directions transverse to the five-brane,
together with the self-dual tensor $H$. 

For a five-brane embedded in $S^1/Z_2\times X \times M_4$, to preserve
Lorentz invariance in $M_4$, $3+1$ dimensions of the five-brane must
be left uncompactified. The remaining two spatial dimensions are then
wrapped on a two-cycle of the Calabi--Yau three-fold. To preserve
supersymmetry, the two-cycle must be a holomorphic
curve~\cite{w,bbs,vb}. Thus, from the point of view of a
five-dimensional effective theory on $S^1/Z_2\times M_4$, since two of
the five-brane directions are compactified, it appears as a flat
three-brane (or equivalently domain wall) located at some point
$x^{11}=x$ on the orbifold. Thus, at low energy, the degrees of
freedom on the brane must fall into four-dimensional supersymmetric
multiplets.

\vspace{0.4cm}

An important question is how much supersymmetry is preserved in the
low-energy theory. One way to address this problem is directly from
the symmetries of the Green--Schwarz action, following the discussion
for similar brane configurations in~\cite{bbs}. Locally, the
11-dimensional spacetime $S^1/Z_2\times X\times M_4$ admits
eight independent Killing spinors $\eta$, so should be described by a
theory with eight supercharges. (Globally, only half of the spinors
survive the non-local orbifold quotienting condition
$\G_{11}\eta(-x^{11})=\eta(x^{11})$, so that, for instance, the
eleven-dimensional bulk fields lead to $\cN=1$, not $\cN=2$,
supergravity in four dimensions.) The Green--Schwarz form of the
five-brane action is then invariant under supertranslations generated
by $\eta$, as well as local kappa-transformations. In general the
fermion fields $\theta$ transform as (see for instance
ref.~\cite{CKvP})
\begin{equation}
   \d\theta = \eta + P_+\k
\end{equation}
where $P_+$ is a projection operator. If the brane configuration is
purely bosonic then $\theta=0$ and the variation of the bosonic fields
is identically zero. Furthermore, if $H=0$ then the projection
operator takes the simple form
\begin{equation}
   P_\pm = \frac{1}{2}\left( 1 \pm \frac{1}{6!\sqrt{g}}\e^{m_1\ldots m_6}
          \pt_{m_1}X^{I_1}\dots\pt_{m_6}X^{I_6}\G_{I_1\ldots I_6} \right)
\end{equation}
where $\s^m$, $m=0,\dots ,5$ label the coordinates on the five-brane
and $g$ is the determinant of the induced metric
\begin{equation}
   g_{mn} = \pt_m X^I \pt_n X^J g_{IJ}\; .
\end{equation}

If the brane configuration is invariant for some combination of
supertranslation $\eta$ and kappa-transformation, then we say it is
supersymmetric. Now $\k$ is a local parameter which can be chosen at
will. Since the projection operators satisfy $P_++P_-=1$, we see that
for a solution of $\d\theta=0$, one is required to set $\kappa=-\eta$,
together with imposing the condition 
\begin{equation}
   P_-\eta = 0
\label{branesusy}
\end{equation} 
For a brane wrapped on a two-cycle in the Calabi--Yau space, spanning
$M_4$ and located at $x^{11}=x$ in the orbifold interval, we
can choose the parameterization 
\begin{equation}
   X^\m = \s^\m \qquad 
   X^A = X^A(\s,\bar{\s}) \qquad
   X^{11} = x
\end{equation}
where $\s=\s^4+i\s^5$. The condition~\eqref{branesusy} then reads 
\begin{equation}
   - \left(i/\sqrt{g}\right) \pt X^A \bar{\pt} X^B \G^{(4)}\G_{AB} \,\eta
        = \eta
\end{equation}
where we have introduced the four-dimensional chirality operator
$\G^{(4)}=\G_0\G_1\G_2\G_3$. Recalling that on the Calabi--Yau three-fold the
Killing spinor satisfies $\G^{\bar{b}}\eta=0$, it is easy to show that
this condition can only be satisfied if the embedding is holomorphic,
that is $X^a=X^a(\s)$, independent of $\bar{\s}$. The condition then
further reduces to 
\begin{equation}
   \G^{(4)}\eta = i \eta
\label{4dchiral}
\end{equation}
which, given that the spinor has definite chirality in eleven dimensions as
well as on the Calabi--Yau space, implies that $\G^{11}\eta=\eta$,
compatible with the global orbifold quotient condition. Thus, finally,
we see that only half of the eight Killing spinors, namely those
satisfying~\eqref{4dchiral}, lead to preserved supersymmetries on the
five-brane. Consequently the low-energy four-dimensional theory
describing the five-brane dynamics will have $\cN=1$ supersymmetry. 

\vspace{0.4cm}

The simplest excitations on the five-brane surviving in the low-energy
four-dimensional effective theory are the moduli describing the
position of the five-brane in eleven dimensions. There is a single
modulus $X^{11}$ giving the position of the brane in the orbifold
interval. In addition, there is the moduli space of holomorphic curves
$\cC_2$ in $X$ describing the position of the brane in the
Calabi--Yau space. This moduli space is generally complicated, and we will
not address its detailed structure here. (As an example, 
the moduli space of genus one curves in K3 is K3
itself~\cite{vb}.) However, we note that these moduli are scalars in 
four dimensions, and we expect them to arrange themselves as a set of chiral
multiplets, with a complex structure presumably inherited from that of
the Calabi--Yau manifold. 

Now let us consider the reduction of the self-dual three-form degrees
of freedom. (Here we are essentially repeating a discussion given
in~\cite{wbranes,KLMVW}.) The holomorphic curve is a Riemann surface
and, so, is characterized by its genus $g$. One recalls that the number
of independent harmonic one-forms on a Riemann surface is given by
$2g$. In addition, there is the harmonic volume two-form
$\Omega$. Thus, if we decompose the five-brane world-volume as
$\cC_2\times M_4$, we can expand $H$ in zero modes as 
\begin{equation}
   H=da\wedge\O+F^u\wedge\l_u+h
\end{equation}
where $\l_u$ are a basis $u=1,\dots ,2g$ of harmonic one-forms on
$\cC_2$, while the four-dimensional fields are a scalar $a$, $2g$
$U(1)$ vector fields $F^u=dA^u$ and a three-form field strength
$h=db$. However, not all these fields are independent
because of the self-duality condition $H=*H$. Rather, one easily concludes
that 
\begin{equation}
   h=*da
\end{equation}
and, hence, that the four-dimensional scalar $a$ and two-form
$b$ describe the same degree of freedom. To analyze the vector
fields, we introduce the matrix ${T_u}^v$ defined by 
\begin{equation}
   *\l_u = {T_u}^v\l_v
\end{equation}
If we choose the basis $\l_u$ such that the moduli space metric
$\int_{\cC_2}\l_u\wedge (*\l_v)$ is the unit matrix, $T$ is antisymmetric and,
of course, $T^2=-1$. The self-duality constraint implies for the
vector fields that
\begin{equation}
 F^u={T_v}^u*F^v\; .
\end{equation}
If we choose a basis for $F^u$ such that
\begin{equation}
 T={\rm diag}\left(\left(\begin{array}{cc}0&1\\-1&0\end{array}\right),\dots ,
   \left(\begin{array}{cc}0&1\\-1&0\end{array}\right)\right)
\end{equation}
with $g$ two by two blocks on the diagonal, one easily concludes that only
$g$ of the $2g$ vector fields are independent. In conclusion, for a genus
$g$ curve $\cC_2$, we have found one scalar and $g$ $U(1)$ vector
fields from the two-form on the five-brane worldvolume. The
scalar has to pair with another scalar to form a chiral $\cN =1$
multiplet. The only other universal scalar available is the zero mode
of the transverse coordinate $X^{11}$ in the orbifold direction. 

Thus, in general, the $\cN=1$ low-energy theory of a single five-brane 
wrapped on a genus $g$ holomorphic curve $\cC_2$ has gauge group $U(1)^g$
with $g$
$U(1)$ vector multiplets and a universal chiral multiplet with bosonic
fields $(a,X^{11})$. Furthermore, there is some number of additional chiral multiplets
describing the moduli space of the curve $\cC_2$ in the
Calabi--Yau three-fold. 

\vspace{0.4cm}

It is well known that when two regions of the five-brane world-volume in
M--theory come into close proximity, new massless states
appear~\cite{wittq,strom}. These are associated with membranes
stretching between the two nearly overlapping five-brane surfaces. 
In general, this can lead to
enhancement of the gauge symmetry. Let us now
consider this possibility, heretofore ignored in our discussion. In
general, one can consider two types of brane degeneracy where parts of
the five-brane world-volumes are in close proximity. The first, and
simplest, is to have $N$ distinct but coincident five-branes, all
wrapping the same cycle $\cC_2$ in the Calabi--Yau space and 
all located at the same
point in the orbifold interval. Here, the new massless states come from
membranes stretching between the distinct five-brane world-volumes. The
second, and more complicated, situation is where there is a degeneracy
of the embedding of a single five-brane, such that parts of the curve
$\cC_2$ become close together in the Calabi--Yau space. 
In this case, the new
states come from membranes stretching between different parts of the
same five-brane world-volume. These two situations were studied in~\cite{us}.
Summarizing the two cases, we found that for $N$ five-branes wrapping the
same curve $\cC_2$ of genus $g$, we expect that the symmetry is enhanced from
$N$ copies of $U(1)^g$ to $U(N)^g$. Alternatively in the second case,
even for a single brane, we can get enhancement if the embedding
degenerates. In general, $U(1)^g$ enhances to a product of unitary
groups such that the total rank is equal to $g$. The maximal
enhancement is presumably to $SU(g+1)$, and the other allowed groups
correspond to different ``Higgsings'' of $SU(g+1)$ by fields in the
adjoint repesentation. For example, if $g=2$, then $SU(3)$ could be
broken to either $SU(2)\times U(1)$ or $U(1)\times U(1)$. In
all cases, the total rank of the symmetry group is conserved. Finally,
we note that in the case where the Calabi--Yau space itself
degenerates to become a singular orbifold, and the five-branes are
wrapped at the singularity, we could expect more exotic enhancement, in
particular, to gauge groups other than unitary groups. In this paper,
however, we will restrict ourselves to the case of smooth Calabi--Yau
spaces.

\subsection{Low Energy Effective Actions}

Next, we would like to discuss the five-dimensional effective actions that
result from the reduction of Ho\v rava--Witten theory on a background
that includes five-branes.
It has already been explained in section 3.2 how the vacua without
five-branes found in this
paper can be used to construct a sensible five-dimensional theory.
Essentially the same arguments apply here. We begin with the
five-dimensional bulk theory. Clearly, the zero-mode content is
unchanged with respect to the case without five-branes. Thus
we have $\cN =1$ supergravity coupled to $h^{1,1}-1$ vector
multiplets and $h^{2,1}+1$ hypermultiplets. What about the gauging of
the hypermultiplet coset space? Inserting the massless
modes~\eqref{massless} into eq.~\eqref{bser} and calculating $G^{(1)}$
via eq.~\eqref{sol} one finds 
\begin{equation}
 G^{(1)} = \frac{1}{2V}(*\o_{\iz})\sum_{m=0}^n\a^{(m),\iz}\e (z)
\end{equation}
in the interval
\begin{equation}
 z_n\leq |z|\leq z_{n+1}
\end{equation}
for fixed $n$, where $n=0,\dots ,N$, and as in eqn.~\eqref{alpha} we
have introduced the parameters
\begin{equation}
   \a^{(m)}_{i_0} = \frac{\sqrt{2}\e_S}{\rho} \b^{(m)}_{i_0}
\end{equation}
to conform with the notation of \cite{losw,add1}.
Hence, we still have a non-zero mode that must be taken into account in the
dimensional reduction. Its form, however, depends on the interval one
is considering. Consequently, the five-dimensional action again contains a
term of the form~\eqref{gauging}, but with $\a_{\iz}$ being replaced by
$\sum_{m=0}^n\a_{\iz}^{(m)}$ for the interval $z_n\leq |z|\leq z_{n+1}$.
In other words, we have gauging in the bulk between each two five-branes
but the gauge charge differs from interval to interval. Since the
bulk potential~\eqref{potential} is directly related to the gauging, it is
subject to a similar replacement of charges. In summary, we conclude that
the bulk theory between any pair of neighboring five-branes in the interval
$z_n\leq |z|\leq z_{n+1}$ is as given in ref.~\cite{losw,add1}, but with
$\a_{\iz}$ replaced by $\sum_{m=0}^n\a_{\iz}^{(m)}$.

We now turn to the orbifold planes. They are described by four-dimensional
$\cN =1$ theories at $x^{11}=0,\p\r$ coupled to the bulk. 
The zero mode spectrum on these
planes is, of course, unchanged with respect to the situation without
five-branes. It consists of gauge multiplets corresponding
to the unbroken gauge groups $G^{(1)}$ and $G^{(2)}$, as dictated by
the choice of the internal gauge bundle, and corresponding gauge matter
multiplets. The height of the boundary potentials (see 
eqn.~\eqref{YM}) is now set by the charges $\a_{i_0}^{(0)}$ and
$\a_{i_0}^{(N+1)}$ which, due the presence of additional five-brane
charges, are no longer necessarily equal and opposite. 

Finally, we should consider the worldvolume theories of the three-branes
that originate from wrapping the  five-branes around supersymmetric
cycles. Applying the results of the previous subsection to each of the $N$
five-branes, we have $N$ additional four-dimensional $\cN =1$ theories at
$x^{11}=x_1,\dots ,x_N$ which couple to the five-dimensional bulk. The
field content of such a theory at $x^{11}=x_n$ for $n=1,\dots ,N$ is
generically given by $U(1)^{g_n}$ gauge multiplets, where $g_n$ is the genus
of the holomorphic
curve on which the $n$-th five-brane is wrapped, a universal chiral multiplet
and a number of additional chiral multiplets describing the moduli space
of the holomorphic curve within the Calabi--Yau manifold. By the mechanisms
described at the end of the previous subsection, the $U(1)^{g_n}$ gauge
groups can be enhanced to non-Abelian groups. As the simplest example,
two five-branes located at $x^{11}=x_n$ and $x^{11}=x_{n+1}$ could be
wrapped on the
same Calabi--Yau cycle with genus $g_n$. As long as two five-branes are
separated in the orbifold, that is, $x_{n+1}\neq x_n$, we have two gauge groups
$U(1)^{g_n}$, one group on each brane. However, when the two five-branes coincide, that is, 
for $x_{n+1}=x_n$, these groups are enhanced to
$U(2)^{g_n}$. The precise form of the three-brane world-volume theories
should be obtained by a reduction of the five-brane world-volume theory
on the holomorphic curves, in a target space background of the
undeformed Calabi--Yau space together with the non-zero mode for the
four-form field strength. We expect those three-brane
theories to have a potential depending on the moduli living on the
three-brane and the projection of the bulk moduli to the three-brane
world-volume. This expectation is in analogy with the theories on the orbifold
planes which, as we have seen, possess such a potential. It has been
shown in ref.~\cite{losw,add1} that those boundary potentials provide the
source terms for a BPS double-domain wall solution of the five-dimensional
theory in the absence of additional five-branes. This double domain wall
is the appropriate background for a further reduction to four dimensions.
Again, in analogy, we expect the vacuum of the five-dimensional theory
in the presence of five-branes to be a BPS multi-domain wall. More
precisely, for $N$ five-branes, we expect $N+2$ domain walls with
two world-volumes stretching across the orbifold planes and the remaining
$N$ stretching across the three-brane planes. The r\^ole of the potentials on the
three-brane world-volume theories is to provide the $N$ additional
source terms needed to support such a solution.

Let us finally discuss some consequences for the four-dimensional
effective theory. Clearly, there is a sector of the theory which has just
the conventional field content of four-dimensional $\cN =1$ low-energy
supergravities derived from string theory. More precisely, this is
$h^{1,1}+h^{2,1}$ chiral matter multiplets containing the moduli, gauge multiplets
with gauge group $G^{(1)}\times G^{(2)}\subset E_8\times E_8$ and
corresponding gauge matter. In the presence of five-branes, however,
we have additional sectors of the four-dimensional theory leading
to additional chiral multiplets containing the five-brane moduli and,
even more important, to gauge multiplets with generic gauge group
\begin{equation}
 G=\prod_{n=1}^N U(1)^{g_n}\; .
\end{equation}
At specific points in the five-brane moduli space, one expects enhancement
to a non-Abelian group $G=G_1\times\cdots\times G_M$. As explained above,
in typical cases, the factors $G_m$ can be $U(n)$ and $SU(n)$
groups. We expect the enhancement to preserve the rank, that is, we
have 
\begin{equation}
 {\rm rank}(G)=\sum_{n=1}^{N}g_n\; .
\end{equation}
We recall that $g_n$ is the genus of the curve on which the $n$-th five-brane
is wrapped. As it stands, it appears that the rank could be made arbitrarily
large. However, for a given Calabi--Yau space, we expect a constraint on
the rank which originates from positivity constraints in the the
zero-cohomology condition~\eqref{ccon}. As is, the five-brane sectors and the conventional
sector of the theory only interact via the bulk supergravity
fields. Therefore, at this point, they are most naturally interpreted
as hidden sectors. 

We should, however, point out that the presence of five-branes provides
considerably more flexibility in the choice of $G^{(1)}\times G^{(2)}$,
the ``conventional'' gauge group that originates from the heterotic
$E_8\times E_8$.
This happens because it is much simpler to satisfy the zero cohomology
condition~\eqref{ccon} in the presence of five-branes. Let us give an
an example which is illuminating, although not necessary physically
relevant. Consider a Calabi--Yau space $X$ with topologically nontrivial
$\tr R\wedge R$. In addition, we set both $E_8$ gauge field backgrounds 
to zero, which
implies that the unbroken gauge group is simply $E_8\times E_8$.
Without five-branes, such a background is inconsistent since it is in
conflict with the zero-cohomology condition~\eqref{ccon}. However, if
for each independent four-cycle $\cC_{4i_0}$, we can introduce
$N_{i_0}$ five-branes, each having unit intersection number with the
cycle $\cC_{4i_0}$, such that 
\begin{equation}
 N_{i_0}=-\frac{1}{8\p^2}\int_{\cC_{4i_0}}\tr R\wedge R
\end{equation}
then the zero-cohomology condition is satisfied. Of course, the gauge group
will then be enlarged to $E_8\times E_8\times G$ where the gauge group $G$
originates from the five-branes, as discussed above.

What about the form of the four-dimensional effective action? We have
seen that non-standard embedding without five-branes does not change the
form of the effective action with respect to the standard embedding case.
This could be understood as a direct
consequence of the fact that the five-dimensional effective theory
remains unchanged. Above we have seen, however, that the five-dimensional
effective theory does change in the presence of five-branes. In particular,
its vacuum BPS solution is now a multi-domain wall, as opposed to a
double-domain wall in the case without five-branes. Hence, we expect
the four-dimensional theory obtained as a reduction on this multi-domain
wall to change as well. Let us, as an example of this, calculate the
gauge kinetic functions in four dimensions to linear order in $\e_S$.
Here, we will not do this using the five-dimensional theory but,
equivalently, reduce directly from eleven to four dimensions. We define the
modulus $R$ for the orbifold radius by
\begin{equation}
 R=\frac{1}{2V\p\r}\int_{S^1/Z_2\times X}\sqrt{^7g}\; .
\end{equation}
Note that with this definition, $R$ measures the averaged orbifold size
in units of $2\p\r$. Let us also introduce the $(1,1)$ moduli $a^{i_0}$
in the usual way as
\begin{equation}
 \o_{AB}=a^{\iz}\o_{\iz AB}\; .
\end{equation}
Then, the real parts of the low energy fields $S$ and $T^i$ are given by
\begin{equation}
 {\rm Re}(S)=V\; ,\qquad {\rm Re}(T^{\iz})=VR^{-1}a^{\iz}\; .
\end{equation}
We stress that with these definitions, $S$ and $T^{\iz}$ have the standard
K\"ahler potential, that is, the order $\e_S$ corrections to the
K\"ahler potential vanish~\cite{low1}.
The gauge kinetic functions can be directly read off from the 10--dimensional
Yang--Mills actions~\eqref{YM}. Using the metric from eq.~\eqref{sol}
with \eqref{massless}, \eqref{bser} inserted and the above definition of
the moduli, we find
\bea
 f^{(1)} &=& S + \p\e_S T^{\iz}\sum_{n=0}^{N+1}(1-z_n)^2
             \b_{\iz}^{(n)} \\
 f^{(2)} &=& S + \p\e_S T^{\iz}\sum_{n=1}^{N+1}{z_n}^2
             \b_{\iz}^{(n)} \; ,
\eea
where, in addition, we have the cohomology constraint~\eqref{coh1}.
Recall from eq.~\eqref{gkf} that in the case without five-branes, the
threshold correction on the two orbifold planes are identical but opposite
in sign. Note that here the expressions for these two thresholds are,
in fact, different. The possiblity of such an asymmetry due to five--branes
has also been suggested in ref.~\cite{stieb}. If, for example, there is only
one five-brane with charges $\b_{\iz}^{(1)}$ at $z=z_1$ on the orbifold, we
have 
\begin{equation}
 f^{(1)}-f^{(2)} = 2\p\e_S T^{\iz}\left[\b_{\iz}^{(0)}+(1-z_1)
                   \b_{\iz}^{(1)}\right]\; .
\end{equation}
We see that the gauge thresholds on the orbifold planes depend on both
the position and the charges of the additional five-branes in the bulk.
This gives considerably more freedom than in the case without five-branes.
In particular, for special choices of the charges and the five-brane
position, the difference of the gauge kinetic functions can be small.
Thus, for instance, the hidden gauge coupling at the physical point
need not be as large as it was in the case without five-branes.

\section*{Lecture 3: Holomorphic Vector Bundles and Non-Perturbative Vacua}

As discussed in Lecture 2, the results of \cite{us} indicated 
the importance of heterotic $M$-theories with
non-standard embeddings and non-perturbative vacua, but did not actually
construct such theories. This shortcoming was rectified in
\cite{don1}, where
explicit constructions were carried out within the context of holomorphic
vector bundles on the orbifold planes of heterotic $M$-theory compactified on
elliptically fibered Calabi--Yau three--folds which admit a section. 
The results of \cite{don1} rely
upon recent mathematical work by Friedman, Morgan and Witten \cite{FMW}, 
Donagi \cite{D} and
Bershadsky, Johansen, Pantev and Sadov \cite{BJPS} 
who show how to explicitly construct such vector bundles, and on results
of \cite{cur,ba} who computed the family generation index 
in this context. Extending
these results, we were able to formulate rules for constructing three-family
particle physics theories with phenomenologically interesting gauge groups. As
expected, the appearance of gauge groups other than the $E_{6}$ group of the
standard embedding, as well as the three-family condition, necessitate the
existence of $M5$-branes and, hence, non-perturbative vacua. In 
\cite{don1}, we showed
how to compute the topological class of these five-branes and, given this
class, how to construct the moduli spaces of the associated holomorphic
curves. Our results were presented as a set of rules in \cite{don1}. 
In addition, we
gave one concrete example of a three-family model with gauge group
$SU(5)$, along with its five-brane class and a discussion of the 
moduli space of that class.
 
In~\cite{usnew}, we greatly enlarged the discussion of the results in
\cite{don1},
deriving in detail the rules presented there. In order to make this work more
accessible to physicists, as well as to lay the foundation for the necessary 
derivations and proofs, we presented brief discussions of $1.$ elliptically
fibered Calabi--Yau three--folds, $2.$ spectral cover constructions of both
$U(n)$ and $SU(n)$ bundles, $3.$ Chern classes and $4.$ complex surfaces,
specifically del Pezzo, Hirzebruch and Enriques surfaces. Using this background, we
explicitly derived the rules for the construction of three-family models based
on semi-stable holomorphic vector bundles with structure group $SU(n)$.
Specifically, we constructed the form of the five-brane class $[W]$, as well as
the constraints imposed on this class by the three-family condition, the
restriction that the vector bundle have structure group $SU(n)$ and the
requirement that $[W]$ be an effective class. From these considerations, we
derived a set of rules. As discussed in that
paper, elliptically fibered Calabi--Yau three--folds that admit a section can
only have del Pezzo, Hirzebruch, Enriques and blown-up Hirzebruch surfaces as
a base. We showed, however, that Enriques surfaces can never
lead to effective five-brane curves in vacua with three generations.
Therefore, the base $B$ of the elliptic fibration is restricted to be a 
del Pezzo, Hirzebruch or a blow-up of a Hirzebruch surface. In Appendix B
of~\cite{usnew}, we
presented the generators of all effective classes in $H_{2}(B, \bf Z \rm)$, as
well as the first and second Chern classes $c_{1}(B)$ and $c_{2}(B)$, for
these allowed bases. Combining the rules with the generators
and Chern classes given in Appendix B, we presented a general algorithm for the
construction of non-perturbative vacua corresponding to three-family particle
physics theories with phenomenologically relevant gauge groups. In this
lecture, we review the results of~\cite{usnew}, referring the reader
frequently to Appendix B of that paper for the necessary details.

\section{Holomorphic Gauge Bundles, Five-Branes and \\
          Non-Perturbative Vacua} 

In this section, we will briefly review the generic properties of heterotic
$M$--theory vacua appropriate for a reduction of the theory to
$\cN=1$ supersymmetric theories in both five and four dimensions. 
As discussed in Lecture 1, the $M$--theory vacuum is
given in eleven dimensions by specifying the metric $g_{IJ}$ and the
three-form $C_{IJK}$ with field strength $G_{IJKL}=24\partial_{[I}
C_{JKL]}$ of the supergravity multiplet. Following Ho\v rava 
and Witten~\cite{hw1,hw2} and
Witten~\cite{w}, the space-time structure of these vacua, to lowest
order in the expansion parameter $\kappa^{2/3}$, will be taken
to be 
\begin{equation}
   M_{11}=M_{4} \times S^{1}/Z_{2} \times X
\label{eq:1}
\end{equation}
where $M_{4}$ is four-dimensional Minkowski space, $S^1/Z_2$ is a
one-dimensional orbifold and $X$ is a smooth Calabi--Yau
three-fold. The vacuum space-time structure becomes more complicated
at the next order in $\kappa^{2/3}$, but, as discussed in the previous two
lectures, this metric ``deformation'',
which has been the subject of a number of papers~\cite{w,bd,low1}, can be
viewed as arising as the static vacuum of the five-dimensional
effective theory~\cite{losw,add1} and, hence, need not concern us here. 

The $Z_2$ orbifold projection necessitates the introduction, on each
of the two ten-dimensional orbifold fixed planes, of an $\cN=1$, $E_8$ Yang-Mills
supermultiplet which is required for anomaly cancellation. On each
plane, the gauge field structure of these vacua, called the gauge
bundle, must be a solution of the hermitian Yang--Mills equations for
an $E_8$-valued connection in order to be compatible with four
preserved supercharges in four dimensions. Equivalently, as shown by
Donaldson, Uhlenbeck and Yau~\cite{Don,UhYau}, each gauge bundle must
be a semi-stable, holomorphic bundle with the structure group being the
complexification $E_{8\mathbf{C}}$ of $E_8$. In the following, we will
denote both groups by $E_8$, letting context dictate which group is
being referred to. (In general, we will denote any group $G$ and its
complexification $G_{\mathbf{C}}$ simply as $G$). These semi-stable,
holomorphic gauge bundles are, a priori, allowed to be arbitrary in
all other respects. In particular, there is no requirement that the
spin-connection of the Calabi--Yau three-fold be embedded into an
$SU(3)$ subgroup of the gauge connection of one of the $E_8$ bundles,
the so-called standard embedding. This generalization to arbitrary
semi-stable holomorphic gauge bundles is what is referred to as
non-standard embedding. The terms standard and non-standard embedding
are historical and somewhat irrelevant in the context of $M$-theory,
where no choice of embedding can ever set the entire three-form to
zero. For this reason, we will avoid those terms and simply refer to
arbitrary semi-stable holomorphic $E_8$ gauge bundles. Since these bundles
can be chosen arbitrarily, it is clear that we can restrict the
transition functions to be elements of any subgroup $G$ of $E_8$, such
as $G=U(n)$, $SU(n)$ or $Sp(n)$. We will refer to the restricted bundle as a
semi-stable, holomorphic $G$ bundle, or simply as a $G$ bundle. It is clear
that the $G_1$ bundle on one orbifold plane and the $G_2$ bundle on
the other plane need not, generically, have the same subgroups $G_1$
and $G_2$ of $E_8$. We will denote the semi-stable holomorphic gauge bundle
on the $i$-th orbifold plane by $V_i$ and the associated structure group
by $G_i$. 

In addition, as discussed in~\cite{us,don1}, we will allow for the
presence of five-branes located at points throughout the orbifold
interval. The five-branes will preserve $\cN=1$ supersymmetry
provided they are wrapped on holomorphic two-cycles within $X$ and
otherwise span the flat Minkowski space $M_4$. The inclusion of
five-branes is essential for a complete discussion of $M$--theory
vacua. The reason for this is that, given a Calabi--Yau three-fold
background, the presence of five-branes allows one to construct large
numbers of gauge bundles that would otherwise be disallowed~\cite{us,don1}. 

The requirements of gauge and gravitational anomaly cancellation on
the two orbifold fixed planes, as well as anomaly cancellation on each
five-brane worldvolume, places a further very strong constraint on,
and relationship between, the space-time manifold, the gauge bundles
and the five-brane structure of the vacuum. Specifically, anomaly
cancellation necessitates the addition of four-forms sources to the
four-form field strength Bianchi identity. As discussed in Lecture 2, 
the modified Bianchi identity 
is given by 
\begin{equation}
   (dG)_{11 \bar{I} \bar{J} \bar{K} \bar{L}}= 2 \sqrt{2} \pi
  (\frac{\kappa}{4 \pi})^{\frac{2}{3}} [J^{(0)}\delta(x^{11}) +
\label{eq:2}
\end{equation}
\begin{displaymath}
 J^{(N+1)}\delta(x^{11}-\pi\rho) +
  \Sigma_{n=1}^{N}J^{(n)}(\delta(x^{11}-x_{n})+\delta(x^{11}+x_{n}))]_
 {\bar{I} \bar{J} \bar{K} \bar{L}}
\label{eq:A2}
\end{displaymath}
The sources $J^{(0)}$ and $J^{(N+1)}$ on the orbifold planes are 
\begin{equation}
 J^{(0)}= - \frac{1}{8\pi} (trF^{(1)} \wedge F^{(1)}
  -\frac{1}{2} trR \wedge R )|_{x^{11}=0}
 \label{eq:3}
\end{equation}
and
\begin{equation}
 J^{(N+1)}=-\frac{1}{8 \pi^{2}} ( trF^{(2)} \wedge F^{(2)}
  -\frac{1}{2} trR \wedge R )|_{x^{11}=\pi \rho}
 \label{eq:4}
\end{equation}
respectively. By $tr$ we mean $\frac{1}{30}$-th of the trace over the
generators in the $\bf 248 \rm$ representation of $E_{8}$. 
The two-form $F^{(i)}$ is the field strength of a
connection on the gauge bundle $V_{i}$ of the $i$-th orbifold plane
and $R$ is the curvature two-form on the Calabi-Yau three-fold. We
have also introduced $N$ additional sources $J^{(n)}$, where
$n=1,...,N$. These arise from $N$ five-branes located at
$x^{11}=x_{1},...,x_{N}$ where $0 \leq x_{1} \leq \cdots \leq x_{N}
\leq \pi\rho$. Note that each five-brane at $x=x_{n}$ has to be paired
with a mirror five-brane at $x=-x_{n}$ with the same source since the
Bianchi identity must be even under the $Z_{2}$ orbifold symmetry. Our
normalization is chosen so that the total source of each pair is
$J^{(n)}$.  

Non-zero source terms on the right hand side of the Bianchi identity
(\ref{eq:2}) preclude the simultaneous vanishing of all components of
the three-form $C_{IJK}$. The result of this is that, to next order in
the Ho\v rava--Witten expansion parameter $\kappa^{2/3}$, the
space-time of the supersymmetry preserving vacua gets ``deformed''
away from that given in expression (\ref{eq:1}). As discussed above,
this deformation of the vacuum need not concern us here. In this
lecture, we will focus on yet another aspect of Bianchi identity
(\ref{eq:2}), a topological condition that constrains the cohomology
of the vacuum. This constraint is found as follows. Consider
integrating the Bianchi identity (\ref{eq:2}) over any five-cycle
which spans the orbifold interval together with an arbitrary
four-cycle ${\cal{C}}_{4}$ in the Calabi-Yau three-fold. Since $dG$ is
exact, this integral must vanish. Physically, this is the statement
that there can be no net charge in a compact space, since there is
nowhere for the flux to ``escape''. Performing the integral over the
orbifold interval, we derive, using (\ref{eq:2}), that
\begin{equation}
 \Sigma_{n=0}^{N+1}\int_{{\cal{C}}_{4}}J^{(n)}=0
 \label{eq:5}
\end{equation}
Hence, the total magnetic charge over ${\cal{C}}_{4}$ vanishes. Since
this is true for an arbitrary four-cycle ${\cal{C}}_{4}$ in the
Calabi-Yau three-fold, it follows that the sum of the sources must be cohomologically trivial. That is
\begin{equation}
[\Sigma_{n=0}^{N+1}J^{(n)}]=0
\label{eq:6}
\end{equation}
The physical meaning of this expression becomes more transparent if we
rewrite it using equations (\ref{eq:3}) and (\ref{eq:4}). Using these
expressions, equation (\ref{eq:6}) becomes
\begin{equation}
 -[\frac{1}{8\pi^{2}} tr F^{(1)} \wedge F^{(1)}]
  -[\frac{1}{8\pi^{2}} tr F^{(2)} \wedge F^{(2)}]
   +[\frac{1}{8\pi^{2}} tr R \wedge R]
  +\Sigma_{n=1}^{N}[J^{(n)}]=0
 \label{eq:7}
\end{equation}
It is useful to recall that the second Chern class of an arbitrary $G$ 
bundle $V$, thought of as an $E_{8}$ sub-bundle, is defined to be
\begin{equation}
 c_{2}(V)= -[\frac{1}{2 \cdot 8\pi^{2}} tr_{f} F \wedge F]
 \label{eq:8}
\end{equation}
Similarly, the second Chern class
of the tangent bundle of the Calabi-Yau manifold $X$ is given by
\begin{equation}
 c_{2}(TX)= -[\frac{1}{2 \cdot 8\pi^{2}} tr_{6} R \wedge R]
 \label{eq:9}
\end{equation}
where $tr_{6}$ implies that the trace is taken over the vector
representation of $SO(6)\supset SU(3)$, that is, the usual tangent
space representation. It follows that expression (\ref{eq:7}) can be
written as 
\begin{equation}
 c_{2}(V_{1}) + c_{2}(V_{2}) +[W]= c_{2}(TX)
 \label{eq:10}
\end{equation}
where 
\begin{equation}
 [W]=\Sigma_{n=1}^{N}[J^{(n)}]
 \label{eq:11}
\end{equation}
is the four-form cohomology class associated with the
five-branes. This is a fundamental constraint imposed on the vacuum
structure. We will explore this cohomology condition in great detail
in this lecture. Note that integrating this constraint over an arbitrary
four-cycle ${\cal{C}}_{4}$ yields the expression
\begin{equation}
 n_{1}({\cal{C}}_{4})+n_{2}({\cal{C}}_{4})+n_{5}({\cal{C}}_{4})=n_{R}({\cal{C}}_{4})
 \label{eq:12}
\end{equation}
which states that the sum of the number of gauge instantons on the two
orbifold planes, plus the sum of the five-brane magnetic charges, must
equal the instanton number for the Calabi-Yau tangent bundle, a number
which is fixed once the Calabi-Yau three-fold is chosen.

To summarize, we are considering vacuum states of $M$--theory with the
following structure. 
\begin{itemize}
\item Space-time is taken to have the form 
\begin{equation}
  M_{11}=M_{4} \times S^{1}/Z_{2} \times X 
  \label{eq:last1}
\end{equation}
where $X$ is a Calabi-Yau three-fold.
\item There is a semi-stable holomorphic gauge bundle $V_{i}$ with fiber
group $G_{i} \subseteq E_{8}$ over the Calabi-Yau three-fold on the
$i$-th orbifold fixed plane for $i=1,2$. The structure groups $G_{1}$ and
$G_{2}$ of the two bundles can be any subgroups of $E_{8}$ and need
not be the same.
\item We allow for the presence of five-branes in the vacuum, which
are wrapped on holomorphic two-cycles within $X$.
\item The Calabi-Yau three-fold, the gauge bundles and the five-branes
are subject to the cohomological constraint
\begin{equation}
  c_{2}(V_{1})+c_{2}(V_{2})+[W]=c_{2}(TX) 
  \label{eq:final2}
\end{equation}
where $c_{2}(V_{i})$ and
$c_{2}(TX)$ are the second Chern classes of the gauge bundle $V_{i}$
and the tangent bundle $TX$ respectively and $[W]$ is the class
associated with the five-branes .
\end{itemize}
Vacua of this type will be referred to as non-perturbative heterotic
M-theory vacua.

The discussion given in this section is completely generic, in that it
applies to any Calabi-Yau three-fold and any gauge bundles that can be
constructed over it. However, realistic particle physics theories
require the explicit construction of these gauge bundles. 
In the following, we will review the formalism for the construction of
semi-stable holomorphic gauge bundles with fiber groups $G_{1}$ and $G_{2}$
over the two orbifold fixed planes. For specificity, we
will restrict the structure groups to be
\begin{equation}
 G_{i}= U(n_{i}) \quad \mbox{or} \quad SU(n_{i})
 \label{eq:18}
\end{equation}
for $i=1,2$. Our explicit bundle constructions
will be achieved over the restricted, but rich, set of elliptically
fibered Calabi-Yau three-folds which admit a section. 
Such three-folds have been extensively
discussed within the context of duality between $M$- and
$F$-theory. Independently of this usage, however, elliptically fibered
Calabi-Yau three-folds with a section are known to be the simplest Calabi-Yau spaces
on which one can explicitly construct bundles, compute Chern classes,
moduli spaces and so on. This  latter property makes them a compelling
choice for the construction of concrete particle physics
theories. Having constructed the bundles, one can explicitly calculate
the gauge bundle Chern classes $c_{2}(V_{i})$ for $i=1,2$, as well as
the tangent bundle Chern class $c_{2}(TX)$. Having done so, one can
then find the class $[W]$ of the five-branes using the
cohomology condition (\ref{eq:10}). That is, we will
present a formalism in which the entire structure of non-perturbative
$M$--theory vacua can be calculated. 

As will be discussed in detail below, having constructed a
non-perturbative vacuum, we can compute the number of low energy
families and the Yang-Mills gauge group associated with that
vacuum. We will show that, because of the flexibility introduced by
the presence of five-branes, we will easily construct non-perturbative
vacua with three-families. Similarly, one easily finds
phenomenologically interesting gauge groups, such as $E_{6}$, $SU(5)$
and $SO(10)$, as the $E_{8}$ subgroups commutant with the $G$-bundle structure
groups, such as $SU(3)$, $SU(4)$ and $SU(5)$ respectively, on the observable
orbifold fixed plane. In addition, using the cohomology constraint
(\ref{eq:10}), one can explicitly determine the cohomology class $[W]$
of the five-branes for a specific vacuum. Hence, one can compute the
holomorphic curve associated with the five-branes exactly and
determine all of its geometrical attributes. These include its the number of
its irreducible components, which in turn tells us how many independent
five-branes appear in five-dimensions, and its genus, which will tell
us the minimal gauge group on the five-brane worldvolume when dimensionally
reduced on the holomorphic curve. Furthermore, we are, in general,
able to compute the entire moduli space of the holomorphic curve. This
can tell us about gauge group enhancement on the five-brane
worldvolume, for example. In \cite{don1}, we 
discussed the generic properties of the holomorphic curves associated with
five-branes. We presented a more detailed discussion in~\cite{usnewnew}.

\section{Elliptically Fibered Calabi--Yau Three-Folds} 

As discussed previously, we will consider
non-perturbative vacua where the Calabi--Yau three--fold is an elliptic 
fibration which admits a section. In this section, 
we give an introduction to these spaces, 
summarizing the properties we will need in order to explicitly compute 
important aspects of the vacua. 

An elliptically fibered Calabi--Yau three--fold $X$ consists of a base $B$,
which is a complex two--surface, and an analytic map
\begin{equation}
 \pi:X \longrightarrow B
 \label{eq:add1}
\end{equation}
with the property that for a generic point $b \in B$, the fiber
\begin{equation}
 E_{b} = \pi^{-1}(b)
 \label{eq:add2}
\end{equation}
is an elliptic curve. That is, $E_{b}$ is a Riemann surface of genus one. In
addition, we will require that there exist a global section, denoted $\sigma$,
defined to be an analytic map
\begin{equation}
 \sigma:B \longrightarrow X
 \label{eq:add3}
\end{equation}
that assigns to every point $b \in B$ the zero element $\sigma(b)= p 
\in E_{b}$ discussed below. The requirement that the elliptic fibration have a
section is crucial for duality to $F$-theory and to make contact with the
Chern class formulas in \cite{FMW}. However, this assumption does not seem
fundamentally essential and we will explore bundles without sections in future
work \cite{wod2}.
The Calabi--Yau three-fold must be a complex K\"ahler
manifold. This implies that the base is itself a complex manifold, while we
have already assumed that the fiber is a Riemann surface and so has a
complex structure. Furthermore, the fibration must be holomorphic, that
is, it must have holomorphic transition functions. Finally, 
the condition that the Calabi--Yau three-fold has vanishing first Chern class
puts a further constraint on the types of fibration allowed.

Let us start by briefly summarizing the properties of an elliptic
curve $E$. It is a genus one Riemann surface and so can be embedded in the
two-dimensional complex projective space $ \bf CP \rm^{2}$. A simple way to
do this is by using the homogeneous Weierstrass equation
\begin{equation}
   zy^2 = 4x^3 - g_2 x z^2 - g_3 z^3
\label{Weier}
\end{equation}
where $x$, $y$ and $z$ are complex homogeneous coordinates on
$ \bf CP \rm^{2}$.  It follows that we identify $(\l x,\l y,\l z)$ 
with $(x,y,z)$ for any
non-zero complex number $\l$. The parameters $g_2$ and $g_3$ encode the
different complex structures one can put on the torus. Provided $z\neq
0$, we can rescale to affine coordinates where $z=1$. We then see,
viewed as a map from $x$ to $y$, that there are two branch cuts in the
$x$-plane, linking $x=\infty$ and the three roots of the cubic equation
$4x^3-g_2x-g_3=0$. When any two of these points coincide, the elliptic
curve becomes singular. This corresponds to one of the cycles in the
torus shrinking to zero. Such singular behaviour is characterized by
the discriminant 
\begin{equation}
 \Delta=g_2^3-27g_3^2
 \label{eq:add4}
\end{equation}
vanishing. Finally, we note that the complex structure provides a
natural notion of addition of points on the elliptic curve. The torus
can also be considered as the complex plane modulo a discrete group of
translations. Addition of points in the complex plane then induces a
natural notion of addition of points on the torus. Translated to the
Weierstrass equation, the identity element corresponds to the point
where $x/z$ and $y/z$ become infinite. Thus, in affine coordinates, the
element $p \in E$ is the point $x=y=\infty$. This can be scaled elsewhere in
non-affine coordinates, such as to $x=z=0$, $y=1$.  

The elliptic fibration is defined by giving the elliptic curve $E$ over
each point in the base $B$. If we assume the fibration has a global
section, and in this lecture we do, then on each coordinate 
patch this requires giving the
parameters $g_2$ and $g_3$ in the Weierstrass equation as functions
on the base. Globally, $g_2$ and $g_3$ will be sections of appropriate line
bundles on $B$. In fact, specifying the type of an elliptic fibration over $B$
is equivalent to specifying a line bundle on $B$. Given the elliptic fibration
$\pi :X \longrightarrow B$, we define ${\cal{L}}$ as the line bundle on $B$
whose fiber at $b \in B$ is the cotangent line $T_{p}(E_{b})$ to the elliptic curve
at the origin. That is, ${\cal{L}}$ is the conormal bundle to the section
$\sigma(B)$ in $X$. Conversely, given ${\cal{L}}$, we take $x$ and $y$ to
scale as
sections of ${\cal{L}}^{2}$ and ${\cal{L}}^{3}$ respectively, which means that
$g_2$ and $g_3$ should be sections of ${\cal{L}}^{4}$ and ${\cal{L}}^{6}$. By 
${\cal{L}}^{i}$ we mean the tensor product of the line bundle ${\cal{L}}$ with
itself $i$ times. In conclusion, we see that the elliptic fibration is
characterized by a line bundle ${\cal{L}}$ over the base $B$ together with a
choice of sections $g_2$ and $g_3$ of ${\cal{L}}^{4}$ and ${\cal{L}}^{6}$.

Note that the set of points in the base over which the fibration becomes
singular is given by the vanishing of
the discriminant $\Delta=g_2^3-27g_3^2$. It follows from the above discussion
that $\Delta$ is a section of the line
bundle  $\cL^{12}$. The   zeros of $\Delta$ then naturally define a
divisor, which in this case is a complex curve, in the base. Since $\Delta$ is
a section of $\cL^{12}$, the cohomology class of the discriminant curve
is 12 times the cohomology class of the divisors defined by sections
of $\cL$.

Finally, we come to the important condition that on a Calabi--Yau three--fold
$X$ the first Chern class of the tangent bundle $T_X$ must vanish. The
canonical bundle $K_X$ is the line bundle constructed as the determinant of
the holomorphic cotangent bundle of $X$. The condition that
\begin{equation}
c_{1}(T_X)=0
\label{eq:ron1}
\end{equation}
implies that $K_{X}={\cal{O}}$, where ${\cal{O}}$ is the trivial bundle. This,
in turn, puts a constraint on $\cL$. To see this, note that the adjunction
formula tells us that, since $B$ is a divisor of $X$, the canonical bundle
$K_{B}$ of $B$ is given by
\begin{equation}
K_{B}=K_{X}|_{B} \otimes N_{B/X}
\label{eq:ron2}
\end{equation}
where $N_{B/X}$ is the normal bundle of $B$ in $X$. From the above discussion,
we know that
\begin{equation}
N^{-1}_{B/X}=\cL, \qquad K_{X}|_{B}={\cal{O}}
\label{eq:ron3}
\end{equation}
Inserting this into~\eqref{eq:ron2} and switching to additive notation tells
us that
\begin{equation}
 \cL=K_B^{-1}
 \label{eq:add5}
\end{equation}
This condition means that $K_{B}^{-4}$ and $K_{B}^{-6}$ must have sections 
$g_2$ and $g_3$ respectively. Furthermore, the Calabi--Yau property imposes
restrictions on how the curves where these sections vanish are allowed to
intersect. It is possible to classify the surfaces on which 
$K_{B}^{-4}$ and $K_{B}^{-6}$ have such sections. These are found to be
\cite{MV} the del Pezzo, Hirzebruch and Enriques surfaces, as well as blow-ups
of Hirzebruch surfaces. In this lecture, we will discuss the first three
possibilities in detail.

As noted previously, in order to discuss the anomaly cancellation 
condition, we will need
the second Chern class of the holomorphic tangent bundle of
$X$. Friedman, Morgan and Witten~\cite{FMW} show that it can be
written in terms of the Chern classes of the holomorphic tangent
bundle of $B$ as
\begin{equation}
   c_2(TX) = c_2(B) + 11 c_1(B)^2 + 12 \s c_1(B)
\label{tangc2}
\end{equation}
where the wedge product is understood, $c_{1}(B)$ and $c_{2}(B)$ are the first
and second Chern classes of $B$ respectively 
and $\sigma$ is the two-form Poincare dual to the global section. 
We have used the fact that 
\begin{equation}
  c_{1}(\cL)=c_{1}(K_{B}^{-1})=c_{1}(B)
\label{ron4}
\end{equation}
in writing~\eqref{tangc2}.


\section{Spectral Cover Constructions}

In this section, we follow the construction of holomorphic
bundles on elliptically fibered Calabi--Yau manifolds presented in
~\cite{FMW,D,BJPS}. The idea is to understand the
bundle structure on a given elliptic fiber and then to patch these
bundles together over the base. The authors in ~\cite{FMW,D,BJPS} discuss a
number of techniques for constructing bundles with different gauge
groups. Here we will restrict ourselves to $U(n)$ and $SU(n)$
sub-bundles of $E_8$. These are sufficient to give suitable
phenomenological gauge groups. This restriction allows us to consider only the
simplest of the different constructions, namely that via spectral
covers. In this section, we will summarize the spectral cover construction,
concentrating on the properties necessary for an explicit discussion of
non-perturbative vacua. We note that for structure groups $ G \neq U(n)$ or 
$SU(n)$, the construction of bundles is more complicated than the construction
of rank $n$ vector bundles presented here.

As we have already mentioned, the condition of supersymmetry requires
that the $E_8$ gauge bundles admit a field strength satisfying the
hermitian Yang--Mills equations. Donaldson, Uhlenbeck and
Yau~\cite{Don,UhYau} have shown that this is equivalent to the
topological requirement that the associated bundle be semi-stable, with
transition functions in the complexification of the gauge group. Since
we are considering $U(n)$ and $SU(n)$ sub-bundles, this means
$U(n)_{\CC}=GL(n,\CC)$ and $SU(n)_{\CC}=SL(n,\CC)$ respectively. 
The spectral cover construction is given in terms
of this latter formulation of the supersymmetry condition. Note that the distinction
between semi-stable and stable bundles corresponds to whether the
hermitian Yang-Mills field strength is reducible or not. This refers
to whether, globally, it can be diagonalized into parts coming from different
subgroups of the full gauge group. More precisely, it refers to
whether or not the holonomy commutes with more that just the center of the
group. Usually, a generic solution of the hermitian Yang--Mills
equations corresponds to a stable bundle. However, on some spaces, for
instance on an elliptic curve, the generic case is semi-stable. 

\subsection*{U(n) and SU(n) Bundles Over An Elliptic Curve}

We begin by considering semi-stable bundles
on a single elliptic curve $E$. A theorem of Looijenga~\cite{Looij} states
that the moduli space of such bundles for any simply-connected group
of rank $r$ is an $r$-dimensional complex weighted projective
space. For the simply-connected group $SU(n)$, this moduli space is 
the projective space ${\bf CP  }^{n-1}$. $U(n)$ is not simply-connected. 
$U(n)$ bundles have a discrete integer invariant, their degree or first Chern class,
which we denote by $d$. Let $k$ be the greatest common divisor of $d$ and $n$.
It can be shown that the moduli space of a $U(n)$ bundle of degree $d$ over a
single elliptic curve $E$ is the $k$-th symmetric product of $E$, denoted by
$E^{[k]}$. In this lecture, we will restrict our discussion to $U(n)$ bundles of
degree zero. For these bundles, the moduli space is $E^{[n]}$.

A holomorphic $U(n)_\CC=GL(n,\CC)$ bundle $V$ over an elliptic curve
$E$ is a rank $n$ complex vector bundle. 
As discussed earlier, we will denote $U(n)_\CC$ simply as 
$U(n)$, letting
context dictate which group is being referred to.
To define the bundle, we need
to specify the holonomy; that is, how the bundle twists as one moves around
in the elliptic curve. The holonomy is a map from the fundamental
group $\pi_1$ of the elliptic curve into the gauge group. Since the fundamental
group of the torus is Abelian, the holonomy must map into the maximal
torus of the gauge group. This means we can diagonalize all the transition
functions, so that $V$ becomes the direct sum of line bundles
\begin{equation}
  V=\cN_1\oplus\dots\oplus\cN_n
 \label{eq:add6}
\end{equation}
 Furthermore, the Weyl group permutes the diagonal elements, so
that $V$ only determines the ordering of the $\cN_{i}$ up to permutations.
To reduce from a $U(n)$ bundle to an $SU(n)$ bundle, one imposes the
additional condition that the determinant of the transition functions
be taken to be unity. This implies that the product
\begin{equation}
 \cN_1\otimes\dots\otimes\cN_n=\cO
 \label{eq:add7}
\end{equation}
where $\cO$ is the trivial bundle on $E$.

The semi-stable condition implies that the line bundles $\cN_i$ are of 
the same degree, which can be taken to be zero. We
can understand this from the hermitian Yang--Mills equations. On a
Riemann surface, these equations imply that the field strength is actually
zero. Thus, the first Chern class of each of the bundles $\cN_i$ must vanish, or
equivalently be of degree zero. On an elliptic curve, this condition
means that there is a unique point $Q_i$ on $E$ such that there is a
meromorphic section of $\cN_i$ which vanishes only at $Q_i$ and has a
simple pole only at the origin $p$. We can write this as 
\begin{equation}
 \cN_{i}=\cO(Q_i)\otimes\cO(p)^{-1}
 \label{eq:add8}
\end{equation}
If one further restricts the structure group to be $SU(n)$, then 
condition~\eqref{eq:add7} translates into the requirement that 
\begin{equation}
 \sum_{i=1}^{n} (Q_i-p) =0 
 \label{eq:add9}
\end{equation}
where one uses the natural addition of points on $E$ discussed above.

Thus, on a given elliptic curve, giving a semi-stable $U(n)$ bundle
is equivalent to giving an unordered (because of the Weyl symmetry)
$n$-tuple of points on the curve. An $SU(n)$ bundle has the further
restriction that $\sum_i (Q_i-p)=0$. For an $SU(n)$ bundle,
these points can be represented very explicitly as roots of an equation
in the Weierstrass coordinates describing the elliptic curve. In affine
coordinates, where $z=1$, we write
\begin{equation}
   s = a_0 + a_2 x + a_3 y + a_4 x^2 + a_5 x^2y + \dots + a_n x^{n/2}
\label{speceq}
\end{equation}
(If $n$ is odd the last term is $a_nx^{(n-3)/2}y$.) Solving the equation $s=0$,
together with the Weierstrass equation (hence the appearance of only
linear terms in $y$ in $s$), gives $n$ roots corresponding to the $n$ points
$Q_i$, where one can show that $\sum_i(Q_i-p)=0$ as required.  One notes
that the roots are determined by the coefficients $a_i$ only up to an overall scale
factor. Thus the moduli space of roots $Q_i$ is the projective space
$ \bf CP \rm ^{n-1}$ as anticipated, with the coefficients $a_i$ acting
as homogeneous coordinates. 

In summary, semi-stable $U(n)$ bundles on an elliptic curve are
described by an unordered $n$-tuple of points $Q_i$ on the elliptic
curve. $SU(n)$ bundles have the additional condition that 
$\sum_i(Q_i-p)=0$. In the $SU(n)$ case, these points can be realized as
roots of the equation $s=0$ and give a moduli space of bundles which
is simply $ \bf CP \rm ^{n-1}$, as mentioned above.

\subsection*{The Spectral Cover and the Line Bundle ${\cal{N}}$}

Given that a bundle on an elliptic curve is described by the
$n$-tuple $Q_i$, it seems reasonable that a bundle on an
elliptic fibration determines how the $n$ points vary as
one moves around the base $B$. The set of all the $n$ points over the 
base is called the spectral
cover $C$ and is an $n$-fold cover of $B$ 
with $\pi_{C}: C \longrightarrow B$. The spectral cover alone does not contain
enough information to allow us to construct the bundle $V$. To do this, one
must specify an additional line bundle, denoted by ${\cal{N}}$, on the
spectral cover $C$. One obtains ${\cal{N}}$, given the vector bundle $V$, as
follows. Consider the elliptic fiber $E_{b}$ at any point $b \in B$. It
follows from the previous section that
\begin{equation}
 V|_{E_{b}}= {\cal{N}}_{1b} \oplus ... \oplus {\cal{N}}_{nb}
\label{eq:night1}
\end{equation}
where ${\cal{N}}_{ib}$ for $i=1,..,n$ are line bundles on $E_{b}$.
In particular, we get a decomposition of the fiber $V_{\sigma(b)}$ of $V$
at $p=\sigma(b)$. Let $V|_{B}$ be the restriction of $V$ to the base $B$
embedded in $X$ via the section $\sigma$. We have just shown that the
$n$-dimensional fibers of $V|_{B}$ come equipped with a decomposition into a
sum of lines. As point $b$ moves around the base $B$, these $n$ lines move in
one to one correspondence with the $n$ points $Q_{i}$ above $b$. This data
specifies a unique line bundle \footnote{When C is singular, ${\cal{N}}$ may be 
more generally a rank-$1$ torsion free sheaf on $C$. 
For non-singular $C$ this is the same as a 
line bundle. }
${\cal{N}}$ on $C$ such that the direct image
$\pi_{C*} {\cal{N}}$ is $V|_{B}$ with its given decomposition. 
The direct image $\pi_{C*} {\cal{N}}$ is a vector bundle on $B$ whose fiber at
a generic point $b$, where the inverse image $\pi_{C}^{-1}(b)$ consists of the $n$
distinct points $Q_{i}$, is the direct sum of the $n$ lines
${\cal{N}}|_{Q_{i}}$.

\subsection*{Construction of Bundles}

We are now in a a position to construct the rank $n$ vector bundle starting
with the spectral data \cite{FMW,D,BJPS}. The spectral data consists 
of the spectral cover $C \subset X$ together with the line bundle
${\cal{N}}$ on $C$. The spectral cover 
is a divisor (hypersurface) $C \subset X$ which is of degree $n$ over the 
base $B$; that is, the restriction $\pi_{C} : C \to B$ 
of the elliptic fibration is an 
$n$-sheeted branched cover. Equivalently, the cohomology class of $C$ in 
$H^2(X,\ZZ)$ must be of the form
\begin{equation}
 [C] = n \sigma + \eta 
 \label{eq:day1}
\end{equation}
where $\eta$ is a class in $H^2(B,\ZZ)$ and $\sigma$ is the 
section. This is equivalent to saying that the line bundle ${\cal{O}}_{X}(C)$
on $X$ determined by $C$, whose sections are meromorphic functions on $X$ with
simple poles along $C$, is given by
\begin{equation}
 {\cal{O}}_{X}(C)= {\cal{O}}_{X}(n\sigma) \otimes {\cal{M}}
 \label{eq:hope1}
\end{equation}
where ${\cal{M}}$ is some line bundle on $X$ whose restriction to each fiber
$E_{b}$ is of degree zero. Written in this formulation
\begin{equation}
 \eta = c_{1}({\cal{M}})
 \label{eq:hope2}
\end{equation}
The line bundle ${\cal{N}}$ is, at this point, completely arbitrary.

Given this data, one can construct a rank $n$ vector bundle $V$ on $X$. It is easy 
to describe the restriction $V|_{B}$ of $V$ to the base $B$. It is 
simply the direct image $V|_{B}= \pi_{C*} {\cal{N}}$.
It is also easy to describe the restriction of $V$ to a general elliptic 
fiber $E_b$. Let $C\cap E_{b} = \pi_{C}^{-1}(b) = Q_1 + \ldots +Q_n$ and 
$\sigma \cap E_{b} = p$. Then each 
$Q_i$ determines a line bundle ${\cal{N}}_i$ of degree zero on $E_{b}$
whose sections are the meromorphic functions on $E_{b}$ with first order poles
at $Q_{i}$ which vanish at $p$. The restriction $V|_{E_{b}}$ is then the sum of the 
${\cal{N}}_i$. Now 
the main point is that there is a unique vector bundle $V$ on $X$ with 
these specified restrictions to the base and the fibers.

To describe the entire vector bundle $V$, we use the 
Poincare bundle ${\cal P}$. This is a 
line bundle on the fiber product $X \times _B X'$. Here $X'$ is the ``dual 
fibration'' to $X$. In general, this is another elliptic fibration which is 
locally, but not globally, isomorphic to $X$. However, when $X$ has a 
section (which we assume), then $X$ and $X'$ are globally isomorphic, so we 
can identify them if we wish. (Actually, the spectral cover $C$ lives most 
naturally as a hypersurface in the dual $X'$, not in $X$. When we described it 
above as living in $X$, we were implicitly using the identification of $X$ and 
$X'$.) The fiber product $X \times _B X'$ is four-dimensional. It is 
fibered over $B$, the fiber over $b \in B$ being the ordinary product $E_b \times 
{E'}_b$ of the two fibers. Now, the Poincare bundle ${\cal P}$ is determined by 
the following two properties: $(1)$ its restriction ${\cal P}|_{E_b \times x}$ to 
a fiber $E_b \times x$, for $x \in {E'}_b$, is the line bundle on $E_b$ determined 
by $x$ while $(2)$ its restriction to $\sigma \times _B X'$ is the trivial 
bundle. Explicitly, ${\cal P}$ can be given by the bundle whose sections are
meromorphic functions on $X \times _B X'$ with first order poles on ${\cal{D}}$
and which vanish on $\sigma \times_{B} X'$ and on $X \times_{B} \sigma'$. That
is
\begin{equation}
{\cal P} = {\cal O}_{X \times _B X'}({\cal{D}}-\sigma \times_{B} X'
           -X \times_{B} \sigma')\otimes K_{B}
 \label{eq:day2}
\end{equation}
where ${\cal{D}}$ is the diagonal divisor representing the graph of the
isomorphism $X\to X'$.

Using this Poincare bundle, we can finally describe the entire vector bundle $V$ 
in terms of the spectral data. It is given by
\begin{equation} 
 V = {p_1}_*({p_2}^*{\cal{N}} \otimes {\cal P})
\label{eq:day3}
\end{equation}
Here $p_1,p_2$ are the two projections of the fiber product $X \times _B 
C$ onto the two factors $X$ and $C$. The two properties of the Poincare bundle 
guarantee that the restrictions of this $V$ to the base and the fibers indeed agree 
with the intuitive versions of $V_B$ and $V|_{E_b}$ given above.

In general, this procedure produces $U(n)$ bundles. In order to get $SU(n)$ 
bundles, two additional conditions must hold. First, the condition 
that the line bundle
${\cal{M}}$ in equation ~\eqref{eq:hope1} has degree 
zero on each fiber $E_{b}$ must be
strengthened to require that the restriction of  ${\cal{M}}$ to $E_{b}$ is the
trivial bundle. Hence, ${\cal{M}}$ is the pullback to $X$ of a line bundle on
$B$ which, for simplicity, we also denote by ${\cal{M}}$.
This guarantees that the restrictions to the fibers $V|_{E_b}$ are $SU(n)$ bundles.
The second condition 
is that $V|_{B}$ must be an $SU(n)$ bundle as well. That is, the line 
bundle ${\cal{N}}$ on $C$ is such that the first Chern class $c_1$ 
of the resulting bundle $V$ vanishes. This
condition, and its ramifications,  will be discussed in the next section.

$U(n)$ vector bundles on the orbifold planes of heterotic $M$-theory 
are always sub-bundles of an $E_{8}$ vector bundle. As such, issues arise
concerning their stability or semi-stability which are important and
require considerable analysis. Furthermore, the associated Chern classes
require an extended analysis to compute. For these reasons, we
will limit our discussion to $SU(n)$ bundles, which are easier to study.

\subsection*{Chern Classes and Restrictions on the Bundle}

As discussed above, the global condition that the bundle be $SU(n)$ is that
\begin{equation}
 c_1(V)=(1/2\pi)\tr F=0
 \label{eq:add10}
\end{equation}
This condition is clearly true since, for structure group $SU(n)$,  
the trace must vanish. A formula for
$c_1(V)$ can be extracted from the discussion in Friedman, Morgan and
Witten \cite{FMW}. One finds that
\begin{equation}
   c_1(V) = \pi_{C*} \left( c_1(\cN) + \frac{1}{2}c_1(C) 
              - \frac{1}{2}\pi^*c_1(B) \right)
\end{equation}
where $c_1(B)$ means the first Chern class of the tangent 
bundle of $B$ considered as a complex vector bundle, and
similarly for $C$, while $\pi_{C}$ is the projection from the spectral
cover onto $B$; that is, $\pi_{C}:C\to B$. The operators $\pi_{C}^*$ and
$\pi_{C*}$ are the pull-back and push-forward of cohomology classes between $B$
and $C$. The condition that $c_1(V)$ is zero then implies that
\begin{equation}
   c_1(\cN) =  - \frac{1}{2}c_1(C) + \frac{1}{2}\pi_{C}^*c_1(B) + \gamma
\end{equation}
where $\gamma$ is some cohomology class satisfying the equation
\begin{equation}
 \pi_{C*}\gamma=0
 \label{eq:add11}
\end{equation}
The general solution for $\gamma$ constructed from cohomology classes is
\begin{equation}
   \gamma = \lambda\left( n\sigma - \pi_{C}^* \eta + n \pi_{C}^* c_1(B) \right)
\end{equation}
where $\l$ is a rational number and $\s$ is the global section of the elliptic
fibration. Appropriate values for $\l$ will emerge shortly. 
From~\eqref{eq:day1} we recall that $c_1(C)$, which is given
by
\begin{equation}
 c_{1}(C)= -n\sigma -\pi_{C}^{*}\eta
 \label{eq:add12}
\end{equation}
Combining the last three equations yields
\begin{equation}
\begin{split}
   c_1(\cN) = n\left( \frac{1}{2} + \l \right) \s 
                + \left( \frac{1}{2} - \l \right) \pi_{C}^* \eta
                + \left( \frac{1}{2} + n\l \right) \pi_{C}^* c_1(B)
\end{split}
\end{equation}
Essentially, this means that the bundle $\cN$ is completely determined
in terms of the elliptic fibration and $\cM$. It is important to note, 
however, that there is not
always a solution for $\cN$. The reason for this is that $c_1(\cN)$ must be
integer, a condition that puts a substantial constraint on the allowed
bundles. To see this, note that the section is a horizontal divisor, 
having unit intersection
number with the elliptic fiber. On the other hand, the quantities $\pi_{C}
^*c_1(B)$ and
$\pi_{C}^*\eta$ are vertical, corresponding to curves in the
base lifted to the fiber and so have zero intersection number with the
fiber. Therefore, we cannot choose $\eta$ to cancel $\s$ and, hence, the coefficient
of $\s$ must, by itself, be an integer . This implies that a consistent bundle
$\cN$ will exist if either 
\begin{equation}
 \mbox{n is odd},  \quad \l = m+\frac{1}{2} 
 \label{eq:add13}
\end{equation}
or
\begin{equation} 
   \mbox{n is even},  \quad \l = m, \quad \eta = c_1(B) mod 2
 \label{eq:add14}
\end{equation}
where $m$ is an integer. Here, the $\eta = c_1(B) mod 2$ condition
means that $\eta$ and $c_1(B)$ differ by an even element of
$H^2(B,\ZZ)$. Note that when $n$ is even, we cannot
choose $\eta$ arbitrarily. For $n$ odd, condition~\eqref{eq:add13} is
necessary and sufficient. For $n$ even, condition~\eqref{eq:add14} is
sufficient for the existence of a consistent line bundle ${\cal{N}}$. It is
also sufficient for the examples we consider in this lecture, and it is the
only class of solutions which is easy to describe in general. However, other
solutions do exist. We could, for example, take $n=4$, $\lambda=\frac{1}{4}$
and $\eta=2c_{1}(B)mod4$.

Finally, we can give the explicit Chern classes for the $SU(n)$ vector bundle $V$. 
Friedman, Morgan and Witten calculate
$c_1(V)$ and $c_2(V)$, while Curio and Andreas~\cite{cur,ba} have found
$c_3(V)$. The results are 
\begin{equation}
\begin{align}
   c_1(V) &= 0 \label{c1} \\
   c_2(V) &= \eta \s - \frac{1}{24} c_1(B)^2 \left(n^3 - n\right) 
              + \frac{1}{2} \left(\l^2 - \frac{1}{4}\right) n \eta 
                      \left(\eta - nc_1(B)\right) \label{c2} \\
   c_3(V) &= 2 \l \s \eta \left( \eta - nc_1(B) \right) \label{c3}
\end{align}
\end{equation}
where the wedge product is understood.

\section{Summary of Elliptic Fibrations and Bundles}

The previous two sections are somewhat abstract. For the sake of clarity, we
will here summarize those results which are directly relevant to
constructing physically acceptable non-perturbative vacua.

\begin{itemize}

\item  An elliptically fibered Calabi--Yau three-fold is composed of a two-fold
base $B$ and elliptic curves $E_{b}$ fibered over each point $b \in B$. In
this lecture, we consider only those elliptic fibrations that admit a
global section $\sigma$.

\item  The elliptic fibration is characterized by a single line bundle $\cL$
over $B$. The vanishing of the first Chern class of the canonical bundle
$K_{X}$ of the Calabi--Yau three-fold $X$ implies that 
\begin{equation}
 \cL= K_{B}^{-1}
 \label{eq:add15}
\end{equation}
where $K_{B}$ is the canonical bundle of the base $B$.

\item From the previous condition, it follows that the base $B$ is
restricted to del Pezzo, Hirzebruch and Enriques surfaces, as well as 
blow-ups of Hirzebruch surfaces. 

\item  The second Chern class of the holomorphic tangent bundle of $X$ is given
by
\begin{equation}
   c_2(TX) = c_2(B) + 11 c_1(B)^2 + 12 \s c_1(B)
  \label{eq:add16}
\end{equation}
where $c_1(B)$ and $c_2(B)$ are the first and second Chern classes of $B$.

\item  A general semi-stable $SU(n)$ gauge bundle $V$ is determined by two
line bundles, $\cM$ and $\cN$. The relevant quantities associated 
with $\cM$ and $\cN$ are their first Chern classes
\begin{equation}
 \eta= c_{1}(\cM)
 \label{eq:add17}
\end{equation}
and $c_{1}(\cN)$ respectively. The class $c_{1}(\cN)$, in addition to
depending on $n, \sigma, c_{1}(B)$ and $\eta$, also contains a complex
number $\lambda$.

\item  The condition that $c_{1}(\cN)$ be an integer leads to the
constraints on $\eta$ and $\lambda$ given by
\begin{equation}
 \mbox{n is odd},  \quad \l = m+\frac{1}{2} 
 \label{eq:add18}
\end{equation}
\begin{equation} 
   \mbox{n is even},  \quad \l = m, \quad \eta = c_1(B) mod 2
 \label{eq:add19}
\end{equation}
where $m$ is an integer.

\item  The relevant Chern classes of an $SU(n)$ gauge bundle $V$ are given by
\begin{equation}
\begin{align}
   c_1(V) &= 0 \label{c1n} \\
   c_2(V) &= \eta \s - \frac{1}{24} c_1(B)^2 \left(n^3 - n\right) 
              + \frac{1}{2} \left(\l^2 - \frac{1}{4}\right) n \eta 
                      \left(\eta - nc_1(B)\right) \label{c2n} \\
   c_3(V) &= 2 \l \s \eta \left( \eta - nc_1(B) \right) \label{c3n}
\end{align}
\label{eq:add20}
\end{equation}

\end{itemize}
How can one use the this data to construct realistic particle physics
theories? One proceeds as follows.
\begin{itemize}

\item  Choose a base $B$ from one of the allowed bases; namely, a del Pezzo,
Hirzebruch or Enriques surface, or a blow-up of a Hirzebruch surface. The
associated Chern classes $c_{1}(B)$ and $c_{2}(B)$ can be computed for any of
these surfaces.

\end{itemize}
This allows one to construct the second Chern class of the 
Calabi-Yau tangent bundle and a significant part of the gauge bundle 
Chern classes.

\begin{itemize}

\item  Specify $\eta$ and $\lambda$ subject to the above constraints.

\end{itemize}
These constraints greatly reduce the number of physically relevant
non-perturbative vacua. Given appropriate $\eta$ and $\lambda$, one
can completely determine the relevant gauge bundle Chern classes. 
We will carry this out explicitly in the next section.

\section{Effective Curves and Five-Branes}

Consider a complex manifold $X$ which is an elliptic fibration over 
a base $B$. Effective classes are defined, and their physical meaning
discussed, in Appendix A of ref.~\cite{usnew}.
Let us suppose we have found an effective class in $H_2(B,\ZZ)$.
Then, it naturally also lies in an effective homology class in $H_2(X,\ZZ)$ of
the elliptic fibration. Note that the fibration structure guarantees that if
two curves are in different classes in the base, then they are in different
classes in the full manifold $X$. This implies, among other things, that if
one finds the effective generating class of the Mori cone of $B$, these
classes remain distinct classes of $X$. In addition,
there is at least one other effective class that is not associated with the
base. This is the class $F$ of the fiber itself. 
There may also be other such classes, for example, those related to points where 
the fiber degenerates. However, we will ignore these since they will not 
appear in the homology classes of the five-branes, our main interest. 

The algebraic classes that arise naturally are quadratic polynomials in
classes of the line bundles. The only line bundle classes on a general
elliptically fibered Calabi--Yau three-fold $X$ are the base $B$ and the divisors
$\pi^{-1}({\cal{C}})$, where ${\cal{C}}$ is a curve in $B$. Any quadratic
polynomial in these classes can be written as
\begin{equation}
   W=W_{B} + a_{f}F
\label{decomp}
\end{equation}
where $ W_{B}$ is an algebraic homology class in the base manifold $B$ embedded
in $X$ and $a_{f}$ is some integer. Under what conditions is $W$ an 
effective class? It is clear that $W$ is effective if $W_{B}$ is an 
effective class in the base and $a_f\geq0$. One can also prove that
the converse is true in almost all cases. Specifically, we can prove the
following. First, the converse is true for any del Pezzo and Enriques surface.
Second, the converse is true for a Hirzebruch surface $F_{r}$, with the
exception of when $W_{B}$ happens to contain the negative section $S$ and
$r\geq3$. In this lecture, for simplicity, we will consider only those cases for
which the converse is true. 
Thus, under this restriction, we have that
\begin{equation}
   W \text{ is effective } \Longleftrightarrow
   W_B \text{ is effective in } B \text{ and } a_f\geq 0
\end{equation}
This reduces the question of finding the effective curves in $X$ to 
knowing the generating set of effective curves in the base $B$. For the set of base
surfaces $B$ we are considering, finding such generators is always possible.

Recall from equation~\eqref{eq:10} that the cohomology class associated with the 
five-branes is given by
\begin{equation}
   [W]=c_{2}(TX)-c_{2}(V_{1})-c_{2}(V_{2})
\label{eq:x4}
\end{equation}
For simplicity, in this lecture we will allow for arbitrary semi-stable 
gauge bundles $V_{1}$, which we henceforth call $V$, on the first 
orbifold plane, but always take the gauge bundle $V_{2}$ to be trivial. 
Physically, this corresponds to allowing observable sector gauge groups 
to be subgroups, such as $SU(5)$, $SO(10)$ or $E_{6}$, of $E_{8}$ but 
leaving the hidden sector $E_{8}$ gauge group unbroken. We do this only 
for simplicity. Our formalism also allows an analysis of the general case 
where the hidden sector $E_{8}$ gauge group is broken by a non-trivial 
bundle $V_{2}$. With this restriction, equation~\eqref{eq:x4} simplifies to
\begin{equation}
   [W]=c_{2}(TX)-c_{2}(V)
\label{eq:x5}
\end{equation}
Inserting the expressions~\eqref{eq:add16} and~\eqref{c2n} for the second Chern classes, 
we find that
\begin{equation}
   [W]= W_{B} +a_{f}F
\label{eq:x6}
\end{equation}
where
\begin{equation}
    W_{B}=\sigma( 12c_{1}(B)-\eta)
\label{eq:x7}
\end{equation}
is the part of the class associated with the base $B$ and
\begin{equation}
   a_f = c_2(B) + \left(11 + \frac{n^3-n}{24}\right) c_1(B)^2 
      - \frac{1}{2}n\left(\lambda^{2}-\frac{1}{4}\right)
          \eta\left(\eta-nc_{1}(B)\right)
\label{eq:x8}
\end{equation}
is the part associated with the elliptic fiber. 

Now, to make physical sense, five-branes must be wrapped on a curve
composed of holomorphic submanifolds of $X$ and, hence, $[W]$
must be an effective class. This physical requirement then implies, using the
above theorem, that necessarily
\begin{equation}
   W_B \text{ is effective in } B, \qquad
   a_f \geq 0
\label{eq:x9}
\end{equation}
As we will see, this puts a strong constraint on the allowed 
non-perturbative vacua.

\section{Number of Families and Model Building Rules}

The first obvious physical criterion for constructing realistic particle
physics models is that we should be able to find theories with a small 
number of families, preferably three. We will see that this is, in fact, 
easy to do via the bundle constructions on elliptically fibered 
Calabi--Yau three-folds that we are discussing. We start by deriving 
the three family criterion as discussed, for instance, in Green, Schwarz 
and Witten~\cite{gsw}.

The number of families is related to the number of zero-modes of the
Dirac operator in the presence of the gauge bundle on the Calabi--Yau
three-fold, since we want to count the number
of massless fermions of different chiralities. The original gauginos
are in the adjoint representation of $E_8$. In this lecture, we are considering 
only gauge bundles $V$ with $SU(n)$ fiber groups. To count the number of
families, we need to count the number of fields in the matter
representations of the low energy gauge group, that is, the subgroup of
$E_{8}$ commutant with $SU(n)$, and their complex conjugates
respectively. Explicitly, in this lecture, we will be 
interested in the following breaking patterns
\begin{equation}
\begin{aligned}
   E_8 \supset SU(3) \times E_6 : & \quad
      \mbf{248} = (\mbf{8},\mbf{1}) \oplus (\mbf{1},\mbf{78}) \oplus
           (\mbf{3},\mbf{27}) \oplus (\mbf{\bar{3}},\mbf{\bar{27}}) \\
   E_8 \supset SU(4) \times SO(10) : & \quad
      \mbf{248} = (\mbf{15},\mbf{1}) \oplus (\mbf{1},\mbf{45}) \oplus
           (\mbf{4},\mbf{16}) \oplus (\mbf{\bar{4}},\mbf{\bar{16}}) \\
   E_8 \supset SU(5) \times SU(5) : & \quad
      \mbf{248} = (\mbf{24},\mbf{1}) \oplus (\mbf{1},\mbf{24}) \oplus
           (\mbf{10},\mbf{5}) \oplus (\mbf{\bar{10}},\mbf{\bar{5}}) \oplus
           (\mbf{5},\mbf{\bar{10}}) \oplus (\mbf{\bar{5}},\mbf{10})
\end{aligned}
\label{subgroup} 
\end{equation}
Note, however, that the methods presented here will apply to any breaking
pattern with an $SU(n)$ subgroup.
We see that all the matter representations appear in the fundamental 
representation of the bundle group $SU(n)$. By definition, 
the index of the Dirac operator
measures the difference in the number of positive and  negative
chirality spinors, in this case, on the Calabi--Yau three-fold. Since
six-dimensional chirality is correlated with four-dimensional
chirality, the index gives the number of families. From the fact that
all the relevant fields are in the fundamental representation of
$SU(n)$, we have that the number of generations is
\begin{equation}
   N_{\text{gen}} = \text{index}\,(V,\Ds) 
      = \int_X \text{td}\,(X) \text{ch}\,(V)
      = \frac{1}{2}\int_X c_3(V)
\label{eq:x10}
\end{equation}
where $\text{td}\,(X)$ is the Todd class of $X$. For the case of
$SU(n)$ bundles on elliptically fibered Calabi--Yau three--folds, 
one can show, using equation~\eqref{c3n} above, that the number of 
families becomes
\begin{equation}
   N_{\text{gen}} = \l \eta ( \eta - nc_1(B) )
\label{eq:x11} 
\end{equation}
where we have integrated over the fiber. Hence, to obtain three families the
bundle must be constrained so that
\begin{equation}
  3 =  \lambda \eta \left( \eta - nc_1(B) \right)
\label{eq:x13} 
\end{equation}
It is useful to express this condition in terms of the class $W_{B}$ given in
equation (\ref{eq:x7}) and integrated over the fiber. We find that
\begin{equation}
 3= \lambda \left( W_{B}^{2}- (24-n)W_{B}c_{1}(B)
      + 12(12-n)c_{1}(B)^{2} \right)
\label{eq:x14}
\end{equation}
Furthermore, inserting the three family constraint into (\ref{eq:x8}) gives
\begin{equation}
   a_f = c_2(B) + \left(11 + \frac{1}{24}(n^3-n)\right) c_1(B)^2 
        - \frac{3n}{2\l}\left(\l^2-\frac{1}{4}\right)
\label{eq:x15}
\end{equation}

We are now in a position to summarize all the rules and constraints that are
required to produce particle physics theories with three families. The
conditions obtained in this section are 

\begin{itemize}

\item  The homology class associated with the five-branes is specifically of
the form 
\begin{equation}
   [W]= W_{B} + a_{f}F
\label{eq:x16}
\end{equation}
where
\begin{equation}
   W_{B}=\sigma( 12c_{1}(B)-\eta)
\label{eq:x17}
\end{equation}
\begin{equation}
   a_f = c_2(B) + \left(11 + \frac{1}{24}(n^3-n)\right) c_1(B)^2
        - \frac{3n}{2\l}\left(\l^2-\frac{1}{4}\right)
\label{eq:x18}
\end{equation}
and $c_{1}(B)$ and $c_{2}(B)$ are the first and second Chern
classes of $B$.

\item  The requirement that the five-brane curve be a true submanifold of $X$
constrains $[W]$ to be an effective class. Therefore, we must guarantee that
\begin{equation}
  W_{B} \text{ is effective in }B, \quad
   a_{f}\geq0  \text{ integer } 
 \label{eq:x19}
\end{equation}

\item  The condition that the theory have three families imposes the further
constraint that 
\begin{equation}
 3= \lambda \left( W_{B}^{2}- (24-n)W_{B}c_{1}(B)
     + 12(12-n)c_{1}(B)^{2} \right)
\label{eq:x20}
\end{equation}

\end{itemize}
To these conditions, we can add the remaining relevant constraint from
section $4$. It is

\begin{itemize}

\item  The condition that $c_{1}({\cal{N}})$ be an integer leads to the
constraints on $W_{B}$ and $\lambda$ given by
\begin{equation}
\begin{aligned}
   {} & n \text{ is odd}, \quad \l = m+\frac{1}{2} \\
   {} & n \text{ is even}, \quad \l = m, \quad 
          W_B = {c_1(B) \mod 2}
\end{aligned}
\label{eq:a21}
\end{equation}
where $m$ is an integer. The $n$ even condition is sufficient, but not
necessary.

\end{itemize}
Note that in this last condition, the class $\eta$, which appeared in 
constraint~\eqref{eq:add19}, has been replaced by $W_{B}$. That this replacement 
is valid can be seen
as follows. For $n$ odd, there is no constraint on $\eta$ and, hence, using
~\eqref{eq:x17}, no constraint on $W_{B}$. When $n$ is even, it is sufficient
for $\eta $ to satisfy
$\eta=c_{1}(B) \mod 2$. Since $12c_{1}(B)$ is an even element of
$H^{2}(B, \bf Z \rm)$, it follows that $W_{B} = c_1(B) \mod 2$.

It is important to note that all quantities and constraints have now been
reduced to properties of the base two-fold $B$. Specifically, if we know 
$c_1(B)$, $c_2(B)$, as well as a set of generators of 
effective classes in $B$ in which to expand
$W_{B}$, we will be able to exactly specify all appropriate non-perturbative
vacua. For the del Pezzo, Hirzebruch, Enriques and blown-up Hirzebruch
surfaces, all of these quantities are known.

Finally, from the expressions in ~\eqref{subgroup} we find the following rule.

\begin{itemize}
 
\item  If we denote by $G$ the structure group of the gauge bundle and by $H$ its
commutant subgroup, then
\begin{displaymath}
 G=SU(3) \Longrightarrow  H=E_{6}
\end{displaymath}
\begin{equation}
 G=SU(4) \Longrightarrow  H=SO(10)
\label{eq:x24}
\end{equation}
\begin{displaymath}
 G=SU(5) \Longrightarrow  H=SU(5)
\end{displaymath}
$H$ corresponds to the low energy gauge group of the theory.

\end{itemize}
Armed with the above rules, we now turn to the explicit construction of
phenomenologically relevant non-perturbative vacua.

\section{Three Family Models}

In this section, we will construct two explicit solutions satisfying the
above rules.
In general, we will look for solutions where the class representing the
curve on which the fivebranes wrap is comparatively simple. As discussed
above, the allowed base surfaces $B$ of elliptically fibered Calabi--Yau
three--folds which admit a section are restricted to be the del Pezzo,
Hirzebruch and Enriques surfaces, as well as blow-ups of Hirzebruch surfaces.
Relevant properties of del Pezzo, Hirzebruch and Enriques surfaces, 
including their generators of effective curves, are given in the Appendix B of
ref.~\cite{usnew}. 
However, we now show that Calabi--Yau three--folds of this type  
with an Enriques base
never admit an effective five-brane curve if one requires that there be three
families. Recall that the cohomology class of the spectral cover must be of
the form
\begin{equation}
 [C]=n\sigma+ \eta
 \label{home1}
\end{equation}
and this necessarily is an
effective class in $X$. We may assume that $C$ does not contain $\sigma(B)$.
Otherwise, replace $C$ in the following discussion with its subcover $C'$
obtained by discarding the appropriate multiples of $\sigma(B)$.
This implies that the class
\begin{equation}
 \sigma [C]=n\sigma^{2}+ \sigma\eta
 \label{home2}
\end{equation}
must be effective in the base $B$. Let us restrict $B$ to be an Enriques
surface. Using the adjunction formula, we find that
\begin{equation}
 \sigma^{2}=K_{B}
 \label{home3}
\end{equation}
where $K_{B}$ is the torsion class. Since $nK_{B}$ vanishes for even $n$, it
follows that when $n$ is even
\begin{equation}
 \sigma [C]= \sigma\eta
 \label{home4}
\end{equation}
Clearly, $\sigma\eta$ is effective, since $\sigma[C]$ is. For $n$ odd,
$nK_{B}=K_{B}$ and, hence
\begin{equation}
 \sigma [C]= K_{B}+ \sigma\eta
 \label{home5}
\end{equation}
Using the discussion in Appendix B of~\cite{usnew}, one 
can still conclude that $\sigma\eta$
is either an effective class or it equals $K_{B}$. From the fact that
\begin{equation}
 \sigma c_{1}(B)= K_{B}
 \label{home6}
\end{equation}
it follows, using equation ~\eqref{eq:x17}, that the five-brane class
restricted to the Enriques base is given by
\begin{equation}
 W_{B}=12 K_{B}- \sigma\eta
 \label{home7}
\end{equation}
Since $12K_{B}$ vanishes, this becomes
\begin{equation}
 W_{B}=- \sigma\eta
 \label{home8}
\end{equation}
from which we can conclude that $W_{B}$ is never effective for non-vanishing
class $\sigma \eta$. Since, as explained above, $W_{B}$ must be effective for
the five-branes to be physical, such theories must be discarded. The only
possible loop-hole is when  $\sigma \eta$ vanishes or equals $K_{B}$. 
However, in this case, it
follows from~\eqref{eq:x11} that
\begin{equation}
 N_{\text{gen}} =0
 \label{home9}
\end{equation}
which is also physically unacceptable. We conclude that, on general grounds,
Calabi--Yau three--folds with an Enriques base never admit effective
five--brane curves if one requires that there be three families
\footnote{We thank E. Witten for pointing out to us the likelihood of this
conclusion.}. For this
reason, we henceforth restrict our discussion to the remaining possibilities.
In this lecture, for specificity, the base
$B$ will always be chosen to be a del Pezzo surface. 

We first give two $SU(5)$ examples, each on del Pezzo surfaces; one
where the base component, $W_{B}$, is simple and one where the fiber component
has a small coefficient.

\subsection*{Example 1: $B=dP_{8}$, $H=SU(5)$}

We begin by choosing 
\begin{equation}
H=SU(5)
\label{eq:x25}
\end{equation} 
as the gauge group for our model. Then it follows from (\ref{eq:x24}) that we must 
choose the structure group of the gauge bundle to be
\begin{equation}
G=SU(5)
\label{eq:x26}
\end{equation}
and, hence, $n=5$. Since $n$ is odd, constraint~\eqref{eq:a21} tells us that
$\lambda=m+\frac{1}{2}$ for integer m. Here we will, for simplicity, choose
$m=1$ and, therefore
\begin{equation}
\lambda=\frac{3}{2}
\label{eq:x27}
\end{equation}
At this point, it is necessary to explicitly choose the base surface, which we
take to be
\begin{equation}
B=dP_{8}
\label{eq:x28}
\end{equation}
It follows from Appendix B of ref.~\cite{usnew} 
that for the del Pezzo surface $dP_{8}$, a basis
for $H_{2}(dP_{8}, \bf Z \rm)$ composed entirely of effective classes 
is given by $l$ and $E_{i}$ for $i=1,..,8$ where
\begin{equation}
l \cdot l=1  \qquad  l \cdot E_{i}=0  \qquad E_{i} \cdot E_{j}=-\delta_{ij}
\label{eq:x29}
\end{equation}
There are other effective classes in $dP_{8}$ not obtainable as a linear
combination of $l$ and $E_{i}$ with non-negative integer coefficients, but we
will not need them in this example.
To these we add the fiber class $F$. Furthermore
\begin{equation}
c_{1}(B)= 3l- \sum_{r=1}^{8} E_{i}
\label{eq:x31}
\end{equation}
and
\begin{equation}
c_{2}(B)= 11
\label{eq:x32}
\end{equation}
We now must specify the component of the five-brane class in the base. 
In this example, we choose
\begin{equation}
W_{B}= 2E_{1}+E_{2}+E_{3}
\label{eq:x33}
\end{equation}
Since $E_{1}$, $E_{2}$ and $E_{3}$ are effective, it follows that $W_{B}$ is
also effective, as it must be. Using the above intersection rules, one can
easily show that
\begin{equation}
W_{B}^{2}=-6,  \qquad  W_{B}c_{1}(B)=4,  \qquad
c_{1}(B)^{2}=1
\label{eq:x34}
\end{equation}
Using these results, as well as $n=5$ and $\lambda=\frac{3}{2}$, one can check
that
\begin{equation}
\lambda( W_{B}^{2}- (24-n)W_{B}c_{1}(B)+12(12-n)c_{1}(B)^{2})=3
\label{eq:x35}
\end{equation}
and, therefore, the three family condition is satisfied. Finally, let us
compute the coefficient $a_{f}$ of $F$. Using the above information, we find
that
\begin{equation}
   a_f = c_2(B) + \left(11 + \frac{n^3-n}{24}\right) c_1(B)^2 
        - \frac{3n}{2\l}\left(\l^2-\frac{1}{4}\right) = 17
\label{eq:x36}
\end{equation}
Since this is a positive integer, it follows from the above discussion that
the full five-brane curve $[W]$ is effective in the Calabi--Yau 
three--fold $X$, as it must be. This completes
our construction of this explicit non-perturbative vacuum. It represents a
model of particle physics with three families and gauge group $H=SU(5)$, along
with  explicit five-branes wrapped on a holomorphic curve specified by 
\begin{equation}
[W]=2E_{1}+E_{2}+E_{3} + 17F
\label{eq:x37}
\end{equation}
The properties of the moduli spaces of five-branes were discussed in
\cite{don1, usnewnew}.

\subsection*{Example 2: $B=dP_{8}$, $H=SO(10)$}

As a second example, we choose the gauge group to be
\begin{equation}
H=SO(10)
\label{eq:x49}
\end{equation} 
and, hence,  the structure group 
\begin{equation}
G=SU(4)
\label{eq:x50}
\end{equation}
Then $n=4$. Since $n$ is even, then from constraint~\eqref{eq:a21} we must
have $\lambda=m$ where $m$ is an integer and $W_{B} = c_1(B)\mod 2$. Here we
will choose $m=-1$ so that
\begin{equation}
\lambda=-1
\label{eq:x51}
\end{equation}
We will return to the choice of $W_{B}$ momentarily.
In this example, we will take as a base surface
\begin{equation}
B=dP_{8}
\label{eq:x52}
\end{equation}
Some of the effective generators and the first and second Chern classes of $dP_{8}$
were given in the previous example.
We now must specify the component of the five-brane class in the base. 
In this example, we choose
\begin{equation}
W_{B}= 2E_{1}+2E_{2}+(3l-\sum_{i=1}^{8}E_{i})
\label{eq:x53}
\end{equation}
Since $E_{1}$, $E_{2}$ and $3l-\sum_{i=1}^{8}E_{i}$ are effective, 
it follows that $W_{B}$ is
also effective, as it must be. Furthermore, since 
\begin{equation}
c_{1}(B)= 3l- \sum_{r=1}^{8} E_{i}
\label{eq:x54}
\end{equation}
it follows that 
\begin{equation}
W_{B}= c_{1}(B) \mod 2  
\label{eq:x55}
\end{equation}
since $2E_{1}+2E_{2}$ is an even element of $H^{2}(dP_{9}, \bf Z \rm)$. 
Using the above intersection rules, one can
easily show that
\begin{equation}
W_{B}^{2}=1,  \qquad  W_{B}c_{1}(B)=5,  \qquad
c_{1}(B)^{2}=1
\label{eq:x56}
\end{equation}
Using these results, as well as $n=4$ and $\lambda=-1$, one can check that
\begin{equation}
\lambda \left( W_{B}^{2}- (24-n)W_{B}c_{1}(B)
    +12(12-n)c_{1}(B)^{2} \right) = 3
\label{eq:x57}
\end{equation}
and, therefore, the three family condition is satisfied. Finally, let us
compute the coefficient $a_{f}$ of $F$. Using the above information, we find
that
\begin{equation}
   a_f = c_2(B) + \left(11 + \frac{n^3-n}{24}\right) c_1(B)^2 
        - \frac{3n}{2\l}\left(\l^2-\frac{1}{4}\right) = 29
\label{eq:x58}
\end{equation}
Since this is a positive integer, it follows from the above discussion that
the full five-brane curve $[W]$ is effective, as it must be. This completes
our construction of this explicit non-perturbative vacuum. It represents a
model of particle physics with three families and gauge group $H=SO(10)$, along
with  explicit five-branes wrapped on a holomorphic curve specified by 
\begin{equation}
[W]=2E_{1}+2E_{2}+(3l-\sum_{i=1}^{8}E_{i}) + 29F
\label{eq:59}
\end{equation}

\vspace{0.4cm}


\section*{Acknowledgments:} Supported in part by the DOE under contract 
No. DE-AC02-76-ER-03071. 




\begin{thebibliography}{99}
\bibitem{hw1} P. Ho\v rava and E. Witten, {\em Nucl. Phys.} {\bf B460}
    (1996) 506.
\bibitem{hw2} P. Ho\v rava and E. Witten, {\em Nucl. Phys.} {\bf B475}
    (1996) 94.
\bibitem{w} E. Witten, {\em Nucl. Phys.} {\bf B471} (1996) 135.

\bibitem{losw} A. Lukas, B.~A. Ovrut, K.S. Stelle and D. Waldram,
               {\em The Universe as a Domain Wall}, UPR-797T,
               hep-th/9803235.
\bibitem{add1} A. Lukas, B.Ovrut, K. Stelle and D. Waldram, {\em 
               Heterotic M-theory in Five Dimensions}, hep-th/9806051.

\bibitem{us}  A. Lukas, B.~A. Ovrut and D. Waldram,
    {\em Non--Standard Embedding and Five--Branes in Heterotic M--Theory}, 
    hep-th/9808101.
\bibitem{don1} R. Donagi, A. Lukas, B.A. Ovrut and D. Waldram, {\em
    Non-Perturbative Vacua and Particle Physics in M-Theory},  hep-th/9811168.
\bibitem{usnew}R. Donagi, A. Lukas, B.A. Ovrut and D. Waldram, {\em
    Holomorphic Vector Bundles and Non-Perturbative Vacua in M-Theory},  
    hep-th/9901009. 
\bibitem{usnewnew} R. Donagi, B.A. Ovrut and D. Waldram,{\em
     Moduli Spaces of Fivebranes on Elliptic Calabi-Yau Threefolds},  
     hep-th/hep-th/9904054.
\bibitem{bd} T. Banks and M. Dine, {\em Nucl. Phys.} {\bf B479} (1996)
173.
\bibitem{CYred} A.~C. Cadavid, A. Ceresole, R. D'Auria and S. Ferrara,
                {\em Phys. Lett.} {\bf B357} {1995} 76;
                I. Antoniadis, S. Ferrara and T.~R. Taylor, {\em Nucl. Phys.}
                {\bf B460} (1996) 489.
\bibitem{paschos} B.~A. Ovrut, {\em Talk given at PASCOS-98}.
\bibitem{low1} A. Lukas, B.~A. Ovrut and D. Waldram, {\em On the 
    Four-Dimensional Effective Action of Strongly Coupled Heterotic
    String Theory}, UPR-771T, hep-th/9710208.
\bibitem{hor} P. Ho\v rava, {\em Phys. Rev.} {\bf D54} (1996) 7561.
\bibitem{aq1} I. Antoniadis and M. Quir\'os, {\em Phys. Lett.} {\bf B392}
    (1997) 61.
\bibitem{kap} E.~Caceres, V.~S. Kaplunovsky and I.~M. Mandelberg, {\em
    Nucl. Phys.} {\bf B493} (1997) 73.
\bibitem{ns} H.-P. Nilles and S. Stieberger, {\em Nucl. Phys.} {\bf B499}
             (1997) 3.
\bibitem{lln} T. Li, J.~L. Lopez and D.~V. Nanopoulos, {\em Mod. Phys. Lett.}
    {\bf A12} (1997) 2647.
\bibitem{conrad} J.~O.~Conrad, {\em Brane Tensions and Coupling
    Constants from within M--Theory}, TUM-HEP-289-97, hep-th/9708031.
\bibitem{du1} E. Dudas and C. Grojean, {\em Nucl. Phys.} {\bf B507}
    (1997) 553.
\bibitem{dm} E. Dudas and J. Mourad, {\em Phys. Lett.} {\bf B400} 
    (1997) 71.
\bibitem{llo} Z. Lalak, A. Lukas and B.~A. Ovrut, {\em Soliton Solutions
    of M--theory on an Orbifold}, UPR-774T, hep-th/9709214, to
    be published in Phys. Lett. B.
\bibitem{choi} K. Choi, {\em Phys. Rev.} {\bf D56} (1997) 6588.
\bibitem{hp} H.~P. Nilles, M. Olechowski and M. Yamaguchi, {\em Phys. Lett.}
             {\bf B415} (1997) 24.
\bibitem{aq2} I. Antoniadis and M. Quir\'os, {\em Phys. Lett.} {\bf B416}
              (1998) 327.
\bibitem{lt} Z. Lalak and S. Thomas, {\em Gaugino Condensation, Moduli
    Potential and Supersymmetry Breaking in M--theory Models},
    QMW-PH-97-23, hep-th/9707223.
\bibitem{du2} E. Dudas, {\em Supersymmetry breaking in M--theory and
    Quantization Rules}, CERN-TH/97-236, hep-th/9709043.
\bibitem{efn} J. Ellis, A.~E. Faraggi and D.~V. Nanopoulos, {\em M--theory
    Model-Building and Proton Stability}, CERN-TH-97-237,
    hep-th/9709049.
\bibitem{nan} D.~V.~Nanopoulos, {\em M Phenomenology}, CTP-TAMU-42-97, 
    hep-th/9711080.
\bibitem{ckm} K. Choi, H.~B. Kim and C. Mu\~noz, {\em Four--Dimensional
              Effective Supergravity and Soft Terms in M--Theory},
              KAIST-TH 97/20, FTUAM 97/16, hep-th/9711158.
\bibitem{low2}  A. Lukas, B.~A. Ovrut and D. Waldram, {\em Gaugino
    Condensation in M--theory on $S^1/Z_2$}, UPR-785T, hep-th/9711197.
\bibitem{noy} H.~P. Nilles, M. Olechowski and M. Yamaguchi,
              {\em Supersymmetry Breakdown at a Hidden Wall}, BONN-TH-97-10,
              hep-th/9801030.
\bibitem{low3} A. Lukas, B.~A. Ovrut and D. Waldram, {\em The Ten-dimensional
               Effective Action of Strongly Coupled Heterotic String Theory},
               UPR-789-T, hep-th/9801087.
\bibitem{li} T. Li, {\em Compactification and Supersymmetry Breaking in
             M Theory}, hep-th/9801123.
\bibitem{mike} M. Faux, {\em New Consistent Limits of M Theory}, HUB-EP-98-9,
               hep-th/9801204.
\bibitem{ms} P. Majumdar and S. SenGupta, {\em Gaugino Mass in the Heterotic
             String with Scherk--Schwarz Compactification}, SINP-98-05,
             hep-th/9802111.
\bibitem{nw} N. Wyllard, {\em The Heterotic Prepotential from Eleven
             Dimensions}, GOTEBORG-ITP-98-01, hep-th/9802148. 
\bibitem{bkl} D. Bailin, G.V. Kraniotis and A. Love, {\em Sparticle
              Spectrum and Dark Matter in M Theory}, SUSX-TH-98-003,
              hep-ph/9803274.
\bibitem{sharpe} E. Sharpe, {\em Boundary Superpotentials}, PUPT-1660,
                 hep-th/9611196.
\bibitem{pes} E.~A. Mirabelli and M.~E. Peskin, {\em Transmission of
              Supersymmetry Breaking from a Four--Dimensional Boundary},
              SLAC-PUB-7681, hep-th/9712214.
\bibitem{lo} A. Lukas and B.~A. Ovrut, {\em U--duality Invariant M--theory
             Cosmology}, UPR-770-T, hep-th/9709030, to be published in Phys.
             Lett. B.
\bibitem{benakli2} K. Benakli, {\em Cosmological Solutions in M-theory
                  on $S^1/Z_2$}, hep-th/9804096.
\bibitem{benakli} K. Benakli, 
    {\em Scales and cosmological applications of M theory}, 
    CTP-TAMU-20-98, hep-th/9805181.
\bibitem{low4}  A. Lukas, B.~A. Ovrut and D. Waldram, {\em
    Cosmological Solutions of Ho\v rava-Witten Theory},  UPR-805T,
    hep-th/9806022.
\bibitem{elpp} J. Ellis, Z. Lalak, S. Pokorski and W. Pokorski,
               {\em Five--dimensional Aspects of M--theory and Supersymmetry
               Breaking}, CERN-TH-98-118, hep-ph/9805377.
\bibitem{c0} E. Cremmer, in {\em Superspace and Supergravity}, ed. S.~W.
             Hawking and M. Rocek, pp.~267--282,
             Cambridge University Press (1981).

\bibitem{deA} S.~P.~de~Alwis, {\em Phys. Lett.} {\bf B388} (1996) 291;
    {\em Phys. Lett.} {\bf B392} (1997) 332.
\bibitem{bcf} M. Bodner, A.~C. Cadavid and S. Ferrara, {\em Class. Quantum
              Grav.} {\bf 8} (1991) 789.

 \bibitem{gsw} M.~B. Green, J.~H. Schwarz and E. Witten, {\em Superstring
              Theory}, vol.~2, pp.~386, Cambridge University Press (1987).
\bibitem{wittq} E. Witten, {\em J. Geom. Phys.} {\bf 22} (1997) 1.
\bibitem{llp} I.~V. Lavrinenko, H. L\"u and C.~N. Pope, {\em Nucl. Phys.}
              {\bf B492} (1996) 278.
\bibitem{GST1} M. G\" unaydin, G. Sierra and P.K. Townsend, {\em Nucl. Phys.}
               {\bf B242} (1084) 244.
\bibitem{GST2} M. G\" unaydin, G. Sierra and P.K. Townsend, {\em
               Nucl. Phys.} {\bf B253} (1985) 573.
\bibitem{Sierra} G. Sierra, {\em Phys. Lett.} {\bf 157B} (1985) 379.
\bibitem{andetal} L. Andrianopoli {\em et al.}, {\em Nucl. Phys.}
                 {\bf B476} (1996) 397.
\bibitem{quat} S. Cecotti, S. Ferrara and L. Girardello, {\em Phys. Lett.}
               {\bf B213} (1988) 443; {\em Int. J. Mod. Phys.} {\bf A4}
               (1989) 2475.
\bibitem{strom} A. Strominger, {\em Loop Corrections to the Universal
                Hypermultiplet}, UCSBTH-97-13, hep-th/9706195.

\bibitem{sp} J. Polchinski and A. Strominger, {\em Phys. Lett.} 
             {\bf B388} (1996) 736.
\bibitem{michelson} J. Michelson, {\em Nucl. Phys.} {\bf B495} (1997) 127.
\bibitem{8brane} E. Bergshoeff, M. de Roo, M.~B. Green, G. Papadopoulos
                 and P.~K. Townsend, {\em Nucl. Phys.} {\bf B470} (1996) 113.
\bibitem{romans} L.~J. Romans, {\em Phys. Lett.} {\bf B169} (1986) 374.
\bibitem{dom} H. Lu, C.~N. Pope, E. Sezgin and K.~S. Stelle,
              {\em Phys. Lett.} {\bf B371} (1996) 46;              
              P.~M. Cowdall, H. Lu, C.~N. Pope, K.~S. Stelle and P.~
              K. Townsend, {\em Nucl. Phys.} {\bf B486} (1997) 49.
\bibitem{cfg} S. Cecotti, S. Ferrara and L. Girardello, {\em Phys. Lett}
              {\bf B213} (1988) 443.
\bibitem{mr} J. March-Russell, {\em The Fayet-Iliopoulos term in
    Type-I string theory and M-theory}, IASSNS-HEP-97/128,
    hep-ph/9806426. 
\bibitem{BDDR} P. Binetruy, C. Deffayet, E. Dudas and P. Ramond,
    {\em Pseudo-anomalous $U(1)$ symmetry in the strong coupling limit
    of the heterotic string}, LPTHE-ORSAY 98/41, hep-th/9807079.
\bibitem{w0} E. Witten, {\em Nucl. Phys.} {\bf B268} (1986) 79.
\bibitem{ww} L. Witten and E. Witten, {\em Nucl. Phys.} {\bf B281}
    (1987) 109.
\bibitem{bbs} K. Becker, M. Becker and A. Strominger, {\em Nucl. Phys.}
    {\bf B456} (1995) 130. 
\bibitem{vb} M. Bershadsky, V. Sadov and C. Vafa, {\em Nucl. Phys.}
    {\bf B463} (1996) 420.
\bibitem{DMW} M. J. Duff, R. Minasian and E. Witten, {\em Nucl. Phys.}
    {\bf B465} (1996) 413.
 \bibitem{GH} see for instance, P. Griffiths and J. Harris, {\em
    Principles of Algebraic Geometry}, John Wiley and Sons, New York
   (1978). 

\bibitem{GT} G. W. Gibbons and P. K. Townsend, {\em Phys. Rev. Lett.} 
    {\bf 71} (1993) 3754.
\bibitem{KM} D. M. Kaplan and J. Michelson, {\em Phys. Rev.} 
    {\bf D53} (1996) 3474.
\bibitem{BLNPST} I. Bandos {\em et al.}, {\em Phys. Rev. Lett.} 
    {\bf 78} (1997) 4332.
\bibitem{APPS} M. Aganagic {\em et al.}, {\em Nucl. Phys.} 
    {\bf B496} (1997) 191.
\bibitem{Kallosh} R. Kallosh, {\em Phys. Rev.} {\bf D57} (1998) 3214.
\bibitem{CKvP} P. Claus, R.~Kallosh and A.~Van~Proeyen, 
    {\em Nucl. Phys.} {\bf B518} (1998) 117.
\bibitem{wbranes} E. Witten, {\em Nucl. Phys.} {\bf B500} (1997) 3.
\bibitem{KLMVW} A. Klemm {\em et al.}, {\em Nucl. Phys.} {\bf B477}
    (1996) 746.
\bibitem{stieb} S. Stieberger, 
    {\em (0,2) heterotic gauge couplings and their M theory origin}, 
    CERN-TH-98-228, hep-th/9807124.

\bibitem{FMW} R. Friedman, J. Morgan and E. Witten,
   \textit{Commun. Math. Phys.} \textbf{187}, 679 (1997).
\bibitem{D} R. Donagi, {\em Asian. J. Math.}{\bf 1}, (1997)214.
\bibitem{BJPS} M. Bershadsky, A. Johansen, T. Pantev and V. Sadov, {\em Nuc.
   Phys.} {\bf B505} (1997) 165.

\bibitem{cur} G. Curio, {\em Phys. Lett.}{\bf B435}, (1998)39.
\bibitem{ba} B. Andreas, {\em On Vector Bundles and Chiral Matter in $N=1$
   Heterotic Compactifications}, hep-th/9802202.
\bibitem{Don} S. Donaldson, {\em Proc. London Math. Soc.}{\bf 3} (1985)1.
\bibitem{UhYau} K. Uhlenbeck and S.-T. Yau, {Comm. Pure App. Math.}{\bf 39}
(1986)257, {\bf 42} (1986)703.
\bibitem{wod2} R. Donagi, B. Ovrut and D. Waldram, manuscript in preparation.
\bibitem{MV} D. R. Morrison and C. Vafa,
    \textit{Nucl. Phys.} \textbf{B476} 437 (1996).
\bibitem{Looij} E. Looijenga, {\em Invent. Math.}{\bf 38} (1977)17, {\bf 61}
   (1980)1.







\end{thebibliography}
\end{document}